\documentclass[11pt, oneside]{amsart}   	
\usepackage{amsaddr}
\usepackage{geometry}                		
\usepackage[parfill]{parskip}    		
\usepackage{graphicx}				
\usepackage{amssymb}
\usepackage{amsmath}
\usepackage{amsfonts}
\usepackage{mathbbol}
\numberwithin{equation}{section}
\newcommand\abs[1]{\left|#1\right|}

\title{Mass gap in weakly coupled Abelian Higgs on a unit lattice}
\author{Abhishek Goswami}
\address{Department of Mathematics, SUNY at Buffalo, Buffalo, NY 14260, USA}
\curraddr{}
\email{goswami3@buffalo.edu}

\date{\today}							

\begin{document}

\begin{abstract}
The proof of Higgs mechanism in a weakly coupled lattice gauge theory in $d \geqslant 2$ is revisited. 
A new power series cluster expansion is applied and the mass gap is shown to exist for the observable $F_{\mu\nu}$.
\end{abstract}
\maketitle

\section{Introduction}

We study Higgs field weakly coupled to an Abelian gauge field on a Euclidean unit lattice and establish mass gap. 
In Standard Model, the presence of Higgs field is responsible for the mass generation of $W^{\pm}$ and \textit{Z} gauge bosons 
as well as fermions. This principle is known as Higgs mechanism. The recent discovery of a Higgs like particle at LHC thus,
confirms this essential feature of the Standard Model. 
We demonstrate Higgs mechanism via exponential decay of correlations between physical observable
$F_{\mu\nu}$ (electromagnetic field strength). This decay implies that the gauge boson in the theory has acquired mass
and the theory is said to have a mass gap. 
 
Mass generation is a long distance problem and thus, we want to study it without worrying about the ultraviolet (short distance) problem.
We develop on the work of Balaban, Imbrie, Jaffe and Brydges \cite{BBIJ}. We apply a new power series cluster expansion developed by
Balaban, Feldman, Kn$\ddot{\text{o}}$rrer and Trubowitz \cite{BFKT}. This is different from the decoupling expansion of \cite{BBIJ}.
 
Our results fit nicely in the program began by Balaban, Imbrie and Jaffe \cite{BIJ2},\cite{BIJ3} to develop a complete analysis
of the ultraviolet problem. They employ a block averaging renormalization group technique which eventually takes us from an
$\varepsilon = L^{-N}$ lattice, where \textit{L} is a positive number and \textit{N} is a positive integer, to a unit lattice,
similar to our model.

The strong coupling regime of lattice gauge theory is widely studied. However, the weak coupling is more relevant at least in $d = 3$
to understand the continuum limit. 
 
\textit{Notation}. Let $\Lambda \subset \mathbb{Z}^{d}$ with $d \geqslant 2$
be a large finite unit lattice of dimension $d$. The Higgs doublet $\phi (x) = \phi_{1}(x) + i \phi_{2}(x)$ is defined on a site $x \in \Lambda$.
A bond $b \in \Lambda$ links two adjacent sites $(x, x + e_{\mu})$. $\Lambda^{\ast}$ denotes the set of all bonds $b \in \Lambda$.
The gauge field $A_{b} \equiv A_{\mu}(x)$ is defined on the bond between $(x, x + e_{\mu})$.
$A_{\mu}(x)$ is the generator of the Abelian group U(1) such that $\text{U}(x+e_{\mu}, x) = e^{i e_{0} A_{\mu}(x)}$
is a phase factor a particle acquires while going from $x$ to $x+e_{\mu}$. $e_{0}$ is the gauge coupling and   
$A_{\mu}(x) \in [\frac{- \pi}{e_{0}}, \frac{\pi}{e_{0}}]$ (compact or C), 
$A_{\mu}(x) \in \mathbb{R}$ (non compact or NC).  A plaquette $p \in \Lambda$ is a unit square formed by
the bonds $(x, x+e_{\nu}), (x+e_{\nu}, x+e_{\mu}+e_{\nu}), (x+e_{\mu}+e_{\nu}, x+e_{\mu})
\hspace{0.05 cm} \text{and} \hspace{0.05 cm} (x+e_{\mu}, x)$. $\Lambda^{\ast\ast}$ denotes the set of all plaquettes $p \in \Lambda$.  

\textit{Action}. For our gauge field action we use non compact formalism,
\begin{equation}
S_{NC}(A) =\frac{1}{2} \sum_{p \subset \Lambda^{\ast\ast}}\hspace{0.1cm} (dA)^{2}(p). 
\end{equation}
 $(dA)(p) = \sum_{b \in \partial p} A_{b}$  is the field strength. 
The action for minimally coupled Higgs field with classical vacuum energy scaled to zero is given by
\begin{equation}
S_{h}(\phi,A) = \frac{1}{2} \sum_{\langle xy\rangle \subset \Lambda} |e^{i e_{0}A_{\langle xy\rangle}}\phi(y) - \phi(x)|^{2}
  + \sum_{x \in \Lambda} \hspace{0.1 cm}\big(\lambda|\phi(x)|^{4} - \frac{1}{4} \mu^{2}|\phi(x)|^{2} + E\big).
\end{equation}
The weak coupling regime is $(\lambda, e_{0}^{2}) \ll 1$ with $\frac{e_{0}^{2}}{\lambda} = \mathcal{O}(1)$. 
The total action for a non compact weakly coupled Abelian Higgs theory is 
\begin{equation}
S(\phi,A) = S_{NC}(A) + S_{h}(\phi, A).
\end{equation}
For a theory defined by action $S(\phi, A)$, the partition function is given by 
\begin{equation}
\mathrm{Z}^{\text{NC}} = \int_{\mathbb{R}} \mathcal{D}\phi \mathcal{D}A \hspace{0.1 cm} e^{- S(\phi, A)}
\end{equation}
where, $\mathcal{D}\phi = \prod_{x\in \Lambda} d\phi_{1}(x) d\phi_{2}(x)$ and $\mathcal{D}A = \prod_{b \in \Lambda^{\ast}} dA_{b}$ 
are product Lebesgue measure on $\mathbb{R}$. 

\textbf{Proposition 1} \cite{BBIJ} (\textit{Gauge fixing}) Let $\theta : \Lambda \rightarrow \mathbb{R}$ be the gauge function
and $\mathcal{D}\theta_{\Lambda} = \prod_{x \in \Lambda} d\theta_{x}$ denote product Lebesgue measure on $\mathbb{R}$.
Introduce a gauge fixing function $G(A)$ satisfying
\begin{equation}
\int e^{- G(A + d\theta)}\mathcal{D}\theta_{\Lambda \backslash x_{0}} = 1
\end{equation}
with $\theta (x_{0}) = 0$ and $d\theta = 0$ on boundary $\partial \Lambda$. 
Let $G(A)$ be given by
\begin{equation}
G(A) = \sum_{x \in \Lambda} \frac{\alpha}{2} (\delta A)^{2} (x) + c
\end{equation}
$\delta$ is the adjoint of differential operator \textit{d} with respect to $l^{2}$ inner product,
then \textit{c} is a finite constant independent of \textit{A}.

\textit{Proof} It is easy to see that 
$c = \text{ln}\hspace{0.05 cm} \int e^{- \frac{\alpha}{2} ||\delta(A + d\theta)||^{2}} \mathcal{D}\theta_{\Lambda \backslash x_{0}}$
satisfies the required condition. To show that \textit{c} is finite and independent of \textit{A} let $\Delta = \delta d$. Then $\Delta$
is invertible on the orthogonal complement of all constants. Make a change of variable as $\omega = \Delta \theta$ and rewrite
\begin{equation}
c = \text{ln}\hspace{0.05 cm} \int e^{- \frac{\alpha}{2} ||\delta A + \omega||^{2}} \mathcal{D}\omega + \text{const.}
\end{equation}
Then the translation $\omega \rightarrow \omega - \delta A$ completes the proof.

Gauge fixing makes the non compact integral in Eq 1.4 converge. Since addition of gauge fixing term changes 
$\int \mathcal{D}A \hspace{0.05 cm} e^{-\langle dA, dA\rangle}$ to $\int \mathcal{D}A \hspace{0.05 cm} e^{-\langle A, \Delta A\rangle}$, 
where, due to Dirichlet boundary condition, $\Delta$  is positive definite, 
$\langle A, \Delta A\rangle \geqslant (\text{const}) ||A||^{2}$, which makes the integral converge. 
With the above gauge fixing function the non compact partition function is defined as
\begin{equation}
\mathrm{Z}^{\text{NC}} = \int_{\mathbb{R}} \mathcal{D}\phi  \hspace{0.05 cm}\mathcal{D}A \hspace{0.1 cm} e^{- S(\phi, A) - G(A)}.
\end{equation}
As action $S(\phi, A)$ and the Lebesgue measures are gauge invariant 
we replace $e^{-G(A)}$ in the partition function by its gauge average over $\theta$. Define the compact gauge average of $e^{-G(A)}$ as
\begin{equation}
\langle\langle e^{-G(A)} \rangle\rangle =  \Big(\frac{2\pi}{e_{0}}\Big)^{-|\Lambda|}
\int^{\frac{\pi}{e_{0}}}_{-\frac{\pi}{e_{0}}} \mathcal{D}\theta_{\Lambda} \hspace{0.1 cm} e^{-G(A + d\theta)}.
\end{equation}
Let J be a function defined on plaquettes which we allow to be complex and assume $|\text{J}| < 1$. 
Consider the observable $e^{-e_{0} \langle dA, \text{J}\rangle}$, where, $\langle dA , \text{J}\rangle$ denotes $l^{2}$ inner product.
The non compact expectation of an observable $e^{-e_{0} \langle dA, \text{J}\rangle}$ is defined as
\begin{equation}
\mathrm{Z}^{\text{NC}}\hspace{0.05 cm} \langle e^{-e_{0} \langle dA, \text{J}\rangle} \rangle^{\text{NC}} = 
\int_{\mathbb{R}} \mathcal{D}\phi  \hspace{0.05 cm}\mathcal{D}A \hspace{0.1 cm} 
e^{- S(\phi, A) -e_{0} \langle dA, \text{J}\rangle} \langle\langle e^{-G(A)} \rangle\rangle.
\end{equation}
 
\textit{Compact formalism}. Our non compact formalism is equivalent to a compact formalism. For the compact, 
we do not use Wilson action which is
\begin{equation}
S_{C} (A) = \frac{1}{2} \sum_{p \subset \Lambda^{\ast\ast}}\hspace{0.1cm} \text{cos}\hspace{0.05 cm} [(dA)(p)] + \text{const}
\hspace{0.4 cm} \text{with} \abs{A_{b}} \leqslant \frac{\pi}{e_{0}}.
\end{equation}
Instead we use Villain type which is
\begin{equation}
e^{- S_{C}(A)} = \sum_{v : dv=0} e^{-\frac{1}{2} \sum_{p \in \Lambda^{\ast\ast}} (dA(p) + v(p))^{2}}
\hspace{0.4 cm} \text{with} \abs{A_{b}} \leqslant \frac{\pi}{e_{0}},
\end{equation}
where $v$ is an integer valued two form, $v : \Lambda^{\ast\ast} \ni p \rightarrow \frac{2\pi}{e_{0}}\mathbb{Z}$. 
We explain the connection between this and non compact formalism  shortly in theorem 1.
The compact partition function for an Abelian Higgs theory is defined as
\begin{equation}
\mathrm{Z}^{\text{C}} = \sum_{v : dv=0}  \int_{-\frac{\pi}{e_{0}}}^{\frac{\pi}{e_{0}}}\mathcal{D}A \hspace{0.05 cm} \int_{\mathbb{R}} \mathcal{D}\phi \hspace{0.05 cm} e^{-\frac{1}{2} \sum_{p \in \Lambda^{\ast\ast}} (dA(p) + v(p))^{2} - S_{h}(\phi, A)},
\end{equation}
where $\mathcal{D}A = \prod_{b \in \Lambda^{\ast}} dA_{b}$ is product Lebesgue measure on $[\frac{- \pi}{e_{0}}, \frac{\pi}{e_{0}}]$. 
The compact expectation of an observable $e^{-e_{0} \langle dA, \text{J}\rangle}$ is defined as
\begin{equation}
\mathrm{Z}^{\text{C}}\hspace{0.05 cm} \langle e^{-e_{0} \langle dA, \text{J}\rangle} \rangle^{\text{C}} = 
 \sum_{v : dv=0} \int_{-\frac{\pi}{e_{0}}}^{\frac{\pi}{e_{0}}}\mathcal{D}A \hspace{0.05 cm}  \int_{\mathbb{R}} \mathcal{D}\phi \hspace{0.1 cm}
 e^{-\frac{1}{2} \sum_{p \in \Lambda^{\ast\ast}} (dA(p) + v(p))^{2} - S_{h}(\phi, A) -e_{0} \langle dA+v, \text{J}\rangle}.
\end{equation} 
\textbf{Theorem 1} \cite{BBIJ} Given an observable $e^{-e_{0} \langle dA, \text{J}\rangle}$ and using the definition of expectation as above,
the compact and non compact formalism are equivalent in following sense 
\begin{equation}
\begin{aligned}
\mathrm{Z}^{\text{NC}} = \Big(\frac{2\pi}{e_{0}}\Big)^{-|\Lambda| + 1} \mathrm{Z}^{\text{C}}  \hspace{0.3 cm}  \text{and} 
 \hspace{0.3 cm}  \langle e^{-e_{0} \langle dA, \text{J}\rangle} \rangle^{\text{NC}} = 
 \langle e^{-e_{0} \langle dA + v, \text{J}\rangle} \rangle^{\text{C}}.
\end{aligned}
\end{equation}
 
\textit{Proof} First break the integral over $\mathbb{R}$ in Eq 1.10 into compact intervals $\Big[\frac{- \pi}{e_{0}}, \frac{\pi}{e_{0}}\Big]$.
Let $n : \Lambda^{\ast} \rightarrow \Big(\frac{2\pi}{e_{0}}\Big) \mathbb{Z}$ be an integer valued one form. 
Eq 1.10 becomes
\begin{equation}
\mathrm{Z}^{\text{NC}}\hspace{0.05 cm} \langle e^{-e_{0} \langle dA, \text{J}\rangle} \rangle^{\text{NC}} = 
\sum_{n}\int_{-\frac{\pi}{e_{0}}}^{\frac{\pi}{e_{0}}}\mathcal{D}A \hspace{0.05 cm}  \int_{\mathbb{R}} \mathcal{D}\phi \hspace{0.1 cm} 
e^{- S(\phi, A+n) -e_{0} \langle d(A+n), \text{J}\rangle} \langle\langle e^{-G(A+n)} \rangle\rangle.
\end{equation}
Rewrite
\begin{equation}
\sum_{n} = \sum_{v} \sum_{n : dn = v}
\end{equation}
Note that in the Higgs action, $e^{i e_{0} (A + n)} = e^{i e_{0}A}$ since $n = \Big(\frac{2\pi}{e_{0}}\Big) \mathbb{Z}$.
Replace the condition $dn = v$ with equivalent condition $dv = 0$ and rewrite the non compact expectation as
\begin{equation}
\mathrm{Z}^{\text{NC}}\hspace{0.05 cm} \langle e^{-e_{0} \langle dA, \text{J}\rangle} \rangle^{\text{NC}} = 
\sum_{v : dv=0} \int^{\frac{\pi}{e_{0}}}_{-\frac{\pi}{e_{0}}} \mathcal{D}A \hspace{0.05 cm} \int_{\mathbb{R}} \mathcal{D}\phi 
\hspace{0.1 cm} e^{ -\frac{1}{2} \sum_{p}|dA + v|^{2}  - S_{h}(\phi, A) -e_{0} \langle dA + v, \text{J}\rangle} 
\sum_{n : dn = v} \langle\langle e^{-G(A + n)} \rangle\rangle 
\end{equation}
Let $s : \Lambda \rightarrow \Big(\frac{2\pi}{e_{0}}\Big) \mathbb{Z}$ be a zero form with $s(x_{0}) = 0$. 
For a fixed \textit{v} any two values, \textit{n} and $\tilde{n}$ satisfying
$dn = d\tilde{n} = v$ differ by \textit{ds}. The construction of such integer valued forms and a choice of point
$x_{0}$ is discussed in \cite{KK}. The gauge function $\theta$ also satisfies $\theta(x_{0}) = 0$. Thus,
\begin{equation}
\begin{aligned}
\sum_{dn = v}  \langle\langle e^{-G(A + n)} \rangle\rangle &= \sum_{s}  \langle\langle e^{-G(A + \tilde{n} + ds)} \rangle\rangle \\
&= \sum_{s} \Big(\frac{2\pi}{e_{0}}\Big)^{-|\Lambda|}
\int^{\frac{\pi}{e_{0}}}_{-\frac{\pi}{e_{0}}} \mathcal{D}\theta_{\Lambda} \hspace{0.1 cm} e^{-G(A + \tilde{n} + d(\theta + s))} \\
&= \Big(\frac{2\pi}{e_{0}}\Big)^{-|\Lambda| + 1} \int_{\mathbb{R}}
\mathcal{D}\theta_{\Lambda \backslash x_{0}} \hspace{0.1 cm} e^{-G(A + \tilde{n} + d\theta)} \\ 
&= \Big(\frac{2\pi}{e_{0}}\Big)^{-|\Lambda| + 1}
\end{aligned}
\end{equation}
where, a factor of $\frac{2\pi}{e_{0}}$ comes from evaluating the integral at $x_{0}$ and final step is due to the gauge fixing condition 
of Eq 1.5. \begin{equation}
\begin{aligned}
\mathrm{Z}^{\text{NC}}\hspace{0.05 cm} \langle e^{-e_{0} \langle dA, \text{J}\rangle} \rangle^{\text{NC}} = 
\Big(\frac{2\pi}{e_{0}}\Big)^{-|\Lambda| + 1} \sum_{v : dv=0} \int^{\frac{\pi}{e_{0}}}_{-\frac{\pi}{e_{0}}} \mathcal{D}A \hspace{0.05 cm} \int_{\mathbb{R}} \mathcal{D}\phi \hspace{0.1 cm} e^{ -\frac{1}{2} \sum_{p}|dA + v|^{2}  - S_{h}(\phi, A) - e_{0} \langle dA + v, \text{J}\rangle}. 
\end{aligned}
\end{equation}
The gauge field \textit{A} on the right hand side takes value in the compact interval $\Big[\frac{- \pi}{e_{0}}, \frac{\pi}{e_{0}}\Big]$.
Setting $e^{-e_{0} \langle dA, \text{J}\rangle} = 1$ gives 
$\mathrm{Z}^{\text{NC}} = \Big(\frac{2\pi}{e_{0}}\Big)^{-|\Lambda| + 1} \mathrm{Z}^{\text{C}}$ and hence
$\langle e^{-e_{0} \langle dA, \text{J}\rangle} \rangle^{\text{NC}} = \langle e^{-e_{0} \langle dA + v, \text{J}\rangle} \rangle^{\text{C}}$.
 
\textit{Classical Higgs Mechanism}. To show that the gauge field has acquired mass $m_{A}$ we make standard change of variables as  
\begin{equation}
\phi (x) = \rho(x) e^{i\theta(x)}, \quad A \rightarrow A - \frac{1}{e_{0}}d\theta.
\end{equation}
$\rho$ is the length of the Higgs field. Due to the absolute value of Higgs in the action (Eq 1.2), there is no explicit 
$\theta$ dependence since $\abs{e^{i e_{0} (A_{\langle xy\rangle} - \frac{1}{e_{0}}d\theta)}e^{i \theta(y)}\rho(y) - e^{i \theta(x)}\rho(x)} 
= \abs{e^{i \theta(x)}} \abs{e^{i e_{0} A_{\langle xy\rangle}} \rho(y) - \rho(x)}$.
We thus take average over $\theta(x)$ as $\prod_{x \in \Lambda} \int_{-\pi}^{\pi} d\theta(x) = (2\pi)^{|\Lambda|}$ and drop this constant.
The potential term in new variables is
\begin{equation}
V(\rho) = \lambda \rho(x)^{4} - \frac{1}{4} \mu^{2} \rho(x)^{2} + E
\end{equation}
with minimum at $\rho_{0} =  \frac{\mu}{\sqrt{8\lambda}}$. Substituting $\rho \rightarrow \rho_{0} + \rho$ the total action is
\begin{equation}
\begin{aligned}
S(\rho, A) &= \frac{1}{2} \sum_{p \subset \Lambda^{\ast\ast}} (dA + v)^{2}(p) + \frac{1}{2}m_{A}^{2}\sum_{b \subset \Lambda^{\ast}}A_{b}^{2}  
+ \frac{1}{2}  \sum_{\langle xy\rangle \subset \Lambda} (\rho(y) - \rho(x))^{2} + \frac{1}{2}\mu^{2}\sum_{x \in \Lambda} \rho^{2}(x) \\
&+ \rho_{0}^{2}  \sum_{\langle xy\rangle \subset \Lambda} (1 -\text{cos}\hspace{0.05 cm} e_{0}A_{\langle xy\rangle}
 - \frac{1}{2}e_{0}^{2}A_{\langle xy\rangle}^{2})
+ \rho_{0} \sum_{\langle xy\rangle \subset \Lambda} (\rho(y) + \rho(x))(1 - \text{cos}\hspace{0.05 cm} e_{0}A_{\langle xy\rangle}) \\ 
&+ \sum_{\langle xy\rangle \subset \Lambda} \rho(y)\rho(x)(1 - \text{cos}\hspace{0.05 cm} e_{0}A_{\langle xy\rangle})  
+ \sum_{x \in \Lambda}\Big(\lambda \rho^{4}(x) + \sqrt{2\lambda} \mu\rho^{3}(x) - \text{log}\left[ 1+\frac{\rho(x)}{\rho_{0}}\right]\Big). 
\end{aligned}
\end{equation}
The gauge field is now massive with mass
\begin{equation}
m_{A} = \frac{\mu \hspace{0.05 cm} e_{0}}{\sqrt{8\lambda}}.
\end{equation}
The term $\text{log}\hspace{0.05 cm} \rho$ comes from the Jacobian of the change of variables $(\phi_{1}, \phi_{2}) \rightarrow  (\rho, \theta)$.
We have dropped a constant term $\text{log} (2\pi \rho_{0})|\Lambda|$. This just changes the normalization. 

\textit{Mass gap}. From the equivalence established in theorem 1 and using the change of variables as above, 
the generating functional for a non compact Abelian Higgs theory is given by 
\begin{equation}
\mathrm{Z}[\text{J}] =  \Big(\frac{2\pi}{e_{0}}\Big)^{-|\Lambda| + 1} \sum_{v: dv=0} \int^{\frac{\pi}{e_{0}}}_{-\frac{\pi}{e_{0}}} \mathcal{D}A 
\int_{-\rho_{0}}^{\infty}  \mathcal{D}\rho \hspace{0.1 cm} e^{- S(\rho, A) - e_{0} \langle dA+v , \text{J} \rangle}.
\end{equation}
Let $p_{1}$ and $p_{2}$ be two plaquettes. The \textit{truncated correlation} between $dA(p_{1})$ and $dA(p_{2})$
is given by
\begin{equation}
\langle dA(p_{1}) dA(p_{2})\rangle^{T} = 
\left. \frac{\partial^{2} \hspace{0.1 cm} \text{log} \hspace{0.1 cm} \mathrm{Z}[\text{J}]}
{\partial\text{J}_{p_{1}}\partial\text{J}_{p_{2}}}\right\rvert_{\text{J} = 0}
\end{equation}
 where, $\langle dA(p_{1}) dA(p_{2})\rangle^{T} = \langle dA(p_{1}) dA(p_{2})\rangle - \langle dA(p_{1})\rangle \langle dA(p_{2})\rangle$.

\textbf{Theorem 2} Given an Abelian Higgs theory on a unit lattice with masses $\mu$, $m_{A}$ fixed and sufficiently large.
Let the coupling constants $e_{0}$ and $\lambda = \frac{\mu^{2} e_{0}^{2}}{8 m_{A}^{2}}$ be sufficiently small depending on the masses.
The correlation of $dA(p)$ defined on two plaquettes $p_{1}$ and $p_{2}$ has an exponential decay as
\begin{equation}
\abs{\langle dA(p_{1}) dA(p_{2})\rangle - \langle dA(p_{1})\rangle \langle dA(p_{2})\rangle}  \leqslant e^{- m\hspace{0.05cm} d(p_{1},p_{2})}
\end{equation}
for some positive constant \textit{m}.  

\textit{Remark} 1. This says that the gauge field  $A_{\mu}$ has a mass at least as big as \textit{m}, i.e. mass gap.
Let $F_{\mu\nu}(x) = dA(x, x+ e_{\mu}, x+ e_{\mu} + e_{\nu}, x + e_{\nu})$. Alternatively with $F_{\mu\nu}(x)$
\begin{equation}
\abs{\langle F_{\mu\nu}(x_{1}) F_{\mu\nu}(x_{2})\rangle - \langle F_{\mu\nu}(x_{1}\rangle \langle F_{\mu\nu}(x_{2})\rangle}  
\leqslant e^{- m\hspace{0.05cm} d(x_{1}, x_{2})}.
\end{equation} 
\textit{Remark} 2. Theorem 2 is new but similar to the results of Balaban, Imbrie, Jaffe and Brydges \cite{BBIJ}
who considered integer charge observables. An observable $F(\phi, A)$ is integer charge observable if for every $b \in \Lambda^{\ast}$,
$F(\phi, A +  \frac{2\pi}{e_{0}} 1_{b}) = F(\phi, A)$, where, $1_{b}$ is the characteristic function of the bond b.

\textit{Remark} 3. Let $\gamma_{1}$ and $\gamma_{2}$ be two closed curves composed of lattice bonds. A Wilson loop variable 
is given by $W_{\gamma_{i}}(A) = \prod_{b \in \gamma_{i}} e^{i e_{0} A_{b}}$. Our methods also show that for some $m > 0$,
the correlation of $W_{\gamma_{1}}(A)$ and $W_{\gamma_{2}}(A)$ has an exponential decay as
\begin{equation}
\abs{\langle W_{\gamma_{1}}(A) W_{\gamma_{2}}(A))\rangle - \langle W_{\gamma_{1}}(A)\rangle \langle W_{\gamma_{1}}(A)\rangle}
\leqslant e^{- m\hspace{0.05cm} d(\gamma_{1},\gamma_{2})}.
\end{equation}
\textit{Remark} 4. In physics literature, the expectation value of Higgs field, $\langle\phi\rangle$ is taken to be non zero to
demonstrate Higgs mechanism. While it is convenient to assume $\langle\phi\rangle \neq 0$ for computations, whether 
there exists such states in this simplified model is a difficult dynamical question mathematically. Our results show there is
mass generation, whether or not $\langle\phi\rangle = 0$.

\textit{Outline of Proof.}
In section 2, we develop framework to construct the small field integral. Section 3 discusses
the application of power series cluster expansion \cite{BFKT} to this integral. In section 4, we discuss the large field estimates.
Then in section 5, we combine various overlapping regions of the lattice into connected components and
rewrite the generating functional as sum over those components. Finally, we establish mass gap in section 6.
\section{The expansion}
 $\Lambda$ consists all the sites x, $\Lambda^{\ast}$ is the set of all the bonds b and $\Lambda^{\ast\ast}$ is  
the set of all the plaquettes p. The Higgs field is $\rho : \Lambda \rightarrow [-\rho_{0}, \infty)$, the gauge field is
$A : \Lambda^{\ast} \rightarrow \Big[-\frac{\pi}{e_{0}}, \frac{\pi}{e_{0}}\Big]$ and the vortex field is
$v : \Lambda^{\ast\ast}\rightarrow \frac{2\pi}{e_{0}}\mathbb{Z}$.
We represent all the fields on lattice as $\Phi = \{\rho, A\} : \Lambda \rightarrow \mathbb{R}$. Let
\begin{equation}
\text{T} = \langle\rho, (- \Delta + \mu^{2}) \rho\rangle + \langle A, (\delta d + m_{A}^{2}) A\rangle
\end{equation}
$\Delta$ and $\delta d$ are operators on $l^{2} (\Lambda)$ and $l^{2}(\Lambda^{\ast})$ respectively, with
\begin{equation}
\langle \rho, -\Delta\rho \rangle = \sum_{\langle x,y\rangle \subset \Lambda} (\rho(y) - \rho(x))^{2}  
\hspace{0.4 cm} \text{and}  \hspace{0.4 cm}
\langle A, -\delta d \hspace{0.1 cm} A \rangle =  \sum_{p \subset \Lambda^{\ast\ast}} (dA)^{2}.
\end{equation}
Rewrite the action (1.23) as a sum of Gaussian part and higher order interaction part,
\begin{equation}
S(\Phi) = \frac{1}{2}\langle\Phi, \text{T}\Phi\rangle + V(\Phi).
\end{equation}
Let $\xi, \xi^{\prime} \in \Lambda \cup \Lambda^{\ast}$. Define the covariance operator $C^{-1}$ as 
\begin{equation}
C(\xi,\xi^{\prime}) = \begin{cases}
(-\Delta + \mu^{2})^{-1}   & \hspace{0.1 cm}  \xi,\xi^{\prime} \in \Lambda, \\
 (-\delta d + m_{A}^{2})^{-1}  & \hspace{0.1 cm} \xi,\xi^{\prime} \in \Lambda^{\ast}.
\end{cases}
\end{equation} 
Thus, $\text{T} = C^{-1}$. The generating functional is 
\begin{equation}
\mathrm{Z}[\text{J}] = \Big(\frac{2\pi}{e_{0}}\Big)^{-|\Lambda| + 1} \sum_{v: dv=0} 
\int \mathcal{D}\Phi \hspace{0.05 cm} e^{-\frac{1}{2} \langle\Phi, C^{-1}\Phi\rangle - V(\Phi)} e^{-e_{0} \langle dA+v , \text{J} \rangle}.
\end{equation}
To construct our localized small field integral, we expand the generating functional into regions where the field $\Phi$ is bounded
 and regions where it is not. The following four steps details this expansion.
 \begin{enumerate}
  \item Let $p_{\lambda} = |\text{log} \hspace{0.05 cm} \lambda|^{2d + 1}$ and $r_{\lambda} = |\text{log} \hspace{0.05 cm} \lambda|^{2}$.
  Divide the lattice $\Lambda$ into blocks $\Box$ of length $[r_{\lambda}]$, where $[\cdot]$ denotes the integer part.
  Define the characteristic function
  \begin{equation}
\chi_{\Box}(\Phi) = \begin{cases} 
1  & \hspace{0.1 cm} \sup_{\xi \in \Box} |\Phi(\xi)| < p_{\lambda},  \\
0  & \hspace{0.1 cm} \text{otherwise}
\end{cases}
\hspace{0.5 cm} \zeta_{\Box}(\Phi) = 1- \chi_{\Box}(\Phi). 
\end{equation}
Then the decomposition of unity is
 \begin{equation}
1 = \prod_{\Box} (\chi_{\Box} + \zeta_{\Box}) = \sum_{\Lambda_{0}\subset \Lambda} 
\prod_{\Box \in \Lambda_{0}} \chi_{\Box} \prod_{\Box \in \Lambda_{0}^{c}} \zeta_{\Box}
= \sum_{\Lambda_{0}\subset \Lambda} \chi_{\Lambda_{0}} \zeta_{\Lambda_{0}^{c}}.
\end{equation}
Thus, a $\Box \in \Lambda^{c}_{0}$  if there is at least one $\xi \in \Box$ with $|\Phi(\xi)| > p_{\lambda}$.
Insert (decomposition of unity) 1 in the generating functional rewrite 
\begin{equation}
\begin{aligned}
\mathrm{Z}[\text{J}]  = \Big(\frac{2\pi}{e_{0}}\Big)^{-|\Lambda| + 1} \sum_{\Lambda_{0} \subset \Lambda}\sum_{v: dv=0}
\int \mathcal{D}\Phi \hspace{0.05 cm} e^{-\frac{1}{2} \langle\Phi, C^{-1}\Phi\rangle - V(\Phi)} e^{-e_{0} \langle dA+v , \text{J} \rangle} 
\chi_{\Lambda_{0}}(\Phi) \zeta_{\Lambda_{0}^{c}}(\Phi).
\end{aligned}
\end{equation}
Set $v = 0$ in $\Lambda_{0}$.

\item Conditioning

Contract $\Lambda_{0}$ by $[r_{\lambda}]$ to get $\Lambda_{1}$.
Let $1_{\Lambda_{1}}$ and $1_{\Lambda^{c}_{1}}$ be  characteristic functions restricting 
operator to $\Lambda_{1}$ and $\Lambda^{c}_{1}$ respectively. Define $\text{T}_{\Lambda_{1}} = 1_{\Lambda_{1}}\text{T}1_{\Lambda_{1}}$ 
and $\text{T}_{\Lambda_{1}\Lambda^{c}_{1}} = 1_{\Lambda_{1}}\text{T}1_{\Lambda^{c}_{1}}$. Then rewrite
\begin{equation}
\frac{1}{2}\langle\Phi, \text{T}\Phi\rangle = \frac{1}{2}\langle\Phi, \text{T}_{\Lambda_{1}}\Phi\rangle + 
\langle\Phi, \text{T}_{\Lambda_{1}\Lambda^{c}_{1}}\Phi\rangle + \frac{1}{2}\langle\Phi, \text{T}_{\Lambda^{c}_{1}}\Phi\rangle.
\end{equation}
The term $\langle\Phi, \text{T}_{\Lambda_{1}}\Phi\rangle = \langle\Phi_{\Lambda_{1}}, \text{T}\Phi_{\Lambda_{1}}\rangle$ 
represents the interactions entirely in the small field region, the term 
$\langle\Phi, \text{T}_{\Lambda^{c}_{1}}\Phi\rangle = \langle\Phi_{\Lambda^{c}_{1}}, \text{T}\Phi_{\Lambda^{c}_{1}}\rangle$
represents the interactions in the large field region and the term 
$\langle\Phi, \text{T}_{\Lambda_{1}\Lambda^{c}_{1}}\Phi\rangle = \langle\Phi_{\Lambda_{1}}, \text{T}\Phi_{\Lambda^{c}_{1}}\rangle$ 
represents the interaction of the large field region $\Lambda^{c}_{1}$ over the boundary $\partial\Lambda_{1}$. 
Absorb the source term $e_{0}\langle dA+v, \text{J}\rangle$ into the potential $V(\Phi)$ and rewrite  
\begin{equation}
\begin{aligned}
\mathrm{Z}[\text{J}]  = \Big(\frac{2\pi}{e_{0}}\Big)^{-|\Lambda| + 1} &\sum_{\Lambda_{0} \subset \Lambda}
\sum_{v: dv=0}\int\mathcal{D}\Phi_{\Lambda^{c}_{1}} \hspace{0.1 cm}
e^{-\frac{1}{2}\langle\Phi, \text{T}_{\Lambda^{c}_{1}}\Phi\rangle 
 - V(\Lambda^{c}_{1}, \Phi)}\zeta_{\Lambda^{c}_{0}}(\Phi)  \chi_{\Lambda_{0} - \Lambda_{1}}(\Phi)
\\ & \int\mathcal{D}\Phi_{\Lambda_{1}} e^{-\frac{1}{2}\langle\Phi_{\Lambda_{1}}, \text{T}\Phi_{\Lambda_{1}}\rangle
 - \langle\Phi_{\Lambda_{1}}, \text{T}\Phi_{\Lambda^{c}_{1}}\rangle - V(\Lambda_{1}, \Phi)}\chi_{\Lambda_{1}}(\Phi).
\end{aligned}
\end{equation} 
In the above equation, make the transformation
\begin{equation}
\Phi_{\Lambda_{1}} \rightarrow \Phi_{\Lambda_{1}} - C_{\Lambda_{1}}\text{T}_{\Lambda_{1}\Lambda^{c}_{1}} \Phi_{\Lambda^{c}_{1}}
\end{equation}
 the generating functional becomes 
\begin{equation}
\begin{aligned}
\mathrm{Z}[\text{J}]  &= \Big(\frac{2\pi}{e_{0}}\Big)^{-|\Lambda| + 1} 
\sum_{\Lambda_{0} \subset \Lambda}\sum_{v: dv=0}\int\mathcal{D}\Phi_{\Lambda^{c}_{1}} \hspace{0.05 cm}
e^{-\frac{1}{2}\langle\Phi, (\text{T}_{\Lambda^{c}_{1}} - \text{T}_{\Lambda^{c}_{1}\Lambda_{1}}
C_{\Lambda_{1}} \text{T}_{\Lambda_{1}\Lambda^{c}_{1}}) \Phi \rangle - 
V(\Lambda^{c}_{1}, \Phi)}\zeta_{\Lambda^{c}_{0}}(\Phi) \chi_{\Lambda_{0} - \Lambda_{1}}(\Phi)  \\ & \int\mathcal{D}\Phi_{\Lambda_{1}}
e^{-\frac{1}{2}\langle\Phi, \text{T}_{\Lambda_{1}}\Phi\rangle  
- V(\Lambda_{1}, \Phi_{\Lambda_{1}} - C_{\Lambda}\text{T}_{\Lambda_{1}\Lambda^{c}_{1}} \Phi_{\Lambda^{c}_{1}})}
\chi_{\Lambda_{1}}(\Phi_{\Lambda_{1}} - C_{\Lambda_{1}}\text{T}_{\Lambda_{1}\Lambda^{c}_{1}} \Phi_{\Lambda^{c}_{1}}).
\end{aligned}
\end{equation} 
This step is equivalent to conditioning the small field integral on values in the large field region.
 
 \item  
The operator \textit{C} couples sites everywhere on lattice $\Lambda$. Therefore, it is important to construct a localized
operator. Define the kernel of operator $C^{\text{loc}}$ as 
\begin{equation}
C^{\text{loc}}(x,y) = \begin{cases}
    C (x,y)  & \text{if} \hspace{0.4 cm} |x-y| < r_{\lambda}, \\
     0 & \text{otherwise}.
\end{cases}
\end{equation}
We adopt the representation for  
$C^{\frac{1}{2}}$ and $C^{\frac{1}{2},\text{loc}}$ as discussed in \cite{D1}. Define
\begin{equation}
C^{\frac{1}{2}} = \frac{1}{\pi}\int_{0}^{\infty}\frac{d r}{\sqrt{r}}\hspace{0.1 cm} C_{r},  \hspace{0.5 cm}
C^{\frac{1}{2},\text{loc}} = \frac{1}{\pi}\int_{0}^{\infty}\frac{d r}{\sqrt{r}}\hspace{0.1 cm} C^{\text{loc}}_{r} 
\end{equation}
with $C_{r} = (\text{T} + r)^{-1}$. Define $\delta C^{\frac{1}{2}} (x,y) = (C^{\frac{1}{2}} - C^{\frac{1}{2},\text{loc}}) (x,y)$ as
\begin{equation}
\delta C^{\frac{1}{2}}  (x,y) = \begin{cases}
    0  & \text{if} \hspace{0.4 cm} |x-y| < r_{\lambda}, \\
     C^{\frac{1}{2}}(x,y) & \text{otherwise}.
\end{cases}
\end{equation}
To work in unit covariance, make the change of variables $\Phi = C^{\frac{1}{2},\text{loc}}_{\Lambda_{1}} \Phi^{\prime}$  
\begin{equation}
\langle\Phi, C^{-1}_{\Lambda_{1}} \Phi\rangle =  
\langle\Phi^{\prime}, C^{\frac{1}{2},\text{loc}}_{\Lambda_{1}} C^{-1}_{\Lambda_{1}} C^{\frac{1}{2},\text{loc}}_{\Lambda_{1}} \Phi^{\prime}\rangle 
 = ||\Phi^{\prime}||_{\Lambda_{1}}^{2} + V_{\varepsilon}(\Lambda_{1}, \Phi^{\prime}).  
\end{equation}
 
\item 
This change of variables introduces non locality in characteristic function as 
$\chi_{\Lambda_{1}}(C^{\frac{1}{2},\text{loc}}_{\Lambda_{1}} \Phi^{\prime} - C_{\Lambda_{1}}\text{T}_{\Lambda_{1}\Lambda^{c}_{1}} \Phi_{\Lambda^{c}_{1}})$.
To construct the localized small field integral we therefore, split the small field region $\Lambda_{1}$ into a new small field region and
an intermediate field region.  Let $p_{0,\lambda} = |\text{log} \hspace{0.05 cm} \lambda|^{a}$ (with $2d < a < 2d+1$) and 
$\Omega_{0} \subset \Lambda_{1}$.
 We split the region $\Lambda_{1}$ into a region $\Omega_{0}$ and $\Lambda_{1}/\Omega_{0}$ 
 by using decomposition of unity as before. Define
\begin{equation}
\hat{\chi}_{\Box}(\Phi^{\prime}) = \begin{cases}
1 & \hspace{0.1 cm} \sup_{\xi \in \Box} |\Phi^{\prime}(\xi)| < p_{0, \lambda}, \\
0 & \hspace{0.1 cm} \text{otherwise}
\end{cases}
\hspace{0.5 cm} \hat{\zeta}_{\Box}(\Phi^{\prime}) = 1- \hat{\chi}_{\Box}(\Phi^{\prime}). 
\end{equation}
Then the decomposition of unity is
 \begin{equation}
1 = \prod_{\Box \subset \Lambda_{1}} (\hat{\chi}_{\Box} + \hat{\zeta}_{\Box}) = \sum_{\Omega_{0}\subset \Lambda_{1}} 
\prod_{\Box \in \Omega_{0}} \hat{\chi}_{\Box} \prod_{\Box \in \Lambda_{1} - \Omega_{0}} \hat{\zeta}_{\Box}
= \sum_{\Omega_{0}\subset \Lambda_{1}} \hat{\chi}_{\Omega_{0}} \hat{\zeta}_{\Lambda_{1} - \Omega_{0}}.
\end{equation} 
Due to unit covariance in $\Omega_{0}$, no conditioning is required along the boundary $\partial\Omega_{0}$.
In potential everything is analytic in field but the characteristic function $\chi_{\Lambda_{1}}$ is a mess which we have to clean.
As a first step, we split the potential term as
\begin{equation}
\begin{aligned}
V(\Lambda_{1}, C^{\frac{1}{2},\text{loc}}_{\Lambda_{1}} \Phi^{\prime} - C_{\Lambda_{1}}\text{T}_{\Lambda_{1}\Lambda^{c}_{1}}
 \Phi_{\Lambda^{c}_{1}}) = 
V(\Omega_{0}, C^{\frac{1}{2},\text{loc}}_{\Lambda_{1}} \Phi^{\prime} - C_{\Lambda_{1}}\text{T}_{\Lambda_{1}\Lambda^{c}_{1}}
 \Phi_{\Lambda^{c}_{1}}) &+ \\
V(\Lambda_{1} - \Omega_{0}, C^{\frac{1}{2},\text{loc}}_{\Lambda_{1}} \Phi^{\prime} - C_{\Lambda_{1}}\text{T}_{\Lambda_{1}\Lambda^{c}_{1}} 
\Phi_{\Lambda^{c}_{1}}).
\end{aligned}
\end{equation}
and factorize $\chi_{\Lambda_{1}}$ as
\begin{equation}
\begin{aligned}
\chi_{\Lambda_{1}}(C^{\frac{1}{2},\text{loc}}_{\Lambda_{1}} \Phi^{\prime} - C_{\Lambda_{1}}\text{T}_{\Lambda_{1}\Lambda^{c}_{1}}
 \Phi_{\Lambda^{c}_{1}})
 &=  \chi_{\Lambda_{1} - \Omega_{0}}(C^{\frac{1}{2},\text{loc}}_{\Lambda_{1}} \Phi^{\prime} - C_{\Lambda_{1}}\text{T}_{\Lambda_{1}\Lambda^{c}_{1}}
 \Phi_{\Lambda^{c}_{1}}) \\
&  \chi_{\Omega_{0}}(C^{\frac{1}{2},\text{loc}}_{\Lambda_{1}} \Phi^{\prime} - C_{\Lambda_{1}}\text{T}_{\Lambda_{1}\Lambda^{c}_{1}}
 \Phi_{\Lambda^{c}_{1}}).   
\end{aligned}
\end{equation} 
In $\Omega_{0},
|C^{\frac{1}{2},\text{loc}}_{\Lambda_{1}} \Phi^{\prime}| \leqslant c\hspace{0.1 cm} ||\Phi^{\prime}||_{\infty,\Omega_{0}}
\leqslant c \hspace{0.1 cm} \text{p}_{0,\lambda}$ (see lemma 2.2) and for $\xi \in \Omega_{0}$,

$\abs{C^{\frac{1}{2},\text{loc}}_{\Lambda_{1}} \Phi^{\prime}(\xi) - C_{\Lambda_{1}}\text{T}_{\Lambda_{1}\Lambda^{c}_{1}} 
\Phi_{\Lambda^{c}_{1}}} < p_{\lambda}$ since the term $\text{T}_{\Lambda_{1}\Lambda^{c}_{1}}$ forces 
$\Phi_{\Lambda^{c}_{1}} = p_{\lambda}$, this implies
\begin{equation}
\chi_{\Omega_{0}}(C^{\frac{1}{2},\text{loc}}_{\Lambda_{1}} \Phi^{\prime} - C_{\Lambda_{1}}\text{T}_{\Lambda_{1}\Lambda^{c}_{1}} 
\Phi_{\Lambda^{c}_{1}}) = 1. 
\end{equation}
 The non locality in $\chi_{\Lambda_{1} - \Omega_{0}}(C^{\frac{1}{2},\text{loc}}_{\Lambda_{1}} \Phi^{\prime} - C_{\Lambda_{1}}\text{T}_{\Lambda_{1}\Lambda^{c}_{1}} \Phi_{\Lambda^{c}_{1}})$ 
 prevents us to carry out integration over the region $\Omega_{0}$.
As a second and final step we contract the region $\Omega_{0}$ by $2 [r_{\lambda}]$ to get a new small field region $\Omega_{1}$,
that is, $d(\Omega_{0}^{c}, \Omega_{1}) = 2 [r_{\lambda}]$. Note that $\chi_{\Lambda_{1} - \Omega_{0}}(C^{\frac{1}{2},\text{loc}}_{\Lambda_{1}} \Phi^{\prime} - C_{\Lambda_{1}}\text{T}_{\Lambda_{1}\Lambda^{c}_{1}} \Phi_{\Lambda^{c}_{1}})$  and 
$V(\Lambda_{1} - \Omega_{0}, C^{\frac{1}{2},\text{loc}}_{\Lambda_{1}} \Phi^{\prime} - C_{\Lambda_{1}}\text{T}_{\Lambda_{1}\Lambda^{c}_{1}} 
\Phi_{\Lambda^{c}_{1}})$ do not depend on $\Phi^{\prime}_{\Omega_{1}}$ since $C^{\frac{1}{2},\text{loc}}_{\Lambda_{1}}$
 couples the fields only up to a distance of $[r_{\lambda}]$. After the above procedure, the generating functional becomes
\begin{equation}
\begin{aligned}
\mathrm{Z}[\text{J}] &= \Big(\frac{2\pi}{e_{0}}\Big)^{-|\Lambda| + 1}  
\sum_{\Lambda_{0} \subset \Lambda} \sum_{\Omega_{0} \subset \Lambda_{1}}\sum_{v: dv=0}
(\text{det}\hspace{0.1 cm}C^{\frac{1}{2},\text{loc}}_{\Lambda_{1}}) \\ &   \int\mathcal{D}\Phi_{\Lambda^{c}_{1}}
 e^{-\frac{1}{2}\langle\Phi, (\text{T}_{\Lambda^{c}_{1}} - \text{T}_{\Lambda^{c}_{1}\Lambda_{1}}
C_{\Lambda_{1}} \text{T}_{\Lambda_{1}\Lambda^{c}_{1}}) \Phi \rangle - 
V(\Lambda^{c}_{1}, \Phi)}\zeta_{\Lambda^{c}_{0}}(\Phi)  \chi_{\Lambda_{0} - \Lambda_{1}}(\Phi) \\ & 
 \int\mathcal{D}\Phi^{\prime}_{\Lambda_{1} - \Omega_{1}} e^{-\frac{1}{2}||\Phi^{\prime}||_{\Lambda_{1} - \Omega_{1}}^{2} - 
 V(\Lambda_{1} - \Omega_{0}, C^{\frac{1}{2},\text{loc}}_{\Lambda_{1}} \Phi^{\prime} - C_{\Lambda_{1}}\text{T}_{\Lambda_{1}\Lambda^{c}_{1}}
  \Phi_{\Lambda^{c}_{1}})}\\ & \hat{\zeta}_{\Lambda_{1} - \Omega_{0}}(\Phi^{\prime})  \hspace{0.1 cm} 
  \chi_{\Lambda_{1} - \Omega_{0}}(C^{\frac{1}{2},\text{loc}}_{\Lambda_{1}} \Phi^{\prime} - C_{\Lambda_{1}}\text{T}_{\Lambda_{1}\Lambda^{c}_{1}}
   \Phi_{\Lambda^{c}_{1}})
 \\ & \hspace{0.1 cm}
 \int\mathcal{D}\Phi^{\prime}_{\Omega_{1}} e^{-\frac{1}{2}||\Phi^{\prime}||_{\Omega_{1}}^{2} - 
 V(\Omega_{0}, C^{\frac{1}{2},\text{loc}}_{\Lambda_{1}} \Phi^{\prime} - C_{\Lambda_{1}}\text{T}_{\Lambda_{1}\Lambda^{c}_{1}} 
 \Phi_{\Lambda^{c}_{1}}) + V_{\varepsilon}(\Lambda_{1}, \Phi^{\prime})}
 \hat{\chi}_{\Omega_{1}}(\Phi^{\prime}).
\end{aligned}
\end{equation}
The above expression is the required expansion of the generating functional.
\end{enumerate}

Consider the term $\text{det}\hspace{0.1 cm}C^{\frac{1}{2},\text{loc}}_{\Lambda_{1}}$.  As we will see in lemma 2.4, 
for some $W_{1}(\Lambda_{1}) = \sum_{\Box \subset \Lambda_{1}} W_{1}(\Box)$
\begin{equation}
\begin{aligned}
\text{det}\hspace{0.1 cm}C^{\frac{1}{2},\text{loc}}_{\Lambda_{1}}  
 &= (\text{det}\hspace{0.1 cm}C^{\frac{1}{2}}_{\Lambda_{1}}) \hspace{0.1 cm} 
 e^{W_{1}(\Lambda_{1})}.
\end{aligned}
\end{equation}
As we will see in lemma 2.7, for some $W_{2}(\Lambda) = \sum_{\Box \subset \Lambda} W_{2}(\Box)$ 
\begin{equation}
\begin{aligned}
\text{det}\hspace{0.1 cm}C^{\frac{1}{2}}_{\Lambda_{1}}  
&= (\text{det}\hspace{0.1 cm}C^{\frac{1}{2}})\hspace{0.1 cm} e^{W_{2}(\Lambda)}
\end{aligned}
\end{equation}
and we write $W_{2} = W_{2}(\Lambda^{c}_{0}) + W_{2}(\Lambda_{0} - \Omega_{0}) + W_{2}(\Omega_{0})$.
Consider the source term $e_{0} \langle dA, \text{J}\rangle$ with $|\text{J}| < 1$. Rewrite 
\begin{equation}
\begin{aligned}
\langle dA, \text{J}\rangle = \sum_{p} dA(p)\hspace{0.5 mm} \text{J}(p) &= \sum_{p} \sum_{b \in \partial p} A(b)\hspace{0.5 mm} \text{J}(b) 
= \sum_{b} A(b) \sum_{p : \partial p \ni b} \text{J}(b)  = \langle A, \delta \text{J}\rangle   
\end{aligned}
\end{equation}
and define
\begin{equation}
\begin{aligned}
V_{1}(\Omega_{0}, \Phi^{\prime}_{\Lambda_{1}}, \Phi_{\Lambda^{c}_{1}}, \text{J})
&= - W_{1}(\Lambda_{1}) - W_{2}(\Omega_{0})  + 
 V(\Omega_{0}, C^{\frac{1}{2},\text{loc}}_{\Lambda_{1}} \Phi^{\prime} - C_{\Lambda_{1}}\text{T}_{\Lambda_{1}\Lambda^{c}_{1}}
  \Phi_{\Lambda^{c}_{1}}) \\
& + V_{\varepsilon}(\Lambda_{1}, \Phi^{\prime}) -
\langle C^{\frac{1}{2},\text{loc}}_{\Lambda_{1}} \Phi^{\prime} - C_{\Lambda_{1}}\text{T}_{\Lambda_{1}\Lambda^{c}_{1}} \Phi_{\Lambda^{c}_{1}},
\delta \text{J}\rangle. 
\end{aligned}
\end{equation}
Define the normalized Gaussian measure with unit covariance as
\begin{equation}
d\mu_{\text{I}}(\Phi^{\prime}_{\Omega_{1}})  = \frac{\mathcal{D}\Phi^{\prime}_{\Omega_{1}} 
e^{-\frac{1}{2}||\Phi^{\prime}||^{2}_{\Omega_{1}}}}{\mathrm{Z}_{0}(\Omega_{1})}, 
\hspace{0.5 cm} \mathrm{Z}_{0}(\Omega_{1}) = \int \mathcal{D}\Phi^{\prime}_{\Omega_{1}} e^{-\frac{1}{2}||\Phi^{\prime}||^{2}_{\Omega_{1}}}.
\end{equation}
The localized small field integral is 
\begin{equation}
\Xi(\Omega_{1}, \Phi^{\prime}_{\Lambda_{1} - \Omega_{1}}, \Phi_{\Lambda_{1}^{c}}, \text{J}) = 
\int d\mu_{\text{I}}(\Phi^{\prime}_{\Omega_{1}}) \hspace{0.1 cm}
e^{-V_{1}(\Omega_{0}, \Phi^{\prime}_{\Lambda_{1}}, \Phi_{\Lambda^{c}_{1}}, \text{J})} \hat{\chi}(\Phi^{\prime}_{\Omega_{1}}).
\end{equation}
Rewrite the generating functional
\begin{equation}
\begin{aligned}
\mathrm{Z}[\text{J}] &= \Big(\frac{2\pi}{e_{0}}\Big)^{-|\Lambda| + 1}   (\text{det}\hspace{0.1 cm}C^{\frac{1}{2}})\hspace{0.1 cm}
\sum_{\Lambda_{0} \subset \Lambda} \sum_{\Omega_{0} \subset \Lambda_{1}}\sum_{v: dv=0}
 \\ &   \int\mathcal{D}\Phi_{\Lambda^{c}_{1}}
 e^{-\frac{1}{2}\langle\Phi, (\text{T}_{\Lambda^{c}_{1}} - \text{T}_{\Lambda^{c}_{1}\Lambda_{1}}
C_{\Lambda_{1}} \text{T}_{\Lambda_{1}\Lambda^{c}_{1}}) \Phi \rangle - 
V(\Phi_{\Lambda^{c}_{1}})}\zeta_{\Lambda^{c}_{0}}(\Phi) \chi_{\Lambda_{0} - \Lambda_{1}}(\Phi)
\hspace{0.05 cm}  e^{W_{2} (\Lambda_{0}^{c})} \\ &  
 \int\mathcal{D}\Phi^{\prime}_{\Lambda_{1} - \Omega_{1}} e^{-\frac{1}{2}||\Phi^{\prime}||_{\Lambda_{1} - \Omega_{1}}^{2} - 
 V(\Lambda_{1} - \Omega_{0}, C^{\frac{1}{2},\text{loc}}_{\Lambda_{1}} \Phi^{\prime} - C_{\Lambda_{1}}\text{T}_{\Lambda_{1}\Lambda^{c}_{1}}
  \Phi_{\Lambda^{c}_{1}})} \hat{\zeta}_{\Lambda_{1} - \Omega_{0}}(\Phi^{\prime}) \\ & \hspace{0.05 cm} 
  \chi_{\Lambda_{1} - \Omega_{0}}(C^{\frac{1}{2},\text{loc}}_{\Lambda_{1}} \Phi^{\prime} - C_{\Lambda_{1}}\text{T}_{\Lambda_{1}\Lambda^{c}_{1}}
   \Phi_{\Lambda^{c}_{1}}) \hspace{0.05 cm} e^{W_{2}(\Lambda_{0} - \Omega_{0})}  
 \hspace{0.1 cm}  \mathrm{Z}_{0}(\Omega_{1}) \hspace{0.1 cm} 
 \Xi(\Omega_{1}, \Phi^{\prime}_{\Lambda_{1} - \Omega_{1}}, \Phi_{\Lambda^{c}_{1}}).
\end{aligned}
\end{equation}
This is our basic expansion of the generating functional including the small field integral. Now, we prove some lemmas.

\textbf{Lemma 2.1} (\textit{Estimate on covariance}.) Let \textit{C} be the positive, self adjoint operator as defined in Eq 2.4.
Then for some $\gamma > 0$, $|C(x,y)| \leqslant c \hspace{0.05 cm} e^{-\gamma |x-y|}$.

\textit{Proof} Let \textit{x} and \textit{y} be two sites. For some vector \textbf{a} define
an operator $e_{\textbf{a}}(x)$ by its action on some function \textit{f} as $e_{\textbf{a}}(x) f = e^{\textbf{a} x} f$.
Let $\delta_{x}(y) = \delta_{xy}$ be the lattice delta function. 
Let $|\textbf{a}|$ be small and $|| \cdot ||$ be the $l^{2}$ norm. Then

\begin{equation}
\begin{aligned}
e_{-\textbf{a}} \partial_{\mu} e_{\textbf{a}} &= \partial_{\mu} + a_{\mu} \\
e_{-\textbf{a}} (- \Delta + \mu^{2}) e_{\textbf{a}} &= - \Delta - 2\hspace{0.05 cm}\textbf{a}\cdot\triangledown + |\textbf{a}|^{2} + \mu^{2} \\
- \Delta + \mu^{2} &= e_{\textbf{a}} (- \Delta - 2\hspace{0.05 cm}\textbf{a}\cdot\triangledown + |\textbf{a}|^{2} + \mu^{2}) e_{-\textbf{a}} \\
(- \Delta + \mu^{2})^{-1} &= e_{\textbf{a}} (- \Delta - 2\hspace{0.05 cm}\textbf{a}\cdot\triangledown + |\textbf{a}|^{2} + \mu^{2})^{-1} e_{-\textbf{a}} \\
|(- \Delta + \mu^{2})^{-1} (x,y)| &= |e^{\textbf{a} x} (- \Delta - 2\hspace{0.05 cm} 
\textbf{a}\triangledown + |\textbf{a}|^{2} + \mu^{2})^{-1}(x,y) e^{- \textbf{a} y}| \\
&= e^{\textbf{a} (x-y)} |\langle \delta_{x}, (- \Delta - 2\hspace{0.05 cm}\textbf{a}\cdot\triangledown + |\textbf{a}|^{2} + \mu^{2})^{-1} \delta_{y} \rangle| \\
&\leqslant e^{\textbf{a} (x-y)} ||\delta_{x}|| \hspace{0.05 cm} ||\delta_{y}|| \hspace{0.05 cm}
||(- \Delta - 2\hspace{0.05 cm}\textbf{a}\cdot\triangledown + |\textbf{a}|^{2} + \mu^{2})^{-1}||  \\
&\leqslant c \hspace{0.05 cm} e^{\textbf{a} (x-y)}
\end{aligned}
\end{equation}
setting $\textbf{a} = - \gamma \frac{(x-y)}{|x-y|}$ gives the result. Note that 
\begin{equation}
\begin{aligned}
(- \Delta + |\textbf{a}|^{2} + \mu^{2} - 2\hspace{0.05 cm}\textbf{a}\cdot\triangledown)^{-1} &=
(- \Delta + |\textbf{a}|^{2} + \mu^{2})^{-1} \sum_{n=0}^{\infty} 
[(- \Delta + |\textbf{a}|^{2} + \mu^{2})^{-1} 2\hspace{0.05 cm}\textbf{a}\cdot\triangledown]^{n} \\
||(- \Delta + |\textbf{a}|^{2} + \mu^{2} - 2\hspace{0.05 cm}\textbf{a}\cdot\triangledown)^{-1}||  
&\leqslant  \mu^{-2} \sum_{n=0}^{\infty} \left[ 2^{d+1} \frac{|\textbf{a}|}{\mu^{2}}\right]^{n}
\end{aligned}
\end{equation}
which converges if $|\textbf{a}| < \frac{\mu^{2}}{2^{d+1}}$. Note that $|\textbf{a}| = \gamma$.
Similarly, we can prove the estimate for term $(-\delta d + m_{A}^{2})^{-1}$ as long as $\abs{\textbf{a}}$ is small to keep the expression
$(-\delta d + |\textbf{a}|^{2} + m_{A}^{2} - 2\hspace{0.05 cm}\textbf{a}\cdot d)^{-1}$ positive, where $d$ denotes the 
exterior derivative.

\textit{Remark} 5.  Let  $\xi, \xi^{\prime} \in \Lambda \cup \Lambda^{\ast}$. Let \textit{K} be a positive, self adjoint operator with
$\abs{K(\xi, \xi^{\prime})} \leqslant e^{-\gamma d(\xi, \xi^{\prime})}$ and \textit{f} be some function. Then
\begin{equation}
\begin{aligned}
(Kf)(\xi) &= \sum_{\xi^{\prime}} K(\xi, \xi^{\prime}) f(\xi^{\prime}) \\
\abs{(Kf)(\xi)} &\leqslant \sum_{\xi^{\prime}} \abs{K(\xi, \xi^{\prime})} ||f||_{\infty} \\
&\leqslant \sum_{\xi^{\prime}} e^{-\gamma d(\xi, \xi^{\prime})}  ||f||_{\infty} \leqslant c \hspace{0.05 cm}  ||f||_{\infty}.
\end{aligned}
\end{equation}
Note that $||Kf||_{\infty} \leqslant \sup_{\xi} \abs{(Kf)(\xi)}$. In the following lemmas operator \textit{K} can be
identified with $C, C^{\frac{1}{2}}, \delta C$.

\textbf{Lemma 2.2} For $\Phi : \Lambda \rightarrow  \mathbb{R}$,
\begin{enumerate}
  \item Let $\delta C^{\frac{1}{2}}$ be the operator as defined in Eq 2.15. Then
  \begin{equation}
|(\delta C^{\frac{1}{2}}   \hspace{0.05 cm}\Phi) (\xi) | \leqslant  c \hspace{0.1 cm} e^{-\gamma^{\prime} r_{\lambda}} ||\Phi||_{\infty}.
\end{equation}

 \item $C^{\frac{1}{2}}$ and $ C^{\frac{1}{2},\text{loc}}$ are invertible and 
 \begin{equation}
|(C^{-\frac{1}{2}}\hspace{0.05 cm} \Phi) (\xi)|, |(C^{-\frac{1}{2},\text{loc}}\hspace{0.05 cm}\Phi) (\xi)|
\leqslant c \hspace{0.1 cm} ||\Phi||_{\infty}.
\end{equation}

\end{enumerate}
\textit{Proof} From definition, 
\begin{equation}
\delta C^{\frac{1}{2}}  = \frac{1}{\pi}\int_{0}^{\infty}\frac{d r}{\sqrt{r}}\hspace{0.1 cm} (C_{r} - C^{\text{loc}}_{r}) = 
\frac{1}{\pi}\int_{0}^{\infty}\frac{d r}{\sqrt{r}}\hspace{0.1 cm} \delta C_{r}.
\end{equation}
Therefore,
\begin{equation}
|(\delta C^{\frac{1}{2}} \hspace{0.05 cm} \Phi) (\xi)| \leqslant  \sum_{\xi^{\prime}} 
\frac{1}{\pi}\int_{0}^{\infty}\frac{d r}{\sqrt{r}}\hspace{0.1 cm}|\delta C_{r} (\xi, \xi^{\prime})| \hspace{0.05 cm} ||\Phi||_{\infty} \leqslant 
 \sum_{\xi^{\prime}}  \frac{1}{\pi}\int_{0}^{\infty}\frac{d r}{\sqrt{r}}\hspace{0.1 cm} |(\text{T}+r)^{-1}(\xi, \xi^{\prime})| \hspace{0.05 cm} ||\Phi||_{\infty}.  
\end{equation}
To estimate the integral, we use the estimate on covariance, $(- \Delta + \mu^{2} + r)^{-1}(x,y) = 
\langle \delta_{x}, (- \Delta + \mu^{2} + r)^{-1}\delta_{y}\rangle \leqslant c \hspace{0.05 cm} e^{-\gamma |x-y|}$ and
$(-\delta d + m_{A}^{2} + r)^{-1}(b, b^{\prime}) = \langle \delta_{b}, (-\delta d + m_{A}^{2} + r)^{-1} \delta_{b^{\prime}} \rangle 
\leqslant c \hspace{0.05 cm} e^{-\gamma d(b,b^{\prime})}$, where $d(b,b^{\prime})$ is the 
infimum of distance between the sites containing bonds $b$ and $b^{\prime}$.
Note that (Eq 2.1) T is a differential operator plus a mass term. Let  $m = \text{min}(\mu^{2}, m_{A}^{2})$, denote the mass term in T. Since 
$d(\xi, \xi^{\prime}) > r_{\lambda}$, rewrite
\begin{equation}
|(\text{T}+r)^{-1}(\xi, \xi^{\prime})| \leqslant |(\text{T}+r)^{-1} (\xi, \xi^{\prime})|^{\frac{1}{4}} (m+r)^{-\frac{3}{4}}
\leqslant c \hspace{0.05 cm} e^{-\frac{\gamma}{8} r_{\lambda}} \hspace{0.05 cm} e^{-\frac{\gamma}{8}d (\xi, \xi^{\prime})} (m+r)^{-\frac{3}{4}}.
\end{equation}
Thus,
\begin{equation}
\begin{aligned}
\sum_{\xi^{\prime}}  \frac{1}{\pi}\int_{0}^{\infty}\frac{d r}{\sqrt{r}}\hspace{0.1 cm} |(\text{T}+r)^{-1}(\xi, \xi^{\prime})| 
&\leqslant \frac{1}{\pi}c \hspace{0.05 cm} e^{-\frac{\gamma}{8} r_{\lambda}} \sum_{\xi^{\prime}} e^{-\frac{\gamma}{8}d (\xi, \xi^{\prime})}
 \int_{0}^{\infty}\frac{d r}{\sqrt{r}\hspace{0.1 cm} (m+r)^{\frac{3}{4}}} \\
&\leqslant c \hspace{0.1 cm} e^{- \gamma^{\prime} r_{\lambda}}. 
\end{aligned}
\end{equation}
The above inequality is also true for $C^{\frac{1}{2},\text{loc}}$. For the second part, 
we note that operator $C$ is invertible with $C^{-1} = \text{T}$, it follows
 $C^{-\frac{1}{2}} = \text{T} C^{\frac{1}{2}}$  and thus,
\begin{equation}
|(C^{-\frac{1}{2}} \hspace{0.05 cm} \Phi)(\xi)| = |( \text{T} C^{\frac{1}{2}}\hspace{0.05 cm}\Phi) (\xi)| 
\leqslant  \mathcal{O}(m^{2}) ||C^{\frac{1}{2}} \Phi||_{\infty} \leqslant c \hspace{0.05 cm} ||\Phi||_{\infty}. 
\end{equation}
Writing $C^{\frac{1}{2},\text{loc}} = C^{\frac{1}{2}} - \delta C^{\frac{1}{2}}$, then
\begin{equation}
C^{-\frac{1}{2},\text{loc}} = C^{-\frac{1}{2}} \sum_{n=0}^{\infty} (\delta C^{\frac{1}{2}} C^{-\frac{1}{2}})^{n},
\end{equation}
the result follows easily.

\textbf{Lemma 2.3} For $\Phi : \Lambda \rightarrow  \mathbb{R}$,
with change of variables, $\Phi = C^{\frac{1}{2},\text{loc}}_{\Lambda_{1}} \Phi^{\prime}$, we have
\begin{equation}
\langle\Phi, C^{-1}_{\Lambda_{1}} \Phi\rangle =  
\langle\Phi^{\prime}, C^{\frac{1}{2},\text{loc}}_{\Lambda_{1}} C^{-1}_{\Lambda_{1}} C^{\frac{1}{2},\text{loc}}_{\Lambda_{1}} \Phi^{\prime}\rangle 
  = ||\Phi^{\prime}||_{\Lambda_{1}}^{2} + V_{\varepsilon}(\Lambda_{1}, \Phi^{\prime}). 
\end{equation}
Then $V_{\varepsilon}(\Lambda_{1}, \Phi^{\prime})$ has a local expansion in $\Box$
\begin{equation}
V_{\varepsilon} = \sum_{\Box \subset \Lambda_{1}} V_{\varepsilon} (\Box) \hspace{1 cm} \text{with}  \hspace{1 cm}
\abs{V_{\varepsilon}(\Box)} \leqslant c \hspace{0.05 cm} [r_{\lambda}]^{d} \hspace{0.05 cm} e^{- \gamma^{\prime} r_{\lambda}}
||\Phi^{\prime}||_{\infty}^{2}.
\end{equation}
\textit{Proof} Using the identity $1 = \sum_{\Box} 1_{\Box}$, where,
\begin{equation}
1_{\Box}(\xi) = 1_{\Box}(x) \hspace{1 cm} \text{if}  \hspace{1 cm} \xi = x  \hspace{1 cm} \text{or \hspace{0.5 cm} if} 
 \hspace{1 cm} \xi = (x, x + e_{\mu})
\end{equation}
and $C^{\frac{1}{2},\text{loc}} = C^{\frac{1}{2}} - \delta C^{\frac{1}{2}}$, since $C^{\frac{1}{2}}$ is self adjoint, 
rewrite $V_{\varepsilon}(\Box)$ as
\begin{equation}
\begin{aligned}
V_{\varepsilon}(\Box) &= \langle\delta C^{\frac{1}{2}} \Phi^{\prime}, 1_{\Box} C^{-1} \delta C^{\frac{1}{2}}\Phi^{\prime}\rangle  - 
2 \langle C^{-\frac{1}{2}}\Phi^{\prime}, 1_{\Box} \delta C^{\frac{1}{2}}\Phi^{\prime}\rangle \\
&= \langle C^{\frac{1}{2}} \Phi^{\prime}, 1_{\Box} \text{T} \delta C^{\frac{1}{2}}\Phi^{\prime}\rangle  - 
2 \langle C^{-\frac{1}{2}}\Phi^{\prime}, 1_{\Box} \delta C^{\frac{1}{2}}\Phi^{\prime}\rangle
\end{aligned}
\end{equation}
Then, 
\begin{equation}
\begin{aligned}
\abs{V_{\varepsilon}(\Box)} &\leqslant 
|\langle\delta C^{\frac{1}{2}} \Phi^{\prime}(\xi), 1_{\Box} \text{T} \delta C^{\frac{1}{2}}\Phi^{\prime}(\xi)\rangle|  + 
2 |\langle C^{-\frac{1}{2}}\Phi^{\prime}(\xi), 1_{\Box} \delta C^{\frac{1}{2}}\Phi^{\prime}(\xi)\rangle|  \\
&\leqslant \text{vol.} (\Box) |\delta C^{\frac{1}{2}} \Phi^{\prime}(\xi)|  |\text{T}\delta C^{\frac{1}{2}}\Phi^{\prime}(\xi)| + 
2 \hspace{0.05 cm} \text{vol.} (\Box) |C^{-\frac{1}{2}}\Phi^{\prime}(\xi)|  |\delta C^{\frac{1}{2}}\Phi^{\prime}(\xi)| \\
&\leqslant [r_{\lambda}]^{d} \hspace{0.05 cm} ||\delta C^{\frac{1}{2}} \Phi^{\prime}||_{\infty} \mathcal{O}(m^{2})
||C^{\frac{1}{2}} \Phi^{\prime}||_{\infty} + 2\hspace{0.05 cm} [r_{\lambda}]^{d} \hspace{0.05 cm}
||C^{-\frac{1}{2}}\Phi^{\prime}||_{\infty}  ||\delta C^{\frac{1}{2}}\Phi^{\prime}||_{\infty} \\
&\leqslant c \hspace{0.05 cm} [r_{\lambda}]^{d} \hspace{0.05 cm} e^{- \gamma^{\prime} r_{\lambda}} ||\Phi^{\prime}||_{\infty}^{2}.
\end{aligned}
\end{equation}
The change of variables carries with it the term $\text{det}\hspace{0.1 cm}C^{\frac{1}{2},\text{loc}}_{\Lambda_{1}}$ in partition function.
In the region $\Lambda_{1}$, using $C^{\frac{1}{2},\text{loc}}_{\Lambda_{1}} = C^{\frac{1}{2}}_{\Lambda_{1}} - 
\delta C^{\frac{1}{2}}_{\Lambda_{1}}$, rewrite
\begin{equation}
\begin{aligned}
\text{det}\hspace{0.1 cm}C^{\frac{1}{2},\text{loc}}_{\Lambda_{1}} &=  
\text{det}\hspace{0.1 cm}(C^{\frac{1}{2}}_{\Lambda_{1}} - \delta C^{\frac{1}{2}}_{\Lambda_{1}}) \\ 
&= \text{det}\hspace{0.1 cm}(C^{\frac{1}{2}}_{\Lambda_{1}}) 
\text{det}\hspace{0.1 cm}(I - C^{-\frac{1}{2}}_{\Lambda_{1}} \delta C^{\frac{1}{2}}_{\Lambda_{1}}) \\
&= \text{det}\hspace{0.1 cm}(C^{\frac{1}{2}}_{\Lambda_{1}}) 
e^{\text{Tr}\hspace{0.6 mm}\text{log}(I - C^{-\frac{1}{2}}_{\Lambda_{1}} \delta C^{\frac{1}{2}}_{\Lambda_{1}})} \\ 
&= \text{det}\hspace{0.1 cm}(C^{\frac{1}{2}}_{\Lambda_{1}}) 
e^{-\sum_{n=1}^{\infty} \frac{1}{n} \text{Tr} \big[(C^{-\frac{1}{2}}_{\Lambda_{1}}\delta C^{\frac{1}{2}}_{\Lambda_{1}})^{n}\big]}.
\end{aligned}
\end{equation}
 \textbf{Lemma 2.4} Let $W_{1}(\Lambda_{1}) = -\sum_{n=1}^{\infty} \frac{1}{n} \text{Tr} \big[(C^{-\frac{1}{2}}_{\Lambda_{1}}\delta C^{\frac{1}{2}}_{\Lambda_{1}})^{n}\big]$. Then $W_{1}$ has a local expansion in $\Box$, 
\begin{equation}
\begin{aligned}
W_{1} = \sum_{\Box \subset \Lambda_{1}} W_{1}(\Box), \hspace{1 cm} \text{with} \hspace{1 cm}
 |W_{1}(\Box)| \leqslant c \hspace{0.05 cm}[r_{\lambda}]^{d} e^{-\gamma^{\prime} r_{\lambda}}
\end{aligned}
\end{equation}
\textit{Proof.}  From the definition of $W_{1}$,
\begin{equation}
\begin{aligned}
W_{1}(\Box) = -\sum_{n=1}^{\infty} \frac{1}{n} \text{Tr} \big[1_{\Box}(C^{-\frac{1}{2}}_{\Lambda_{1}}\delta C^{\frac{1}{2}}_{\Lambda_{1}})^{n}\big]
= -\sum_{n=1}^{\infty} \frac{1}{n} \sum_{x \in \Box} \big[(C^{-\frac{1}{2}}_{\Lambda_{1}} \delta C^{\frac{1}{2}}_{\Lambda_{1}})^{n} \delta_{x} \big] (x).
\end{aligned}
\end{equation}
From the bounds on $C^{-\frac{1}{2}}$ and $\delta C^{\frac{1}{2}}$, we have
\begin{equation}
\left| \big[(C^{-\frac{1}{2}}_{\Lambda_{1}} \delta C^{\frac{1}{2}}_{\Lambda_{1}})^{n} \delta_{x} \big] (x)\right|
\leqslant (c \hspace{0.05 cm} e^{-\gamma^{\prime} r_{\lambda}})^{n} ||\delta_{x}||
\leqslant (c \hspace{0.05 cm} e^{-\gamma^{\prime} r_{\lambda}})^{n}
\end{equation}
and so $|W_{1}(\Box)| \leqslant \sum_{n=1}^{\infty} \frac{1}{n} \sum_{x} (c \hspace{0.05 cm} e^{-\gamma^{\prime} r_{\lambda}})^{n}
= [r_{\lambda}]^{d} \sum_{n=1}^{\infty} \frac{1}{n} (c \hspace{0.05 cm} e^{-\gamma^{\prime} r_{\lambda}})^{n}$ and summing over
n gives the result.
Next we want to write $\text{det} (C^{\frac{1}{2}}_{\Lambda_{1}})$ in terms of global term $\text{det} (C^{\frac{1}{2}})$ as
\begin{equation}
\text{det} (C^{\frac{1}{2}}_{\Lambda_{1}}) = \text{det} (C^{\frac{1}{2}})  e^{W_{2}}
\end{equation}
where,
\begin{equation}
\begin{aligned}
W_{2} &=   \text{Tr}\hspace{0.6 mm}\text{log}\hspace{0.5 mm}(C^{\frac{1}{2}}_{\Lambda_{1}}) - 
  \text{Tr}\hspace{0.6 mm}\text{log} \hspace{0.5 mm} (C^{\frac{1}{2}}) \\
&=  - \frac{1}{2} \text{Tr}\hspace{0.6 mm}\text{log}\hspace{0.5 mm}(\text{T}_{\Lambda_{1}}) + 
\frac{1}{2}  \text{Tr}\hspace{0.6 mm}\text{log} \hspace{0.5 mm} (\text{T}).
\end{aligned}
\end{equation}
First we prove a determent identity which is due to Balaban \cite{B}.

\textbf{Lemma 2.5} (\textit{Determinant identity}) Let \textit{K} be an invertible positive self adjoint operator. Then $\text{det} \hspace{0.1 cm} K= 
e^{\text{Tr\hspace{0.02 cm} log} \hspace{0.1 cm}K}$ where for 
any $R_{0} > 0$
\begin{equation}
 \text{log} \hspace{0.1 cm}K= K\int_{R_{0}}^{\infty} \frac{dx}{x} (K+x)^{-1} - \int_{0}^{R_{0}}dx (K+x)^{-1} +  \text{log} \hspace{0.1 cm} R_{0}.
\end{equation}
\textit{Proof} We start with resolvent equation for the function  $\text{log} \hspace{0.1 cm}K$,
\begin{equation}
 \text{log} \hspace{0.1 cm}K = \frac{1}{2\pi i} \int_{\Gamma} dz \hspace{0.1 cm} \text{log} \hspace{0.1 cm} z \hspace{0.1 cm} (z-K)^{-1}.
\end{equation}
 Construct $\Gamma$ as a simple closed curve traversed counterclockwise. Let $\theta = \text{arg}\hspace{0.1 cm} z$. Then $\Gamma$ consists of 
\begin{itemize}
  \item $\Gamma_{R} = \{z \in \mathbb{C} : |z| = R,\hspace{0.1 cm} - \pi + \epsilon \leqslant \theta \leqslant \pi - \epsilon\}$.
  \item $\Gamma_{+} = \{z \in \mathbb{C} : r \leqslant |z| \leqslant R,\hspace{0.1 cm} \theta = \pi - \epsilon \}$.
  \item $\Gamma_{r} = \{z \in \mathbb{C} : |z| = r,\hspace{0.1 cm} - \pi + \epsilon \leqslant \theta \leqslant \pi - \epsilon\}$.
  \item $\Gamma_{-} = \{z \in \mathbb{C} : r \leqslant |z| \leqslant R,\hspace{0.1 cm} \theta = - \pi + \epsilon \}$.
\end{itemize}
For \textit{R} sufficiently large and \textit{r} sufficiently small, $\Gamma$ encloses the spectrum of operator \textit{K}. The function 
$\text{log} \hspace{0.1 cm}z$ is analytic inside $\Gamma$.

Now we cut $\Gamma$ at $r < R_{0} < R$ and write it as $\Gamma = \Gamma_{<} + \Gamma_{>}$ where
$\Gamma_{<} = \Gamma \cap \{z : |z| \leqslant R_{0}\} $ and $ \Gamma_{>} = \Gamma \cap \{z : |z| \geqslant R_{0}\}$.
In the integral over $\Gamma_{>}$, we use the following identity in resolvent equation 
\begin{equation}
(z - K)^{-1} = z^{-1} + z^{-1}K(z-K)^{-1}.
\end{equation}
The integral of the first term is
\begin{equation}
\lim_{\epsilon \rightarrow 0} \frac{1}{2\pi i}\int_{\Gamma_{>}} \frac{dz}{z} \text{log} \hspace{0.1 cm}z =
\frac{1}{2\pi i}\int_{|z| = R} \frac{dz}{z} \text{log} \hspace{0.1 cm}z + \frac{1}{2\pi i}\int_{R}^{R_{0}}\frac{dx}{x} \text{log} \hspace{0.1 cm}x +
\frac{1}{2\pi i}\int_{R_{0}}^{R}\frac{dx}{x} \text{log} \hspace{0.1 cm}x
\end{equation}
where we have parametrized $z = -x$. Writing $\text{log} \hspace{0.1 cm}x = \text{log} \hspace{0.1 cm}|x| + i \pi$ above the real axis
and $\text{log} \hspace{0.1 cm}x = \text{log} \hspace{0.1 cm}|x| - i \pi$ below the real axis 
and noting that $z = R e^{i \theta}$ implies $dz/z = i \hspace{0.1 cm} d\theta$,
we get
\begin{equation}
\begin{aligned}
\lim_{\epsilon \rightarrow 0} \frac{1}{2\pi i}\int_{\Gamma_{>}} \frac{dz}{z} \text{log} \hspace{0.1 cm}z &=
\frac{1}{2\pi i}\int_{|z| = R} \frac{dz}{z} \text{log} \hspace{0.1 cm}z - 
\int_{R_{0}}^{R}\frac{dx}{x} \\  
&= \frac{1}{2\pi i}\int_{-\pi}^{\pi}i \hspace{0.1 cm}d\theta\hspace{0.1 cm} (\text{log} \hspace{0.1 cm}R + i \theta)
 - \text{log} \hspace{0.1 cm}R + \text{log} \hspace{0.1 cm}R_{0} \\  
 &= \frac{1}{2 \pi i} \text{log} \hspace{0.1 cm}R (i 2\pi) - \text{log} \hspace{0.1 cm}R  +  \text{log} \hspace{0.1 cm}R_{0} 
 \\ &=  \text{log} \hspace{0.1 cm}R_{0}.
\end{aligned}
\end{equation}
Similarly, the integral of the second term is
\begin{equation}
\lim_{\epsilon \rightarrow 0} \frac{1}{2\pi i}\int_{\Gamma_{>}} \frac{dz}{z} \text{log} \hspace{0.1 cm}z \hspace{0.1 cm} K(z-K)^{-1} =
\frac{K}{2\pi i} \int_{|z| = R}  \frac{dz}{z} \text{log} \hspace{0.1 cm}z \hspace{0.1 cm} K(z-K)^{-1} + 
K\int_{R_{0}}^{R} \frac{dx}{x}(x+K)^{-1}.
\end{equation}
Take the limit $R \rightarrow \infty$. The first term is $\mathcal{O} (R^{-1}\text{log} \hspace{0.1 cm}R)$ which converges to zero.
For the second term the integrand is $\mathcal{O}(x^{-2})$ which converges to the integral over $[R_{0}, \infty)$. Next we do the
integral over $\Gamma_{<}$ of the resolvent equation and get
\begin{equation}
\lim_{\epsilon \rightarrow 0} \frac{1}{2\pi i}\int_{\Gamma_{<}} dz \hspace{0.1 cm} \text{log} \hspace{0.1 cm}z (z-K)^{-1} = 
\frac{1}{2\pi i} \int_{|z| = r} dz \hspace{0.1 cm} \text{log} \hspace{0.1 cm}z (z-K)^{-1} - 
\int_{r}^{R_{0}} dx (x+K)^{-1}.
\end{equation}
The minus sign is due to fact that in $\Gamma_{<}$ traversing from r to $R_{0}$ is above the real axis whereas in 
$\Gamma_{>}$ it is from R to $R_{0}$ and $r < R_{0} < R$. Take the limit $r \rightarrow 0$. Since zero is not an
eigenvalue the first term is $\mathcal{O}(r \text{log} \hspace{0.1 cm} r)$ and converges to zero. For the second term
the integrand is bounded and it converges to the integral over $[0, R_{0}]$. This completes the proof.
 
Use lemma 2.5 with $R_{0} =1$ to rewrite,
\begin{equation}
\begin{aligned}
\text{Tr\hspace{0.02 cm} log \hspace{0.02 cm}T} &= -\sum_{\Box \subset \Lambda} \int_{0}^{1} dr \hspace{0.05 cm}
\text{Tr\hspace{0.05 cm}} 1_{\Box}(\text{T}+ r)^{-1} 1_{\Box} + \sum_{\Box \subset \Lambda}  \int_{1}^{\infty} \frac{dr}{r} \hspace{0.05 cm}
\text{Tr\hspace{0.05 cm}} 1_{\Box} \text{T} (\text{T}+ r)^{-1} 1_{\Box} \\
&= -\sum_{\Box \subset \Omega_{1}^{c}} \int_{0}^{1} dr \hspace{0.05 cm}
\text{Tr\hspace{0.05 cm}} 1_{\Box}(\text{T}+ r)^{-1} 1_{\Box} + \sum_{\Box \subset \Omega_{1}^{c}} \int_{1}^{\infty} \frac{dr}{r} \hspace{0.05 cm}
\text{Tr\hspace{0.05 cm}} 1_{\Box} \text{T} (\text{T}+ r)^{-1} 1_{\Box} \\
& \hspace{0.4 cm} -\sum_{\Box \subset \Omega_{1}} \int_{0}^{1} dr \hspace{0.05 cm}
\text{Tr\hspace{0.05 cm}} 1_{\Box}(\text{T}+ r)^{-1} 1_{\Box} + \sum_{\Box \subset \Omega_{1}} \int_{1}^{\infty} \frac{dr}{r} \hspace{0.05 cm}
\text{Tr\hspace{0.05 cm}} 1_{\Box} \text{T} (\text{T}+ r)^{-1} 1_{\Box} 
\end{aligned}
\end{equation}

\begin{equation}
\begin{aligned}
\text{Tr\hspace{0.02 cm} log \hspace{0.02 cm}T}_{\Lambda_{1}} &= -\sum_{\Box \subset \Lambda_{1}} \int_{0}^{1} dr \hspace{0.05 cm}
\text{Tr\hspace{0.05 cm}} 1_{\Box}(\text{T}_{\Lambda_{1}} + r)^{-1} 1_{\Box} + \sum_{\Box \subset \Lambda_{1}} 
 \int_{1}^{\infty} \frac{dr}{r} \hspace{0.05 cm}
\text{Tr\hspace{0.05 cm}} 1_{\Box} \text{T}_{\Lambda_{1}} (\text{T}_{\Lambda_{1}}+ r)^{-1} 1_{\Box} \\
&= -\sum_{\Box \subset \Lambda_{1} - \Omega_{1}} \int_{0}^{1} dr \hspace{0.05 cm}
\text{Tr\hspace{0.05 cm}} 1_{\Box}(\text{T}_{\Lambda_{1}}+ r)^{-1} 1_{\Box} + 
\sum_{\Box \subset \Lambda_{1} - \Omega_{1}} \int_{1}^{\infty} \frac{dr}{r} \hspace{0.05 cm}
\text{Tr\hspace{0.05 cm}} 1_{\Box} \text{T}_{\Lambda_{1}} (\text{T}_{\Lambda_{1}}+ r)^{-1} 1_{\Box} \\
& \hspace{0.4 cm} -\sum_{\Box \subset \Omega_{1}} \int_{0}^{1} dr \hspace{0.05 cm}
\text{Tr\hspace{0.05 cm}} 1_{\Box}(\text{T}_{\Lambda_{1}}+ r)^{-1} 1_{\Box} + 
\sum_{\Box \subset \Omega_{1}} \int_{1}^{\infty} \frac{dr}{r} \hspace{0.05 cm}
\text{Tr\hspace{0.05 cm}} 1_{\Box} \text{T}_{\Lambda_{1}} (\text{T}_{\Lambda_{1}}+ r)^{-1} 1_{\Box}. 
\end{aligned}
\end{equation}
Consider the difference of Eq 2.59 and 2.60, 
\begin{equation}
\begin{aligned}
\text{Tr\hspace{0.02 cm} log \hspace{0.02 cm}T} - \text{Tr\hspace{0.02 cm} log \hspace{0.02 cm}T}_{\Lambda_{1}} &=
\sum_{\Box \subset \Omega_{1}} \int_{0}^{1} dr \hspace{0.05 cm}
\text{Tr\hspace{0.05 cm}} (1_{\Box}(\text{T}_{\Lambda_{1}}+ r)^{-1} - (\text{T}+ r)^{-1} 1_{\Box}) \\
& -\sum_{\Box \subset \Omega_{1}^{c}} \int_{0}^{1} dr \hspace{0.05 cm} \text{Tr\hspace{0.05 cm}} 1_{\Box}(\text{T}+ r)^{-1} 1_{\Box} 
+ \sum_{\Box \subset \Lambda_{1} - \Omega_{1}} \int_{0}^{1} dr \hspace{0.05 cm}
\text{Tr\hspace{0.05 cm}} 1_{\Box}(\text{T}_{\Lambda_{1}}+ r)^{-1} 1_{\Box} \\
& + \sum_{\Box \subset \Omega_{1}} \int_{1}^{\infty} \frac{dr}{r} \hspace{0.05 cm}
\text{Tr\hspace{0.05 cm}} (1_{\Box} \text{T} (\text{T}+ r)^{-1} - \text{T}_{\Lambda_{1}} (\text{T}_{\Lambda_{1}}+ r)^{-1} 1_{\Box}) \\
& + \sum_{\Box \subset \Omega_{1}^{c}} \int_{1}^{\infty} \frac{dr}{r} \hspace{0.05 cm}
\text{Tr\hspace{0.05 cm}} 1_{\Box} \text{T} (\text{T}+ r)^{-1} 1_{\Box} \\ 
& - \sum_{\Box \subset \Lambda_{1} - \Omega_{1}} \int_{1}^{\infty} \frac{dr}{r} \hspace{0.05 cm} \text{Tr\hspace{0.05 cm}} 1_{\Box} \text{T}_{\Lambda_{1}} (\text{T}_{\Lambda_{1}}+ r)^{-1} 1_{\Box}.
\end{aligned}
\end{equation}
\textbf{Lemma 2.6} Let 
\begin{equation}
\begin{aligned}
\mathcal{A}(\Box) = \text{Tr\hspace{0.02 cm}} (1_{\Box}\hspace{0.05 cm} \mathcal{A} \hspace{0.05 cm} 1_{\Box}) &= 
\text{Tr\hspace{0.02 cm}} (1_{\Box}(\text{T}_{\Lambda_{1}}+ r)^{-1} - (\text{T}+ r)^{-1} 1_{\Box}) \\
\mathcal{B}(\Box) = \text{Tr\hspace{0.02 cm}} (1_{\Box}\hspace{0.05 cm} \mathcal{B} \hspace{0.05 cm} 1_{\Box}) &= 
\text{Tr\hspace{0.02 cm}} (1_{\Box} \text{T} (\text{T}+ r)^{-1} - \text{T}_{\Lambda_{1}} (\text{T}_{\Lambda_{1}}+ r)^{-1} 1_{\Box}).
\end{aligned}
\end{equation}
Let $m = \text{min} (\mu, m_{A})$. For $\Box \subset \Omega_{1}$ and \textit{m} sufficiently large
\begin{equation}
\begin{aligned}
|\mathcal{A}(\Box)| \leqslant \mathcal{O}(m^{-2[r_{\lambda}]})  \hspace{0.5 cm} \text{and} \hspace{0.5 cm} 
|\mathcal{B}(\Box)| \leqslant   \mathcal{O}(m^{-2[r_{\lambda}]}).
\end{aligned}
\end{equation}
 \textit{Proof} First consider the case T $ = - \Delta + \mu^{2}$. Then
\begin{equation}
(-\Delta + \mu^{2}+ r)^{-1} = (\mu^{2} + r)^{-1} \sum_{n=1}^{\infty} ((\mu^{2} + r)^{-1}\Delta)^{n}.
\end{equation}
Next we construct a \textit{random walk} expansion of the operator $(-\Delta + \mu^{2}+ r)^{-1}$,
\begin{equation}
((\mu^{2} + r)^{-1}\Delta)^{n}(x, x^{\prime}) = (\mu^{2} + r)^{-n} \sum_{x_{1},x_{2},\cdots, x_{n}}\Delta(x, x_{1}) 
\Delta(x_{1}, x_{2})\cdots \Delta(x_{n}, x^{\prime})
\end{equation}
where, $x_{1},x_{2},\cdots, x_{n}$ is a random walk $\omega$ connecting sites $x, x^{\prime}$. Note that $\Delta$ forces
$x_{i}'s \in \omega$ to be nearest neighbors. This random walk expansion exists everywhere on lattice $\Lambda$ and converges
for $\mu$ sufficiently large.
\par
Next we note that in $\Omega_{1}$, $\text{T}_{\Lambda_{1}} = \text{T}$. 
Thus, in $\mathcal{A}(\Box)$ and $\mathcal{B}(\Box)$ only those random walks contribute that join $\Omega_{1}$ with $\Lambda_{1}^{c}$ 
forcing $n \geqslant 2\hspace{0.05 cm} [r_{\lambda}]$. Since $\Delta$ is bounded,
\begin{equation}
\begin{aligned}
\text{Tr\hspace{0.02 cm}}(1_{\Box}\hspace{0.05 cm} \mathcal{A} \hspace{0.05 cm} 1_{\Box}) &=
(\mu^{2} + r)^{-1}\sum_{n=2\hspace{0.05 cm} [r_{\lambda}]}^{\infty} \sum_{x \in \Box} ((\mu^{2} + r)^{-1}\Delta)^{n}(x,x) 
\leqslant [r_{\lambda}]^{d} (\mu^{2} + r)^{- 2\hspace{0.05 cm} [r_{\lambda}]} \\
\text{Tr\hspace{0.02 cm}}(1_{\Box}\hspace{0.05 cm} \mathcal{B} \hspace{0.05 cm} 1_{\Box}) &=
(-\Delta + \mu^{2})(\mu^{2} + r)^{-1}\sum_{n=2\hspace{0.05 cm} [r_{\lambda}]}^{\infty} \sum_{x \in \Box} (\mu^{2} + r)^{-n}\Delta^{n}(x,x) 
\leqslant [r_{\lambda}]^{d} (\mu^{2}+2^{d}) (\mu^{2} + r)^{- 2\hspace{0.05 cm} [r_{\lambda}]}.
\end{aligned}
\end{equation}
The case with T $ =  \delta d + m_{A}^{2}$ can be treated in the same manner, since $ \delta d$ is also bounded.

\textbf{Lemma 2.7} $W_{2}$ (as defined in 2.51) has a local expansion in $\Box$, 
\begin{equation}
\begin{aligned}
W_{2} = \sum_{\Box \subset \Lambda} W_{2}(\Box), \hspace{0.7 cm} \text{with} \hspace{0.7 cm}
 |W_{2}(\Box)| \leqslant  \begin{cases}
\mathcal{O}(m^{-2[r_{\lambda}]}) & \hspace{0.1 cm}  \text{if}  \hspace{0.1 cm} \Box \subset \Omega_{1}, \\
\mathcal{O}([r_{\lambda}]^{d}) & \hspace{0.1 cm} \text{if}  \hspace{0.1 cm} \Box \subset \Omega_{1}^{c}.
\end{cases}
\end{aligned}
\end{equation}
\textit{Proof} From definition of $W_{2}$, for $\Box \subset \Omega_{1}$
\begin{equation}
\begin{aligned}
 W_{2}(\Box) =  \int_{0}^{1} dr \hspace{0.05 cm}  \mathcal{A}(\Box) +
  \int_{1}^{\infty} \frac{dr}{r} \hspace{0.05 cm} \mathcal{B}(\Box) 
\end{aligned}
\end{equation}
using lemma 2.6
\begin{equation}
\begin{aligned}
\abs{\int_{0}^{1} dr \hspace{0.05 cm}  \mathcal{A}(\Box)} &\leqslant  [r_{\lambda}]^{d}  \mu^{-4 [r_{\lambda}] -2} \int_{0}^{1} dr 
\leqslant [r_{\lambda}]^{d}  \mu^{-4 [r_{\lambda}] -2} \\  
\abs{\int_{1}^{\infty} \frac{dr}{r} \hspace{0.05 cm} \mathcal{B}(\Box)} &\leqslant 
[r_{\lambda}]^{d} (2^{d} + \mu^{2}) \hspace{0.05 cm} \abs{\int_{1}^{\infty} \frac{dr}{r} \hspace{0.05 cm} r^{-2}\mu^{-2([r_{\lambda}] + 1)}} 
\leqslant [r_{\lambda}]^{d}  \frac{1}{2}(2^{d} + \mu^{2}) \hspace{0.05 cm} \mu^{-2([r_{\lambda}] +1)}.
\end{aligned}
\end{equation}
For $\Box \subset \Omega_{1}^{c}$,  
\begin{equation}
\begin{aligned}
W_{2}(\Box) =  \int_{0}^{1} dr \hspace{0.05 cm} \text{Tr\hspace{0.05 cm}} 1_{\Box}(\text{T}+ r)^{-1} 1_{\Box} 
+   \int_{1}^{\infty} \frac{dr}{r} \hspace{0.05 cm} \text{Tr\hspace{0.05 cm}} 1_{\Box} \text{T} (\text{T}+ r)^{-1} 1_{\Box}   
\end{aligned}
\end{equation}
from 2.36,
\begin{equation}
\begin{aligned}
\int_{0}^{1} dr \hspace{0.05 cm} \text{Tr\hspace{0.05 cm}} 1_{\Box}(\text{T}+ r)^{-1} 1_{\Box} &= 
\int_{0}^{1} dr \hspace{0.05 cm} \sum_{\xi \in \Box} (\text{T}+ r)^{-1}(\xi, \xi)  \leqslant [r_{\lambda}]^{d} \\
\int_{1}^{\infty} \frac{dr}{r} \hspace{0.05 cm}
\text{Tr\hspace{0.05 cm}} 1_{\Box} \text{T} (\text{T}+ r)^{-1} 1_{\Box} &\leqslant (2^{d} + m^{2}) 
\int_{1}^{\infty} \frac{dr}{r^{3/2}} \sum_{\xi \in \Box} (\text{T}+ r)^{-\frac{1}{2}}(\xi, \xi) \leqslant 
2(2^{d} + m^{2}) [r_{\lambda}]^{d}.
\end{aligned}
\end{equation}
   
\section{Power series representation of 
$\Xi(\Omega_{1}, \Phi^{\prime}_{\Lambda_{1} - \Omega_{1}}, \Phi_{\Lambda_{1}^{c}}, \text{J})$}

Recall that $\Xi(\Omega_{1}, \Phi^{\prime}_{\Lambda_{1} - \Omega_{1}}, \Phi_{\Lambda_{1}^{c}}, \text{J}) = 
\int d\mu_{\text{I}}(\Phi^{\prime}_{\Omega_{1}}) \hspace{0.1 cm}
e^{-V_{1}(\Omega_{0}, \Phi^{\prime}_{\Lambda_{1}}, \Phi_{\Lambda^{c}_{1}}, \text{J})} \hat{\chi}(\Phi^{\prime}_{\Omega_{1}})$,
where, $V_{1}(\Omega_{0}, \Phi^{\prime}_{\Lambda_{1}}, \Phi_{\Lambda^{c}_{1}}, \text{J})$ is given by 2.26
\begin{equation}
\begin{aligned}
V_{1}(\Omega_{0}, \Phi^{\prime}_{\Lambda_{1}}, \Phi_{\Lambda^{c}_{1}}, \text{J})
&= - W_{1}(\Lambda_{1}) - W_{2}(\Omega_{0})  + 
 V(\Omega_{0}, C^{\frac{1}{2},\text{loc}}_{\Lambda_{1}} \Phi^{\prime} - C_{\Lambda_{1}}\text{T}_{\Lambda_{1}\Lambda^{c}_{1}}
  \Phi_{\Lambda^{c}_{1}}) \\
& + V_{\varepsilon}(\Lambda_{1}, \Phi^{\prime}) -
e_{0} \langle C^{\frac{1}{2},\text{loc}}_{\Lambda_{1}} \Phi^{\prime} - C_{\Lambda_{1}}\text{T}_{\Lambda_{1}\Lambda^{c}_{1}} 
\Phi_{\Lambda^{c}_{1}}, \delta \text{J}\rangle. 
\end{aligned}
\end{equation}
Note that $W_{1}(\Lambda_{1})$ and $W_{2}(\Omega_{0})$ are independent of $\Phi^{\prime}$. 
We want to construct a power series representation of 
$\text{ln} \hspace{0.1 cm} \Xi(\Omega_{1}, \Phi^{\prime}_{\Lambda_{1} - \Omega_{1}}, \Phi_{\Lambda_{1}^{c}}, \text{J})$, where
ln denotes natural logarithm.
 
\textit{Notation}.  Let $\xi \in \Lambda \cup \Lambda^{\ast}$. Then for example,  
\begin{equation}
\begin{aligned}
\sum_{x \in \Lambda} \rho(x)^{4} &= \sum_{\xi} \Phi^{\prime} (\xi)^{4} 1_{\xi \in \Lambda}, \\
\sum_{x \in \Lambda, x^{\prime} \in \Lambda, b \in \Lambda^{\ast}} \rho(x)\rho(x^{\prime}) A(b) &= \sum_{\xi, \xi^{\prime}, \xi^{\prime\prime}} 
\Phi^{\prime} (\xi) \Phi^{\prime} (\xi^{\prime}) \Phi^{\prime} (\xi^{\prime\prime}) 1_{\xi \in \Lambda}
1_{\xi^{\prime} \in \Lambda} 1_{\xi^{\prime\prime} \in \Lambda^{\ast}}.
\end{aligned}
\end{equation}
where, $1_{\xi \in \Lambda}$ is the characteristic function that restricts $\xi \in \Lambda \cup \Lambda^{\ast}$ to
$\xi \in \Lambda$.
 
\textit{Power series}. Let $X \subset \Lambda$. Let $f(\Phi, \Psi)$ be smooth and analytic function 
of the fields $\Phi, \Psi$ defined on \textit{X}. Then a power series expansion of \textit{f} is given by
\begin{equation}
\begin{aligned}
f(\Phi, \Psi) = \sum_{n, m \geqslant 0} \sum_{\substack{\xi_{1}, \cdots,  \xi_{n} \in X \\ \eta_{1}, \cdots, \eta_{m} \in X}} 
 a (\xi_{1}, \cdots \xi_{n}, \eta_{1}, \cdots, \eta_{m}) 
 \hspace{0.1 cm} \Phi (\xi_{1}) \cdots \Phi (\xi_{n}) \Psi (\eta_{1}) \cdots \Psi (\eta_{m}).
\end{aligned} 
\end{equation}
Define a n-component vector $\vec{\xi}$ as  
\begin{equation}
\vec{\xi} = \{\xi_{1}, \xi_{2}, \cdots, \xi_{n} \}.
\end{equation}
Then for a n-component vector $\vec{\xi}$, write
\begin{equation}
\Phi (\vec{\xi}) = \Phi (\xi_{1}) \cdots \Phi (\xi_{n}).
\end{equation}
Rewrite power series expansion of \textit{f} as
\begin{equation}
f(\Phi, \Psi) =   \sum_{n, m \geqslant 0} \sum_{\substack{\vec{\xi} = (\xi_{1}, \cdots,  \xi_{n}) \in X \\ 
\vec{\eta} = (\eta_{1}, \cdots, \eta_{m}) \in X}}  
a (\vec{\xi}, \vec{\eta}) \hspace{0.1 cm} \Phi (\vec{\xi}) \Psi (\vec{\eta}).
\end{equation}
Let 
\begin{equation}
a_{\Psi}(\vec{\xi}) = \sum_{\vec{\eta} \in X} a (\vec{\xi}, \vec{\eta}) \hspace{0.1 cm} \Psi (\vec{\eta}) 
\end{equation}
and rewrite,
\begin{equation}
f(\Phi, \Psi) = \sum_{\vec{\xi} \in X} a_{\Psi} (\vec{\xi}) \hspace{0.1 cm} \Phi (\vec{\xi}) 
\end{equation}
such that
\begin{equation}
\begin{aligned}
e^{f(\Phi, \Psi)} &= \sum_{l = 0}^{\infty} \frac{1}{l!} f(\Phi, \Psi)^{l}  
= 1 +  \sum_{l = 1}^{\infty} \frac{1}{l!}  \sum_{\vec{\xi}_{1},\cdots, \vec{\xi}_{l} \in X}
a_{\Psi}(\vec{\xi}_{1}) \cdots a_{\Psi} (\vec{\xi}_{l})
\hspace{0.1 cm} \Phi (\vec{\xi}_{1}) \cdots \Phi (\vec{\xi}_{l}).
\end{aligned}
\end{equation}
First set 
$\Phi_{\Lambda^{c}_{1}} = 0$ and rewrite 
\begin{equation}
V_{1}(\Omega_{0}, \Phi^{\prime}_{\Lambda_{1}}, \text{J}) = V(\Omega_{0}, C^{\frac{1}{2},\text{loc}}_{\Lambda_{1}} \Phi^{\prime})
+ V_{\varepsilon}(\Lambda_{1}, \Phi^{\prime}) - e_{0} \langle C^{\frac{1}{2},\text{loc}}_{\Lambda_{1}} \Phi^{\prime}, \delta \text{J}\rangle. 
\end{equation}
Next note that
\begin{equation}
C^{\frac{1}{2},\text{loc}}_{\Lambda_{1}} \Phi^{\prime} = C^{\frac{1}{2},\text{loc}}_{\Lambda_{1}}
(\Phi^{\prime}_{\Lambda_{1} - \Omega_{1}}, \Phi^{\prime}_{\Omega_{1}}).
\end{equation}
Let
\begin{equation}
\Phi = \Phi^{\prime}_{\Omega_{1}}, \hspace{1 cm} \Psi = (\Psi_{1}, \Psi_{2}) = (\Phi^{\prime}_{\Lambda_{1} - \Omega_{1}}, \text{J}).
\end{equation}
Thus, $V_{1}(\Omega_{0}, \Phi^{\prime}_{\Lambda_{1}}, \text{J}) = V_{1}(\Omega_{0}, \Phi, \Psi)$ and
$\Xi(\Omega_{1}, \Phi^{\prime}_{\Lambda_{1} - \Omega_{1}}, \text{J}) = \Xi(\Omega_{1}, \Psi)$.

We are interested in the case $f(\Phi, \Psi) = V_{1}(\Omega_{0}, \Phi, \Psi)$ and
\begin{equation} 
\Xi(\Omega_{1}, \Psi) = \int d\mu_{\text{I}}(\Phi) \hat{\chi}(\Phi) \hspace{0.1 cm} e^{f(\Phi, \Psi)}. 
\end{equation}
We want to write a power series of $\text{ln} \hspace{0.1 cm} \Xi(\Omega_{1}, \Psi)$ as
\begin{equation}
\begin{aligned}
\text{ln} \hspace{0.1 cm} \Xi(\Omega_{1}, \Psi) &= 
b_{0} +  \sum_{\alpha_{1}} b(\eta_{1}) \Psi_{\alpha_{1}} (\eta_{1}) \hspace{0.1 cm} + 
\sum_{\alpha_{1}, \alpha_{2}} b(\eta_{1}, \eta_{2}) \Psi_{\alpha_{1}} (\eta_{1}) \Psi_{\alpha_{2}} (\eta_{2}) + \cdots \\  
&=  \sum_{m \geqslant 0}  \sum_{\vec{\eta} =  (\eta_{1}, \cdots, \eta_{m})} 
\sum_{\alpha_{1},\cdots, \alpha_{m}} b(\vec{\eta}) \Psi_{\alpha} (\vec{\eta}). 
\end{aligned}
\end{equation}
Note that $\text{ln} \hspace{0.1 cm} \Xi(\Omega_{1}, 0) = b_{0}$.
As a first step to construct a power series written above, we prove theorem 3.1.  

\textbf{Theorem 3.1} Let $f(\Phi, \Psi) = V_{1}(\Omega_{0}, \Phi, \Psi)$.
Then there exists a coefficient system $a(\vec{\xi}, \vec{\eta})$ such that
\begin{equation}
V_{1}(\Omega_{0}, \Phi, \Psi) = \sum_{\substack{n, m \\ n + m > 0}} 
 \sum_{\substack{\xi_{1}, \cdots,  \xi_{n} \\ \eta_{1}, \cdots, \eta_{m}}} a(\xi_{1}, \cdots,  \xi_{n}, \eta_{1}, \cdots, \eta_{m})
 \Phi (\xi_{1})\cdots \Phi (\xi_{n}) \Psi_{\alpha} (\eta_{1})\cdots \Psi_{\alpha} (\eta_{m})
\end{equation}
with
\begin{equation}
\abs{a (\xi_{1}, \cdots \xi_{n}, \eta_{1}, \cdots, \eta_{m})}  \leqslant c \hspace{0.05 cm} (c_{0} \hspace{0.05 cm} e_{0})^{\frac{n+m}{2}} 
\hspace{0.05 cm} e^{- \gamma^{\prime}t(\xi_{1}, \cdots,  \xi_{n},  \eta_{1}, \cdots, \eta_{m})}
\end{equation}
where, \textit{c} and $c_{0}$ are constants and $t(\xi_{1}, \cdots, \xi_{n}, \eta_{1}, \cdots, \eta_{m})$ is the length of minimal tree.
 
\textit{Proof} Recall that $V_{1}(\Omega_{0}, \Phi, \Psi) = V(\Omega_{0}, C^{\frac{1}{2},\text{loc}}_{\Lambda_{1}} \Phi^{\prime})
+ V_{\varepsilon}(\Lambda_{1}, \Phi^{\prime}) - e_{0} \langle C^{\frac{1}{2},\text{loc}}_{\Lambda_{1}} \Phi^{\prime}, \delta \text{J}\rangle$
and
\begin{equation}
\begin{aligned}
V & (\Omega_{0}, C^{\frac{1}{2},\text{loc}}_{\Lambda_{1}} \Phi^{\prime}) =  \rho_{0}^{2}  \sum_{\xi} 
(1 -\text{cos}\hspace{0.05 cm} [e_{0}\hspace{0.05 cm}C^{\frac{1}{2},\text{loc}}_{\Lambda_{1}} \Phi^{\prime}(\xi)] 1_{\xi \in \Omega_{0}^{\ast}}
 - \frac{1}{2}e_{0}^{2}\hspace{0.05 cm} (C^{\frac{1}{2},\text{loc}}_{\Lambda_{1}}\Phi^{\prime}(\xi))^{2} 1_{\xi \in \Omega_{0}^{\ast}}) \\
&+ \rho_{0} \sum_{\xi, \xi^{\prime}} (C^{\frac{1}{2},\text{loc}}_{\Lambda_{1}}\Phi^{\prime} (\xi) 1_{\xi \in \Omega_{0}} + 
C^{\frac{1}{2},\text{loc}}_{\Lambda_{1}}\Phi^{\prime} (\xi^{\prime}) 1_{\xi^{\prime} \in \Omega_{0}})
 (1 - \text{cos}\hspace{0.05 cm} [e_{0} 
 C^{\frac{1}{2},\text{loc}}_{\Lambda_{1}}\Phi^{\prime} (\xi^{\prime\prime})]1_{|\xi - \xi^{\prime}| = 1} 1_{\xi^{\prime\prime} = (\xi, \xi^{\prime})}) \\ 
& + \sum_{\xi, \xi^{\prime}} C^{\frac{1}{2},\text{loc}}_{\Lambda_{1}}\Phi^{\prime}(\xi) \hspace{0.05 cm} 
C^{\frac{1}{2},\text{loc}}_{\Lambda_{1}}\Phi^{\prime} (\xi^{\prime})
(1 - \text{cos}\hspace{0.05 cm}[e_{0}C^{\frac{1}{2},\text{loc}}_{\Lambda_{1}}\Phi^{\prime} (\xi^{\prime\prime})]) 
1_{\xi \in \Omega_{0}} 1_{\xi^{\prime} \in \Omega_{0}} 1_{|\xi - \xi^{\prime}| = 1} 1_{\xi^{\prime\prime} = (\xi, \xi^{\prime})}  \\
& + \sum_{\xi} \Big(\lambda \hspace{0.05 cm} (C^{\frac{1}{2},\text{loc}}_{\Lambda_{1}}\Phi^{\prime}(\xi))^{4}  + 
\sqrt{2\lambda} \mu\hspace{0.05 cm}(C^{\frac{1}{2},\text{loc}}_{\Lambda_{1}}\Phi^{\prime}(\xi))^{3} - 
\text{log}\Big[ 1+\frac{C^{\frac{1}{2},\text{loc}}_{\Lambda_{1}}\Phi^{\prime}(\xi)}{\rho_{0}}\Big]\Big) 1_{\xi \in \Omega_{0}}. 
\end{aligned}
\end{equation}
Note that
\begin{equation}
C^{\frac{1}{2},\text{loc}}_{\Lambda_{1}}\Phi^{\prime} (\xi) =  (C^{\frac{1}{2},\text{loc}}_{\Lambda_{1}}\Phi) (\xi) +
(C^{\frac{1}{2},\text{loc}}_{\Lambda_{1}}\Psi_{1}) (\xi). 
\end{equation}

\begin{enumerate}
  \item First rewrite 
\begin{equation}
\begin{aligned}
\sum_{\xi} (\text{cos} &[e_{0} \hspace{0.05 cm}  C^{\frac{1}{2},\text{loc}}_{\Lambda_{1}}\Phi^{\prime} (\xi)] - 1)
1_{\xi \in\Omega_{0}^{\ast}} =  \sum_{\xi} \sum_{\substack{l : \text{even} \\ l > 0}} (-1)^{l/2}  
 \frac{(e_{0} \hspace{0.05 cm} C^{\frac{1}{2},\text{loc}}_{\Lambda_{1}}\Phi^{\prime} (\xi))^{l}}{l!} 1_{\xi \in \Omega_{0}^{\ast}}
\\ 
&= \sum_{\xi} \sum_{\substack{l : \text{even} \\ l > 0}} (-1)^{l/2}  
 \frac{[e_{0} \hspace{0.05 cm} C^{\frac{1}{2},\text{loc}}_{\Lambda_{1}}\Phi (\xi) + e_{0} \hspace{0.05 cm}
 C^{\frac{1}{2},\text{loc}}_{\Lambda_{1}}\Psi_{1} (\xi)]^{l}}{l!} 1_{\xi \in \Omega_{0}^{\ast}}
\\  
&= \sum_{\xi} \sum_{l : \text{even}} \frac{(-1)^{l/2} }{l!} \sum_{\substack{n, m  \\ n + m = l}} \frac{l!}{n! m!}
(e_{0} \hspace{0.05 cm} C^{\frac{1}{2},\text{loc}}_{\Lambda_{1}}\Phi (\xi))^{n} \hspace{0.05 cm}
(e_{0} \hspace{0.05 cm} C^{\frac{1}{2},\text{loc}}_{\Lambda_{1}}\Psi_{1} (\xi))^{m} 1_{\xi \in \Omega_{0}^{\ast}} \\
&= \sum_{\xi} \sum_{\substack{n, m  \\ n + m : \text{even}}} e_{0}^{n+m} \hspace{0.05 cm} \frac{(-1)^{(n+m)/2}}{n! m!}
\sum_{\substack{\xi_{1}, \cdots,  \xi_{n} \\ \eta_{1}, \cdots, \eta_{m}}}
C^{\frac{1}{2},\text{loc}}_{\Lambda_{1}}(\xi, \xi_{1})\Phi (\xi_{1}) \cdots  
C^{\frac{1}{2},\text{loc}}_{\Lambda_{1}}(\xi, \xi_{n})\Phi (\xi_{n}) \\
& \hspace{2 cm} C^{\frac{1}{2},\text{loc}}_{\Lambda_{1}}(\xi, \eta_{1})\Psi_{1}(\eta_{1}) \cdots  
C^{\frac{1}{2},\text{loc}}_{\Lambda_{1}}(\xi, \eta_{m})\Psi_{1}(\eta_{m}) 
1_{\xi \in \Omega_{0}^{\ast}} 1_{\xi_{j} \in \Lambda_{1}^{\ast}} 1_{\eta_{j} \in \Lambda_{1}^{\ast}} \\
&=  \sum_{\substack{n, m  \\ n + m : \text{even}}} \sum_{\substack{\xi_{1}, \cdots,  \xi_{n} \\ \eta_{1}, \cdots, \eta_{m}}}
a(\xi_{1}, \cdots,  \xi_{n}, \eta_{1}, \cdots, \eta_{m})
\Phi (\xi_{1})\cdots \Phi (\xi_{n}) \Psi_{1} (\eta_{1})\cdots \Psi_{1} (\eta_{m})
\end{aligned}
\end{equation}
where,
\begin{equation}
\begin{aligned}
a(\xi_{1}, \cdots,  \xi_{n}, \eta_{1}, \cdots, \eta_{m}) &= e_{0}^{n+m} \hspace{0.05 cm}  \frac{(-1)^{(n+m)/2}}{n! m!}
\sum_{\xi}  C^{\frac{1}{2},\text{loc}}_{\Lambda_{1}}(\xi, \xi_{1}) \cdots  
C^{\frac{1}{2},\text{loc}}_{\Lambda_{1}}(\xi, \xi_{n}) \\ & C^{\frac{1}{2},\text{loc}}_{\Lambda_{1}}(\xi, \eta_{1}) \cdots 
C^{\frac{1}{2},\text{loc}}_{\Lambda_{1}}(\xi, \eta_{m}) 
1_{\xi \in \Omega_{0}^{\ast}} 1_{\xi_{j} \in \Lambda_{1}^{\ast}} 1_{\eta_{j} \in \Lambda_{1}^{\ast}}.
\end{aligned}
\end{equation}
Let $\tilde{\xi}_{j} = \{\xi_{j}, \eta_{j}\}$.
From lemma 2.2, $C^{\frac{1}{2},\text{loc}}_{\Lambda_{1}}(\xi, \tilde{\xi}_{j}) \leqslant c \hspace{0.05 cm} e^{- \frac{\gamma}{8} d(\xi, \tilde{\xi}_{j})}$,
where $d(\xi, \tilde{\xi}_{j})$ denotes the infimum of the distance between the sites containing the bonds $\xi, \tilde{\xi}_{j}$. Using
\begin{equation}
\sum_{j} d(\xi, \tilde{\xi}_{j}) = \text{length of tree joining} \hspace{0.1 cm} (\xi, \xi_{1}, \cdots, \xi_{n},  \eta_{1}, \cdots, \eta_{m}) 
\geqslant t(\xi_{1}, \cdots, \xi_{n}, \eta_{1}, \cdots, \eta_{m})
\end{equation}
write 
\begin{equation}
\begin{aligned}
\abs{a(\xi_{1}, \cdots,  \xi_{n}, \eta_{1}, \cdots, \eta_{m})} &\leqslant c \hspace{0.05 cm}
\frac{e_{0}^{n+m}}{n! m!} \sum_{\xi} e^{-  \frac{\gamma}{8} d(\xi, \xi_{1}) - \cdots -  \frac{\gamma}{8} d(\xi, \xi_{n}) - 
\frac{\gamma}{8} d(\xi, \eta_{1}) - \cdots - \frac{\gamma}{8} d(\xi, \eta_{m})} \\
&\leqslant c \hspace{0.05 cm} \frac{e_{0}^{n+m}}{n! m!} \hspace{0.05 cm} e^{-\frac{\gamma}{16} t(\xi_{1}, \cdots, \xi_{n}, \eta_{1}, \cdots, \eta_{m})} 
\sum_{\xi} e^{- \frac{\gamma}{16} \sum_{j} d(\xi, \tilde{\xi}_{j})} \\
&\leqslant c \hspace{0.05 cm} \frac{e_{0}^{n+m}}{n! m!}
\hspace{0.05 cm} e^{-\frac{\gamma}{16} t(\xi_{1}, \cdots, \xi_{n}, \eta_{1}, \cdots, \eta_{m})}.
\end{aligned}
\end{equation}
\item Similarly, rewrite
\begin{equation}
\begin{aligned}
\sum_{\xi} \lambda \hspace{0.05 cm}& (C^{\frac{1}{2},\text{loc}}_{\Lambda_{1}}\Phi^{\prime} (\xi))^{4} 1_{\xi \in \Omega_{0}} 
= \sum_{\xi} \lambda \hspace{0.05 cm}[C^{\frac{1}{2},\text{loc}}_{\Lambda_{1}}\Phi (\xi) + 
C^{\frac{1}{2},\text{loc}}_{\Lambda_{1}}\Psi_{1} (\xi)]^{4} 1_{\xi \in \Omega_{0}} \\
&= \sum_{\xi} \lambda \sum_{\substack{n, m  \\ n + m = 4}} \frac{(n+m)!}{n! m!} (C^{\frac{1}{2},\text{loc}}_{\Lambda_{1}}\Phi (\xi))^{n}
\hspace{0.05 cm} (C^{\frac{1}{2},\text{loc}}_{\Lambda_{1}}\Psi_{1} (\xi))^{m} 1_{\xi \in \Omega_{0}} \\
&= \sum_{\xi} \lambda \sum_{\substack{n, m  \\ n + m = 4}} \frac{(n+m)!}{n! m!}
\sum_{\substack{\xi_{1}, \cdots,  \xi_{n} \\ \eta_{1}, \cdots, \eta_{m}}} C^{\frac{1}{2},\text{loc}}_{\Lambda_{1}}(\xi, \xi_{1})\Phi (\xi_{1})
\cdots  C^{\frac{1}{2},\text{loc}}_{\Lambda_{1}}(\xi, \xi_{n})\Phi (\xi_{n}) \\
& \hspace{1 cm} C^{\frac{1}{2},\text{loc}}_{\Lambda_{1}}(\xi, \eta_{1}) \Psi_{1} (\eta_{1}) \cdots 
C^{\frac{1}{2},\text{loc}}_{\Lambda_{1}}(\xi, \eta_{m}) \Psi_{1} (\eta_{m})
1_{\xi \in \Omega_{0}} 1_{\xi_{j} \in \Lambda_{1}} 1_{\eta_{j} \in \Lambda_{1}} \\
&=\sum_{\substack{n, m  \\ n + m = 4}} 
\sum_{\substack{\xi_{1}, \cdots,  \xi_{n} \\ \eta_{1}, \cdots, \eta_{m}}} a(\xi_{1}, \cdots,  \xi_{n}, \eta_{1}, \cdots, \eta_{m})
\Phi (\xi_{1})\cdots\Phi (\xi_{n}) \Psi_{1} (\eta_{1}) \cdots \Psi_{1} (\eta_{m})
\end{aligned}
\end{equation}
where,
\begin{equation}
\begin{aligned}
a(\xi_{1}, \cdots,  \xi_{n}, \eta_{1}, \cdots, \eta_{m}) &= \lambda \hspace{0.05 cm} \frac{(n+m)!}{n! m!} \hspace{0.05 cm} \sum_{\xi}  
C^{\frac{1}{2},\text{loc}}_{\Lambda_{1}}(\xi, \xi_{1}) \cdots  C^{\frac{1}{2},\text{loc}}_{\Lambda_{1}}(\xi, \xi_{n}) \\ 
& C^{\frac{1}{2},\text{loc}}_{\Lambda_{1}}(\xi, \eta_{1}) \cdots 
C^{\frac{1}{2},\text{loc}}_{\Lambda_{1}}(\xi, \eta_{m}) 1_{\xi \in \Omega_{0}} 1_{\xi_{j} \in \Lambda_{1}} 1_{\eta_{j} \in \Lambda_{1}}.
\end{aligned}
\end{equation}
Note that $\lambda = \mathcal{O}(e_{0}^{2})$, using 
$C^{\frac{1}{2},\text{loc}}_{\Lambda_{1}}(\xi, \tilde{\xi}_{j}) \leqslant c \hspace{0.05 cm} e^{-  \frac{\gamma}{8} |\xi - \tilde{\xi}_{j}|}$ 
and $\frac{(n+m)!}{n! m!} \leqslant 2^{n+m} = 16$, from Eq 3.21 and 3.22,
\begin{equation}
\abs{a(\xi_{1}, \cdots, \xi_{n}, \eta_{1}, \cdots, \eta_{m})} \leqslant
c \hspace{0.05 cm} (4 \hspace{0.05 cm} e_{0})^{\frac{n+m}{2}} \hspace{0.05 cm} 
e^{-\frac{\gamma}{16} t(\xi_{1}, \cdots, \xi_{n}, \eta_{1}, \cdots, \eta_{m})}.
\end{equation}
 \item Next note that in the log term, $\abs{\frac{\Phi^{\prime} (\xi) 1_{\xi \in \Omega_{0}}}{\rho_{0}}} < 1$
($-\rho_{0} < \Phi^{\prime} (\xi)1_{\xi \in \Omega_{0}} < p_{\lambda}$ and $p_{\lambda} < \rho_{0}$, since
$\rho_{0} = \mathcal{O}(\lambda^{-1/2})$). Rewrite
\begin{equation}
\begin{aligned}
\sum_{\xi} &\Big(\text{log} \Big[1+\frac{C^{\frac{1}{2},\text{loc}}_{\Lambda_{1}}\Phi^{\prime} (\xi) 1_{\xi \in \Omega_{0}}}{\rho_{0}} \Big]\Big) = 
\sum_{\xi} \sum_{l}  (-1)^{l+1} \frac{(C^{\frac{1}{2},\text{loc}}_{\Lambda_{1}} \Phi^{\prime} (\xi))^{l}}{l \rho_{0}^{l}} 1_{\xi \in  \Omega_{0}} 
  \\ 
&= \sum_{\xi} \sum_{l}  (-1)^{l+1} \frac{[C^{\frac{1}{2},\text{loc}}_{\Lambda_{1}} \Phi (\xi) + 
C^{\frac{1}{2},\text{loc}}_{\Lambda_{1}} \Psi_{1} (\xi)]^{l}}{l \rho_{0}^{l}} 1_{\xi \in  \Omega_{0}}   \\ 
&=\sum_{\xi} \sum_{l > 0}  (-1)^{l+1} \frac{1}{l \rho_{0}^{l}} \sum_{\substack{n, m \\ n + m = l}}  \frac{l!}{n! m!}  
(C^{\frac{1}{2},\text{loc}}_{\Lambda_{1}} \Phi (\xi))^{n} \hspace{0.05 cm}
(C^{\frac{1}{2},\text{loc}}_{\Lambda_{1}} \Psi_{1} (\xi))^{m}1_{\xi \in  \Omega_{0}}  \\
&= \sum_{\xi} \sum_{\substack{n, m \\ n + m > 0}}  (-1)^{n+m+1} \frac{1}{(n+m) \rho_{0}^{n+m}} \frac{(n+m)!}{n! m!}   
\sum_{\substack{\xi_{1}, \cdots,  \xi_{n} \\ \eta_{1}, \cdots, \eta_{m}}} C^{\frac{1}{2},\text{loc}}_{\Lambda_{1}}(\xi, \xi_{1})\Phi (\xi_{1}) 
\cdots   \\
& \hspace{0.5 cm} C^{\frac{1}{2},\text{loc}}_{\Lambda_{1}}(\xi, \xi_{n}) \Phi (\xi_{n}) C^{\frac{1}{2},\text{loc}}_{\Lambda_{1}}(\xi, \eta_{1}) 
\Psi_{1} (\eta_{1}) \cdots C^{\frac{1}{2},\text{loc}}_{\Lambda_{1}}(\xi, \eta_{m}) \Psi_{1} (\eta_{m})
1_{\xi \in \Omega_{0}} 1_{\xi_{j} \in \Lambda_{1}} 1_{\eta_{j} \in \Lambda_{1}}  \\
&= \sum_{\substack{n, m \\ n + m > 0}} \sum_{\substack{\xi_{1}, \cdots,  \xi_{n} \\ \eta_{1}, \cdots, \eta_{m}}}
a(\xi_{1}, \cdots,  \xi_{n}, \eta_{1}, \cdots, \eta_{m})
\Phi (\xi_{1})\cdots \Phi (\xi_{n}) \Psi_{1} (\eta_{1})\cdots \Psi_{1} (\eta_{m})  
\end{aligned}
\end{equation}
where,
\begin{equation}
\begin{aligned}
a(\xi_{1}, \cdots,  \xi_{n}, \eta_{1}, \cdots, \eta_{m}) = (-1)^{n+m+1}& \frac{1}{(n+m) \rho_{0}^{n+m}} \frac{(n+m)!}{n! m!}  
\sum_{\xi}  C^{\frac{1}{2},\text{loc}}_{\Lambda_{1}}(\xi, \xi_{1}) \cdots  
C^{\frac{1}{2},\text{loc}}_{\Lambda_{1}}(\xi, \xi_{n}) \\ & C^{\frac{1}{2},\text{loc}}_{\Lambda_{1}}(\xi, \eta_{1}) \cdots 
C^{\frac{1}{2},\text{loc}}_{\Lambda_{1}}(\xi, \eta_{m}) 1_{\xi \in \Omega_{0}} 1_{\xi_{j} \in \Lambda_{1}} 1_{\eta_{j} \in \Lambda_{1}}.
\end{aligned}
\end{equation}
Since $\rho_{0} = \frac{\mu}{\sqrt{8\lambda}}$ and $\lambda = \mathcal{O}(e_{0}^{2})$, using 
$C^{\frac{1}{2},\text{loc}}_{\Lambda_{1}}(\xi, \tilde{\xi}_{j}) \leqslant c \hspace{0.05 cm} e^{-  \frac{\gamma}{8} |\xi - \tilde{\xi}_{j}|}$
and $\frac{(n+m)!}{n! m!} \leqslant 2^{n+m}$, from Eq 21 and 22,
\begin{equation}
\begin{aligned}
\abs{a(\xi_{1}, \cdots, \xi_{n}, \eta_{1}, \cdots, \eta_{m})} &\leqslant
c \hspace{0.05 cm}\frac{1}{(n+m)}  \Big(\frac{2}{\rho_{0}}\Big)^{n+m} 
 \hspace{0.05 cm} e^{-\frac{\gamma}{16} t(\xi_{1}, \cdots, \xi_{n}, \eta_{1}, \cdots, \eta_{m})} \\
&\leqslant c \hspace{0.05 cm} (2\hspace{0.05 cm} e_{0})^{n+m}  \hspace{0.05 cm} 
e^{-\frac{\gamma}{16} t(\xi_{1}, \cdots, \xi_{n}, \eta_{1}, \cdots, \eta_{m})}.
\end{aligned}
\end{equation}
\item Next consider the term with $\xi, \xi^{\prime}$ as nearest neighbor. First note that
\begin{equation}
\begin{aligned}
C^{\frac{1}{2},\text{loc}}_{\Lambda_{1}}\Phi^{\prime}(\xi) \hspace{0.05 cm} C^{\frac{1}{2},\text{loc}}_{\Lambda_{1}}\Phi^{\prime} (\xi^{\prime})
&= (C^{\frac{1}{2},\text{loc}}_{\Lambda_{1}}\Phi (\xi) + C^{\frac{1}{2},\text{loc}}_{\Lambda_{1}}\Psi_{1}(\xi)) \hspace{0.05 cm} 
(C^{\frac{1}{2},\text{loc}}_{\Lambda_{1}}\Phi (\xi^{\prime}) + C^{\frac{1}{2},\text{loc}}_{\Lambda_{1}}\Psi_{1} (\xi^{\prime})) \\
&= C^{\frac{1}{2},\text{loc}}_{\Lambda_{1}}\Phi (\xi) \hspace{0.05 cm} C^{\frac{1}{2},\text{loc}}_{\Lambda_{1}}\Phi (\xi^{\prime})
+ C^{\frac{1}{2},\text{loc}}_{\Lambda_{1}}\Phi (\xi) \hspace{0.05 cm} C^{\frac{1}{2},\text{loc}}_{\Lambda_{1}}\Psi_{1} (\xi^{\prime}) \\
& + C^{\frac{1}{2},\text{loc}}_{\Lambda_{1}}\Psi_{1}(\xi)\hspace{0.05 cm} C^{\frac{1}{2},\text{loc}}_{\Lambda_{1}}\Phi (\xi^{\prime}) 
+ C^{\frac{1}{2},\text{loc}}_{\Lambda_{1}}\Psi_{1}(\xi)\hspace{0.05 cm} C^{\frac{1}{2},\text{loc}}_{\Lambda_{1}}\Psi_{1} (\xi^{\prime}).
\end{aligned}
\end{equation}
We use $C^{\frac{1}{2},\text{loc}}_{\Lambda_{1}}\Phi (\xi) \hspace{0.05 cm} 
C^{\frac{1}{2},\text{loc}}_{\Lambda_{1}}\Phi (\xi^{\prime})$ to extract the coefficients of power series.
Let $\xi^{\prime\prime} = (\xi, \xi^{\prime})$.
\begin{equation}
\begin{aligned}
&\sum_{\xi, \xi^{\prime}} C^{\frac{1}{2},\text{loc}}_{\Lambda_{1}}\Phi (\xi) \hspace{0.05 cm} 
C^{\frac{1}{2},\text{loc}}_{\Lambda_{1}}\Phi (\xi^{\prime})  
(1 - \text{cos}\hspace{0.05 cm}[e_{0}C^{\frac{1}{2},\text{loc}}_{\Lambda_{1}}\Phi^{\prime} (\xi^{\prime\prime})])   \\
&= \sum_{\substack{\xi, \xi^{\prime}, \xi^{\prime\prime} \\ |\xi - \xi^{\prime}| = 1}} 
\sum_{l : \text{even}} \frac{(-1)^{l/2} }{l!} \sum_{\substack{n, m  \\ n + m = l}} \frac{l!}{n! m!} \hspace{0.05 cm}
C^{\frac{1}{2},\text{loc}}_{\Lambda_{1}}\Phi^{\prime}(\xi) \hspace{0.05 cm} C^{\frac{1}{2},\text{loc}}_{\Lambda_{1}}\Phi (\xi^{\prime})  
(e_{0} \hspace{0.05 cm} C^{\frac{1}{2},\text{loc}}_{\Lambda_{1}}\Phi (\xi^{\prime\prime}))^{n} \hspace{0.05 cm}
(e_{0} \hspace{0.05 cm} C^{\frac{1}{2},\text{loc}}_{\Lambda_{1}}\Psi_{1} (\xi^{\prime\prime}))^{m}  \\
& \hspace{6 cm} 1_{\xi \in \Omega_{0}} 1_{\xi^{\prime} \in \Omega_{0}} 1_{\xi^{\prime\prime} \in \Omega_{0}^{\ast}}   \\
&= \sum_{\substack{\xi, \xi^{\prime}, \xi^{\prime\prime} \\ |\xi - \xi^{\prime}| = 1}} 
 \sum_{\substack{n, m  \\ n + m : \text{even}}} e_{0}^{n+m} \hspace{0.05 cm} \frac{(-1)^{(n+m)/2}}{n! m!}
\sum_{\substack{\xi_{a}, \xi_{b}, \xi_{1}, \cdots,  \xi_{n} \\ \eta_{1}, \cdots, \eta_{m}}}  
 C^{\frac{1}{2},\text{loc}}_{\Lambda_{1}}(\xi, \xi_{a}) \Phi (\xi_{a}) 
C^{\frac{1}{2},\text{loc}}_{\Lambda_{1}}(\xi^{\prime}, \xi_{b}) \Phi (\xi_{b})   \\
& \hspace{1 cm} C^{\frac{1}{2},\text{loc}}_{\Lambda_{1}}(\xi^{\prime\prime}, \xi_{1})\Phi (\xi_{1}) \cdots  
C^{\frac{1}{2},\text{loc}}_{\Lambda_{1}}(\xi^{\prime\prime}, \xi_{n})\Phi (\xi_{n})  
 C^{\frac{1}{2},\text{loc}}_{\Lambda_{1}}(\xi^{\prime\prime}, \eta_{1})\Psi_{1}(\eta_{1}) \cdots  
C^{\frac{1}{2},\text{loc}}_{\Lambda_{1}}(\xi^{\prime\prime}, \eta_{m})\Psi_{1}(\eta_{m})   \\
&\hspace{1 cm} 1_{\xi_{a} \in \Lambda_{1}} 1_{\xi_{b} \in \Lambda_{1}} 1_{\xi_{j} \in \Lambda_{1}^{\ast}} 1_{\eta_{j} \in \Lambda_{1}^{\ast}}  
1_{\xi \in \Omega_{0}} 1_{\xi^{\prime} \in \Omega_{0}} 1_{\xi^{\prime\prime} \in \Omega_{0}^{\ast}}   \\
&=  \sum_{\substack{n, m  \\ n + m : \text{even}}} \sum_{\substack{\xi_{a}, \xi_{b}, \xi_{1}, \cdots,  \xi_{n} \\ \eta_{1}, \cdots, \eta_{m}}}
a(\xi_{a}, \xi_{b}, \xi_{1}, \cdots,  \xi_{n}, \eta_{1}, \cdots, \eta_{m}) 
\Phi (\xi_{a})\cdots \Phi (\xi_{n}) \Psi_{1} (\eta_{1})\cdots \Psi_{1} (\eta_{m})
\end{aligned}
\end{equation}
where,
\begin{equation}
\begin{aligned}
a(\xi_{a}, \xi_{b}, \xi_{1}, \cdots,  \xi_{n}, \eta_{1}, \cdots, \eta_{m})  &= 
\sum_{\substack{\xi, \xi^{\prime}, \xi^{\prime\prime} \\ |\xi - \xi^{\prime}| = 1}} 
e_{0}^{n+m} \hspace{0.05 cm} \frac{(-1)^{(n+m)/2}}{n! m!} \hspace{0.05 cm} C^{\frac{1}{2},\text{loc}}_{\Lambda_{1}}(\xi, \xi_{a})
C^{\frac{1}{2},\text{loc}}_{\Lambda_{1}}(\xi^{\prime}, \xi_{b}) \\ & C^{\frac{1}{2},\text{loc}}_{\Lambda_{1}}(\xi^{\prime\prime}, \xi_{1})
\cdots C^{\frac{1}{2},\text{loc}}_{\Lambda_{1}}(\xi^{\prime\prime}, \xi_{n}) C^{\frac{1}{2},\text{loc}}_{\Lambda_{1}}(\xi^{\prime\prime}, \eta_{1})
\cdots C^{\frac{1}{2},\text{loc}}_{\Lambda_{1}}(\xi^{\prime\prime},\eta_{m}) \\
& 1_{\xi_{a} \in \Lambda_{1}} 1_{\xi_{b} \in \Lambda_{1}} 1_{\xi_{j} \in \Lambda_{1}^{\ast}} 1_{\eta_{j} \in \Lambda_{1}^{\ast}}  
1_{\xi \in \Omega_{0}} 1_{\xi^{\prime} \in \Omega_{0}} 1_{\xi^{\prime\prime} \in \Omega_{0}^{\ast}}.
\end{aligned}
\end{equation}  
\textit{Tree decay}. Let $\tilde{\xi} = \{\xi, \xi^{\prime}, \xi^{\prime\prime}\}$ and 
$\xi_{a} \in \Lambda_{1}, \xi_{b} \in \Lambda_{1}, \xi_{j} \in \Lambda_{1}^{\ast}, \eta_{j} \in \Lambda_{1}^{\ast}$.
Then from triangle inequality, $d(\tilde{\xi}, \xi_{a}) \geqslant | d(\xi, \xi_{a}) - d(\xi, \tilde{\xi}) | \geqslant d(\xi, \xi_{a}) - 1$.
Similarly, $d(\tilde{\xi}, \xi_{b})  \geqslant d(\xi, \xi_{b}) - 1$, $d(\tilde{\xi}, \xi_{j})  \geqslant d(\xi, \xi_{j}) - 1$
and $d(\tilde{\xi}, \eta_{j})  \geqslant d(\xi, \eta_{j}) - 1$. From lemma 2.2, 
$C^{\frac{1}{2},\text{loc}}_{\Lambda_{1}}(\xi, \xi_{j}) \leqslant c \hspace{0.05 cm} e^{- \frac{\gamma}{8} d(\xi, \xi_{j})}$ and using
\begin{equation}
\begin{aligned}
&d(\xi, \xi_{a}) + d(\xi, \xi_{b}) + \sum_{j} d(\xi, \xi_{j}) +\sum_{j}  d(\xi, \eta_{j}) = \\
& \text{length of tree joining} \hspace{0.1 cm} (\xi, \xi_{a}, \xi_{b}, \xi_{1}, \cdots, \xi_{n}, \eta_{1}, \cdots, \eta_{m}) 
\geqslant t(\xi_{a}, \xi_{b}, \xi_{1}, \cdots, \xi_{n}, \eta_{1}, \cdots, \eta_{m})
\end{aligned}
\end{equation}
write 
\begin{equation}
\begin{aligned}
\abs{a(\xi_{a}, \xi_{b}, \xi_{1}, \cdots,  \xi_{n}, \eta_{1}, \cdots, \eta_{m})} &\leqslant c \hspace{0.05 cm}
\frac{e_{0}^{n+m}}{n! m!}  \sum_{\xi} e^{- \frac{\gamma}{8} d(\xi, \xi_{a}) - \cdots - 
 \frac{\gamma}{8} d(\xi, \eta_{m})} e^{n+m+2} \\
&\leqslant c \hspace{0.05 cm} e^{2} \frac{1}{n! m!} \hspace{0.05 cm}  (e \hspace{0.05 cm} e_{0})^{n+m}
 \hspace{0.05 cm} e^{-\frac{\gamma}{16} t(\xi_{a}, \cdots, \eta_{m})} 
\sum_{\xi} e^{- \frac{\gamma}{16} (d(\xi, \xi_{a}) - \cdots - d(\xi, \eta_{m}))} \\
&\leqslant c  \frac{1}{n! m!} \hspace{0.05 cm}  (e \hspace{0.05 cm} e_{0})^{n+m} \hspace{0.05 cm} 
e^{-\frac{\gamma}{16} t(\xi_{a}, \xi_{b}, \xi_{1}, \cdots, \xi_{n}, \eta_{1}, \cdots, \eta_{m})}.
\end{aligned}
\end{equation}
\item For the next term, $C^{\frac{1}{2},\text{loc}}_{\Lambda_{1}}\Phi^{\prime} (\xi) 
(1 - \text{cos}\hspace{0.05 cm} [e_{0}C^{\frac{1}{2},\text{loc}}_{\Lambda_{1}}\Phi^{\prime} (\xi^{\prime})]) $
there are two cases since $C^{\frac{1}{2},\text{loc}}_{\Lambda_{1}}\Phi^{\prime} (\xi) = 
C^{\frac{1}{2},\text{loc}}_{\Lambda_{1}}\Phi (\xi) + C^{\frac{1}{2},\text{loc}}_{\Lambda_{1}}\Psi_{1} (\xi)$. 
We use the first term to extract the coefficients of power series.
\begin{equation}
\begin{aligned}
\sum_{\xi, \xi^{\prime}} & C^{\frac{1}{2},\text{loc}}_{\Lambda_{1}}\Phi (\xi) 
(1 - \text{cos}\hspace{0.05 cm} [e_{0}C^{\frac{1}{2},\text{loc}}_{\Lambda_{1}}\Phi^{\prime} (\xi^{\prime})]) 
1_{\xi \in \Omega_{0}} 1_{\xi^{\prime} \in \Omega_{0}^{\ast}}    \\ 
&= \sum_{\xi, \xi^{\prime}} \sum_{l : \text{even}} \frac{(-1)^{l/2} }{l!} \sum_{\substack{n, m  \\ n + m = l}} \frac{l!}{n! m!} \hspace{0.05 cm}
C^{\frac{1}{2},\text{loc}}_{\Lambda_{1}}\Phi (\xi) \hspace{0.05 cm}  
 (e_{0} \hspace{0.05 cm} C^{\frac{1}{2},\text{loc}}_{\Lambda_{1}}\Phi (\xi^{\prime}))^{n} \hspace{0.05 cm}
(e_{0} \hspace{0.05 cm} C^{\frac{1}{2},\text{loc}}_{\Lambda_{1}}\Psi_{1} (\xi^{\prime}))^{m} 
1_{\xi \in \Omega_{0}} 1_{\xi^{\prime} \in \Omega_{0}^{\ast}}  \\ 
&= \sum_{\xi, \xi^{\prime}} \sum_{\substack{n, m  \\ n + m : \text{even}}} e_{0}^{n+m} \hspace{0.05 cm} \frac{(-1)^{(n+m)/2}}{n! m!}
\sum_{\substack{\xi_{a}, \xi_{1}, \cdots,  \xi_{n} \\ \eta_{1}, \cdots, \eta_{m}}}  
 C^{\frac{1}{2},\text{loc}}_{\Lambda_{1}}(\xi, \xi_{a}) \Phi (\xi_{a}) 
\hspace{0.05 cm} C^{\frac{1}{2},\text{loc}}_{\Lambda_{1}}(\xi^{\prime}, \xi_{1})\Phi (\xi_{1}) \cdots    \\ &
C^{\frac{1}{2},\text{loc}}_{\Lambda_{1}}(\xi^{\prime}, \xi_{n}) \Phi (\xi_{n})   
C^{\frac{1}{2},\text{loc}}_{\Lambda_{1}}(\xi^{\prime}, \eta_{1})\Psi_{1}(\eta_{1}) \cdots  
C^{\frac{1}{2},\text{loc}}_{\Lambda_{1}}(\xi^{\prime}, \eta_{m})\Psi_{1}(\eta_{m}) 
\hspace{0.05 cm} 1_{\xi_{a} \in \Lambda_{1}} 1_{\xi_{j} \in \Lambda_{1}^{\ast}} 1_{\eta_{j} \in \Lambda_{1}^{\ast}}  
1_{\xi \in \Omega_{0}} 1_{\xi^{\prime} \in \Omega_{0}^{\ast}}   \\
&= \sum_{\substack{n, m  \\ n + m : \text{even}}} \sum_{\substack{\xi_{a}, \xi_{1}, \cdots,  \xi_{n} \\ \eta_{1}, \cdots, \eta_{m}}}
a(\xi_{a}, \xi_{1}, \cdots,  \xi_{n}, \eta_{1}, \cdots, \eta_{m}) 
\Phi (\xi_{a})\Phi (\xi_{1})\cdots \Phi (\xi_{n}) \Psi_{1} (\eta_{1})\cdots \Psi_{1} (\eta_{m})
\end{aligned}
\end{equation} 
where,
\begin{equation}
\begin{aligned}
a(\xi_{a}, \xi_{1}, \cdots,  \xi_{n}, \eta_{1}, \cdots, \eta_{m}) &= e_{0}^{n+m} \hspace{0.05 cm} \frac{(-1)^{(n+m)/2}}{n! m!}
\sum_{\xi, \xi^{\prime}} C^{\frac{1}{2},\text{loc}}_{\Lambda_{1}}(\xi, \xi_{a})   C^{\frac{1}{2},\text{loc}}_{\Lambda_{1}}(\xi^{\prime}, \xi_{1}) \cdots  
C^{\frac{1}{2},\text{loc}}_{\Lambda_{1}}(\xi^{\prime}, \xi_{n}) \\ & C^{\frac{1}{2},\text{loc}}_{\Lambda_{1}}(\xi^{\prime}, \eta_{1}) \cdots 
C^{\frac{1}{2},\text{loc}}_{\Lambda_{1}}(\xi^{\prime}, \eta_{m}) 1_{\xi_{a} \in \Lambda_{1}} 1_{\xi_{j} \in \Lambda_{1}^{\ast}} 
1_{\eta_{j} \in \Lambda_{1}^{\ast}} 1_{\xi \in \Omega_{0}} 1_{\xi^{\prime} \in \Omega_{0}^{\ast}}  
\end{aligned}
\end{equation}
and from the paragraph \textit{tree decay} as in Eq 3.32 and 3.33, setting $\tilde{\xi} = \{\xi, \xi^{\prime} \}$, it follows that
\begin{equation}
\abs{a(\xi_{a}, \xi_{1}, \cdots,  \xi_{n}, \eta_{1}, \cdots, \eta_{m})} \leqslant 
c  \frac{1}{n! m!} \hspace{0.05 cm}  (e \hspace{0.05 cm} e_{0})^{n+m} \hspace{0.05 cm} 
e^{-\frac{\gamma}{2} t(\xi_{a}, \xi_{1}, \cdots, \xi_{n}, \eta_{1}, \cdots, \eta_{m})}.
\end{equation}
This completes the proof for $V(\Omega_{0}, C^{\frac{1}{2},\text{loc}}_{\Lambda_{1}} \Phi^{\prime})$.
\item Next rewrite from lemma 2.3
\begin{equation}
\begin{aligned}
V_{\varepsilon}(\Lambda_{1}, \Phi^{\prime}) &= \langle C^{\frac{1}{2}}_{\Lambda_{1}} \Phi^{\prime}, \text{T}_{\Lambda_{1}}
 \delta C^{\frac{1}{2}}_{\Lambda_{1}} \Phi^{\prime}\rangle  - 
2 \langle C^{-\frac{1}{2}}_{\Lambda_{1}}\Phi^{\prime}, \delta C^{\frac{1}{2}}_{\Lambda_{1}}\Phi^{\prime}\rangle \\
&= \sum_{\xi} (C^{\frac{1}{2}}_{\Lambda_{1}} \Phi^{\prime})(\xi) (\text{T}_{\Lambda_{1}} \delta C^{\frac{1}{2}}_{\Lambda_{1}}\Phi^{\prime})(\xi) - 
2 \sum_{\xi} (C^{-\frac{1}{2}}_{\Lambda_{1}} \Phi^{\prime})(\xi)(\delta C^{\frac{1}{2}}_{\Lambda_{1}}\Phi^{\prime})(\xi) \\
&= \sum_{\xi} (C^{\frac{1}{2}}_{\Lambda_{1}} \Phi (\xi)  + C^{\frac{1}{2}}_{\Lambda_{1}}\Psi_{1} (\xi))
(\text{T}_{\Lambda_{1}} \delta C^{\frac{1}{2}}_{\Lambda_{1}}\Phi  (\xi) + 
\text{T}_{\Lambda_{1}} \delta C^{\frac{1}{2}}_{\Lambda_{1}}\Psi_{1}(\xi)) \\ &- 2 \sum_{\xi} (C^{-\frac{1}{2}}_{\Lambda_{1}} \Phi (\xi)
+ C^{-\frac{1}{2}}_{\Lambda_{1}} \Psi_{1}(\xi)) (\delta C^{\frac{1}{2}}_{\Lambda_{1}}\Phi (\xi) + 
\delta C^{\frac{1}{2}}_{\Lambda_{1}}\Psi_{1} (\xi)).
\end{aligned}
\end{equation}
Consider the following part of $V_{\varepsilon}(\Lambda_{1}, \Phi^{\prime})$,
\begin{equation}
\begin{aligned}
\sum_{\xi} & C^{\frac{1}{2}}_{\Lambda_{1}} \Phi (\xi) \text{T}_{\Lambda_{1}} \delta C^{\frac{1}{2}}_{\Lambda_{1}}\Phi (\xi)
 -2 \sum_{\xi} C^{-\frac{1}{2}}_{\Lambda_{1}} \Phi (\xi) \delta C^{\frac{1}{2}}_{\Lambda_{1}}\Phi (\xi) \\
&= \sum_{\xi, \xi_{1}, \xi_{2}} \text{T}_{\Lambda_{1}} (\xi, \xi_{2}) C^{\frac{1}{2}}_{\Lambda_{1}} (\xi, \xi_{1}) 
 \delta C^{\frac{1}{2}}_{\Lambda_{1}} (\xi_{1}, \xi_{2}) \hspace{0.1 cm}   \Phi (\xi_{1})\Phi (\xi_{2})  
 - 2 \sum_{\xi, \xi_{1}} C^{-\frac{1}{2}}_{\Lambda_{1}} (\xi, \xi_{1}) \delta C^{\frac{1}{2}}_{\Lambda_{1}} (\xi, \xi_{1}) 
 \hspace{0.1 cm}\Phi (\xi_{1})^{2}.
\end{aligned}
\end{equation}
We use the above part to extract the coefficients for $V_{\varepsilon}(\Lambda_{1}, \Phi^{\prime})$. First rewrite
\begin{equation}
\sum_{\xi, \xi_{1}, \xi_{2}}  \text{T}_{\Lambda_{1}} (\xi, \xi_{2}) C^{\frac{1}{2}}_{\Lambda_{1}} (\xi, \xi_{1}) 
 \delta C^{\frac{1}{2}}_{\Lambda_{1}} (\xi_{1}, \xi_{2}) \hspace{0.1 cm}   \Phi (\xi_{1})\Phi (\xi_{2})  =
 \sum_{\xi_{1}, \xi_{2}} a (\xi_{1}, \xi_{2}) \hspace{0.1 cm} \Phi (\xi_{1})\Phi (\xi_{2}) 
\end{equation}
where,
\begin{equation}
a (\xi_{1}, \xi_{2}) = \sum_{\xi} \text{T}_{\Lambda_{1}} (\xi, \xi_{2}) C^{\frac{1}{2}}_{\Lambda_{1}} (\xi, \xi_{1}) 
 \delta C^{\frac{1}{2}}_{\Lambda_{1}} (\xi_{1}, \xi_{2}).
\end{equation}
From lemma 2.2, using $\delta C^{\frac{1}{2}}_{\Lambda_{1}} (\xi_{1}, \xi_{2}) \leqslant c \hspace{0.05 cm} e^{- \frac{\gamma}{8} r_{\lambda}}$,
\begin{equation}
\abs{a (\xi_{1}, \xi_{2})} \leqslant \sum_{\xi} c \hspace{0.05 cm} e^{-\frac{\gamma}{8} |\xi - \xi_{1}|} e^{-\frac{\gamma}{8} |\xi_{1} - \xi_{2}|}
\leqslant c \hspace{0.05 cm} e^{- \frac{\gamma}{16} r_{\lambda}} \sum_{\xi} e^{- \frac{\gamma}{8}|\xi - \xi_{1}|} 
e^{-\frac{\gamma}{16} |\xi_{1} - \xi_{2}|} \leqslant c \hspace{0.05 cm} e_{0} e^{-\frac{\gamma}{16} |\xi_{1} - \xi_{2}|}
\end{equation}
where, we have used that $e^{- \frac{\gamma}{16}r_{\lambda}} = \mathcal{O}(e_{0}^{2})$. Next rewrite 
\begin{equation}
\begin{aligned}
 \sum_{\xi, \xi_{1}} C^{-\frac{1}{2}}_{\Lambda_{1}} (\xi, \xi_{1}) \delta C^{\frac{1}{2}}_{\Lambda_{1}} (\xi, \xi_{1}) 
 \hspace{0.1 cm}\Phi (\xi_{1})^{2} &= 
 \sum_{\xi, \xi_{1}, \xi_{2}} C^{-\frac{1}{2}}_{\Lambda_{1}} (\xi, \xi_{1}) \Phi (\xi_{1})  \hspace{0.05 cm}
 \delta C^{\frac{1}{2}}_{\Lambda_{1}} (\xi, \xi_{2}) \Phi (\xi_{2}) \\ 
&=  \sum_{\xi_{1}, \xi_{2}} a (\xi_{1}, \xi_{2}) \hspace{0.1 cm} \Phi (\xi_{1})\Phi (\xi_{2}) 
\end{aligned}
\end{equation}
where,
\begin{equation}
a (\xi_{1}, \xi_{2}) = \sum_{\xi} C^{-\frac{1}{2}}_{\Lambda_{1}} (\xi, \xi_{1})\hspace{0.05 cm} 
\delta C^{\frac{1}{2}}_{\Lambda_{1}} (\xi, \xi_{2})\hspace{0.05 cm}\delta(\xi_{1} - \xi_{2}).
\end{equation} 
From lemma 2.2, using $\delta C^{\frac{1}{2}}_{\Lambda_{1}} (\xi_{1}, \xi_{2}) \leqslant c \hspace{0.05 cm} e^{- \frac{\gamma}{8} r_{\lambda}}$,
\begin{equation}
\abs{a (\xi_{1}, \xi_{2})} \leqslant \sum_{\xi} c \hspace{0.05 cm} e^{-\frac{\gamma}{8} |\xi - \xi_{2}|} \delta(\xi_{1} - \xi_{2}) 
\leqslant c \hspace{0.05 cm} e^{- \frac{\gamma}{16} r_{\lambda}} \sum_{\xi} e^{-\frac{\gamma}{16} |\xi - \xi_{2}|}
\delta(\xi_{1} - \xi_{2}) \leqslant c \hspace{0.05 cm} e_{0} \hspace{0.05 cm} \delta(\xi_{1} - \xi_{2}). 
\end{equation}
The other terms in Eq 3.37 are treated similarly.
\item The source term
\begin{equation}
\begin{aligned}
e_{0} \langle C^{\frac{1}{2},\text{loc}}_{\Lambda_{1}} \Phi^{\prime}, \delta \text{J}\rangle &=  
e_{0} \hspace{0.05 cm} \sum_{\xi} (C^{\frac{1}{2},\text{loc}}_{\Lambda_{1}} \Phi (\xi)  \Psi_{2}(\xi)
+ C^{\frac{1}{2},\text{loc}}_{\Lambda_{1}} \Psi_{1} (\xi)  \Psi_{2}(\xi)) 1_{\xi \in \Omega_{0}^{\ast}} \\
&= e_{0} \hspace{0.05 cm} \sum_{\xi, \xi_{1}} (C^{\frac{1}{2},\text{loc}}_{\Lambda_{1}}(\xi, \xi_{1}) \Phi (\xi_{1}) \Psi_{2}(\xi) +
C^{\frac{1}{2},\text{loc}}_{\Lambda_{1}}(\xi, \xi_{1}) \Psi_{1}(\xi_{1}) \Psi_{2}(\xi))
1_{\xi \in \Omega_{0}^{\ast}} 1_{\xi_{1} \in \Lambda_{1}^{\ast}} \\ 
&= \sum_{\xi, \xi_{1}} a (\xi, \xi_{1}) \hspace{0.05 cm}  (\Phi (\xi_{1}) \Psi_{2}(\xi) + \Psi_{1}(\xi_{1}) \Psi_{2}(\xi))
\end{aligned}
\end{equation}
where,
\begin{equation}
a (\xi, \xi_{1}) = e_{0} \hspace{0.05 cm}  C^{\frac{1}{2},\text{loc}}_{\Lambda_{1}}(\xi, \xi_{1}) 
1_{\xi \in \Omega_{0}^{\ast}} 1_{\xi_{1} \in \Lambda_{1}^{\ast}}
\end{equation}
and using lemma 2.2, $|a (\xi, \xi_{1})| < \text{c} \hspace{0.05 cm}  e_{0} \hspace{0.05 cm} e^{-\frac{\gamma}{8} d(\xi, \xi_{1})}$. 
This completes the proof of theorem 3.1.
\end{enumerate}

\textit{Norm} of $V_{1}$.
Let $w_{4\text{r}, \hat{\kappa}}(\vec{\xi}, \vec{\eta}) =  e^{m t(\text{supp}(\vec{\xi}, \vec{\eta}))} (4\text{r})^{n} \hat{\kappa}^{m}$ 
be the weight system with mass \textit{m} giving weight at least 4r to $\Phi$ and $\hat{\kappa}$ to $\Psi_{\alpha}$.
Let $X \subset \Lambda_{1}$ with $X \cap \Omega_{1} \neq \oslash$. Define 
\begin{equation}
 |a|_{w_{4\text{r}, \hat{\kappa}}} = \sum_{ n, m \geqslant 0}
\max\limits_{\xi \in X}  \hspace{0.05 cm} \max\limits_{\substack{1\leqslant i \leqslant n \\ 1\leqslant j \leqslant m}} 
\sum_{\substack{\xi_{1}, \cdots, \xi_{n}, \eta_{1}, \cdots, \eta_{m} \in X \\ \xi_{i} = \xi \hspace{0.05 cm} or \hspace{0.05 cm}
\eta_{j} = \xi}}  e^{m t(\text{supp}(\vec{\xi}, \vec{\eta}))} 
 (4\text{r})^{n} \hat{\kappa}^{m} \hspace{0.05 cm}  |a(\xi_{1}, \cdots, \xi_{n}, \eta_{1}, \cdots, \eta_{m})|.
 \end{equation}
Then the norm is defined as  $||V_{1}||_{w_{4\text{r}, \hat{\kappa}}} = |a|_{w_{4\text{r}, \hat{\kappa}}}$. Let $4\text{r} = p_{0,\lambda}$,
$\hat{\kappa} = 1$ and $m < \gamma^{\prime}$. Then from theorem 3.1
\begin{equation}
\begin{aligned}
||V_{1}||_{w_{4\text{r}, \hat{\kappa}}} &\leqslant  \sum_{ n+m > 0}
\max\limits_{\xi \in X}  \hspace{0.05 cm} \max\limits_{ \substack{1\leqslant i \leqslant n \\ 1\leqslant j \leqslant m}} 
\sum_{\substack {\xi_{1}, \cdots, \xi_{n}, \eta_{1}, \cdots, \eta_{m} \in X \\ 
\xi_{i} = \xi \hspace{0.05 cm} or \hspace{0.05 cm} \eta_{j} = \xi}} e^{m t(\text{supp}(\vec{\xi}, \vec{\eta}))} 
 p_{0,\lambda}^{n} \hspace{0.1 cm}  c \hspace{0.05 cm} (c_{0} \hspace{0.05 cm} e_{0})^{\frac{n+m}{2}} \hspace{0.05 cm} 
 e^{- \gamma^{\prime} t(\text{supp}(\vec{\xi}, \vec{\eta}))} \\
 &\leqslant  \sum_{ n+m > 0}
\max\limits_{\xi \in X}  \hspace{0.05 cm} \max\limits_{ \substack{1\leqslant i \leqslant n \\ 1\leqslant j \leqslant m}} 
\sum_{\substack {\xi_{1}, \cdots, \xi_{n}, \eta_{1}, \cdots, \eta_{m} \in X \\ \xi_{i} = \xi \hspace{0.05 cm} or \hspace{0.05 cm}
\eta_{j} = \xi}} c \hspace{0.05 cm} (c_{0} \hspace{0.05 cm}p_{0,\lambda}^{2} e_{0})^{\frac{n}{2}} (c_{0} \hspace{0.05 cm} e_{0})^{\frac{m}{2}} 
e^{- (\gamma^{\prime} - m) t(\text{supp}(\vec{\xi}, \vec{\eta}))} \\
&\leqslant \sum_{ n+m > 0}
\max\limits_{\xi \in X}  \hspace{0.05 cm} \max\limits_{ \substack{1\leqslant i \leqslant n \\ 1\leqslant j \leqslant m}} 
\sum_{\substack {\xi_{1}, \cdots, \xi_{n}, \eta_{1}, \cdots, \eta_{m} \in X \\ \xi_{i} = \xi \hspace{0.05 cm} or \hspace{0.05 cm}
\eta_{j} = \xi}} c \hspace{0.05 cm} (c_{0} \hspace{0.05 cm} e_{0}^{1-2 \epsilon})^{\frac{n}{2}} (c_{0} \hspace{0.05 cm} e_{0})^{\frac{m}{2}} 
e^{- (\gamma^{\prime} - m) t(\text{supp}(\vec{\xi}, \vec{\eta}))} \\
& \leqslant c \hspace{0.05 cm} \sum_{ n+m > 0} (c_{0} \hspace{0.05 cm} e_{0}^{1-2 \epsilon})^{\frac{n}{2}}
 (c_{0} \hspace{0.05 cm} e_{0})^{\frac{m}{2}} \\
& \leqslant c \hspace{0.05 cm} (c_{0} \hspace{0.05 cm} e_{0}^{1-2 \epsilon})^{\frac{1}{2}}.
\end{aligned}
\end{equation}
Note that $p_{0,\lambda}^{n} \leqslant e_{0}^{- n \epsilon}$, $(c_{0} \hspace{0.05 cm} e_{0}) < 1$ and 
$(c_{0} \hspace{0.05 cm} e_{0}^{1-2 \epsilon}) < 1$.
 
\textbf{Theorem 3.2} Given a power series of $V_{1}(\Omega_{0}, \Phi, \Psi)$ with
coefficient system $a (\vec{\xi}, \vec{\eta})$, let $w_{\hat{\kappa}}(\vec{\eta}) = e^{m t(\text{supp}(\vec{\eta}))} 
\hat{\kappa}^{n(\vec{\eta})}$ be the weight system of mass \textit{m} giving weight $\hat{\kappa}$ to 
the field $\Psi_{\alpha}$. Let $Z = \text{supp} (\vec{\eta})$. Set $\hat{\kappa} = 1$.
Then there exists a coefficient system $b(\vec{\eta})$ having decay properties as $a (\vec{\xi}, \vec{\eta})$, with
\begin{equation}
|b|_{w_{\hat{\kappa}}} = \sum_{n \geqslant 0}   
\max\limits_{\eta \in X} \max\limits_{1\leqslant i \leqslant n} \sum_{\substack{(\eta_{1}, \cdots, \eta_{n}) \in X 
 \\ \eta_{i} = \eta}} w_{\hat{\kappa}}(\vec{\eta})  \abs{b (\eta_{1}, \cdots, \eta_{n})}
\end{equation}
and a function $H (Z, \Psi)=\sum_{\substack{\vec{\eta} \\ \text{supp} (\vec{\eta}) = Z}} b (\vec{\eta}) \Psi_{\alpha} (\vec{\eta})$
with $H(0)=b_{0}$ and $\text{ln}\hspace{0.1 cm} \Xi(\Omega_{1}, \Psi) = \sum_{Z} H (Z, \Psi)$. 
Define $||H||_{w_{\hat{\kappa}}}=|b|_{w_{\hat{\kappa}}}$, then for $e_{0}$ sufficiently small
\begin{equation}
||H - H(0)||_{w_{\hat{\kappa}}} \leqslant 
\frac{||V_{1}||_{w_{4\text{r}, \hat{\kappa}}}}{1 - 16 ||V_{1}||_{w_{4\text{r}, \hat{\kappa}}}} 
\leqslant c \hspace{0.1 cm} (c_{0} \hspace{0.05 cm} e_{0}^{1-2 \epsilon})^{\frac{1}{2}}.
\end{equation}
\textit{Proof} follows directly from Balaban, Feldman, Kn$\ddot{\text{o}}$rrer and Trubowitz \cite{BFKT}.  

Note that the series $H (Z, \Psi)=\sum_{\substack{\vec{\eta} \\ \text{supp} (\vec{\eta}) = Z}} b (\vec{\eta}) \Psi_{\alpha} (\vec{\eta})$
converges in a unit disc in complex plane containing J and therefore, $\sum_{Z} H (Z, \Psi)$ also converges. From theorem 3.2,
\begin{equation}
\begin{aligned}
|H(Z)| &\leqslant e^{- m t(\text{supp}(\vec{\eta}))} \sum_{n \geqslant 0}   
\max\limits_{\eta \in X} \max\limits_{1\leqslant i \leqslant n} \sum_{\substack{(\eta_{1}, \cdots, \eta_{n}) \in X 
 \\ \eta_{i} = \eta}} w_{\hat{\kappa}}(\vec{\eta})  \abs{b (\eta_{1}, \cdots, \eta_{n})}  \\
&\leqslant ||H||_{w_{\hat{\kappa}}} \hspace{0.1 cm} e^{- m t(Z)} \\
&\leqslant c \hspace{0.1 cm} (c_{0} \hspace{0.05 cm} e_{0}^{1-2 \epsilon})^{\frac{1}{2}} \hspace{0.1 cm} e^{- m t(Z)}.
\end{aligned}
\end{equation}
So far we have assumed $\Phi_{\Lambda^{c}_{1}} = 0$. Now we drop this assumption.

\textbf{Proposition 3.1} (\textit{Shift}) Let $H$ be an analytic function.
Let $\phi = - C_{\Lambda_{1}}\text{T}_{\Lambda_{1}\Lambda^{c}_{1}} \Phi_{\Lambda_{1}^{c}}$ denote shift.
Define $H^{\#}$ by 
\begin{equation}
H^{\#}(\Psi, \phi) = H (\Psi + \phi) = \sum_{\vec{\eta} = (\eta_{1}, \cdots, \eta_{n + m}) \in X} b(\vec{\eta}) (\Psi + \phi)(\vec{\eta}).
\end{equation}
Let $w_{\hat{\kappa}, \hat{\kappa}_{2}}(\vec{\eta}) =  e^{m t(\text{supp}(\vec{\eta}))}  \hat{\kappa}^{n}  \hat{\kappa}_{2}^{m}$  
be the weight system of mass \textit{m} that gives weight $ \hat{\kappa}_{2}$ to $\phi$.
Set   $\hat{\kappa} = 1$ and $\hat{\kappa}_{2} = p_{\lambda}$. Let $Z = \text{supp} (\vec{\eta})$. Then 
$||H^{\#}||_{w_{\hat{\kappa}, \hat{\kappa}_{2}}} = ||H||_{w_{\hat{\kappa}+ \hat{\kappa}_{2}}}$
and $|H(Z)| = |H^{\#}(Z)|$.

\textit{Proof} follows directly from \cite{BFKT}.  

\subsection{From $\vec{\eta}$ to polymer} 

A union of $[r_{\lambda}]$ blocks $\Box_{i}$ is called \textit{connected} if any two blocks centered on nearest $[r_{\lambda}]$  
neighbor sites, have a common hypersurface, face, edge or a site.  
A \textit{polymer} is a connected union of $[r_{\lambda}]$ blocks, $\Box$. For a polymer X, let $|\text{X}|$ denote the number of $\Box$ in X.
Recall that $Z = \text{supp} (\vec{\eta})$. Define the set
\begin{equation}
\mathrm{X}_{Z} = \{\text{all}\hspace{0.1 cm} [r_{\lambda}] \hspace{0.1 cm} \text{blocks}, 
\Box\hspace{0.1 cm} \text{containing}\hspace{0.1 cm} Z  : \hspace{0.1 cm} \mathrm{X}_{Z} \cap \Omega_{1} \neq \oslash\}. 
\end{equation}
Let $\mathrm{X} =\{\mathrm{X}_{i}\} \subset \Omega_{0}$ be a collection of polymers. Rewrite,
\begin{equation}
\text{ln}\hspace{0.1 cm} \Xi(\Omega_{1}, \Psi) = 
\sum_{Z} H^{\#}(Z, \Psi) = \sum_{\mathrm{X}} \sum_{Z : \mathrm{X}_{Z} = \mathrm{X}} H^{\#} (Z, \Psi) 
= \sum_{\mathrm{X}} H^{\#}(\mathrm{X}, \Psi).
\end{equation}
Rewrite $\Xi(\Omega_{1}, \Psi)$ using Mayer expansion as
\begin{equation}
\begin{aligned}
\Xi(\Omega_{1}, \Psi) = e^{\sum_{\mathrm{X}}H^{\#}(\mathrm{X}, \Psi)} =
\prod_{\mathrm{X}}\Big[e^{H^{\#}(\mathrm{X}, \Psi)} -1 + 1\Big] &= 
\sum_{\{\mathrm{X}_{i}\} \rightarrow \mathrm{Y}}
\prod_{i}\Big[e^{H^{\#}(\mathrm{X}_{i}, \Psi)} -1\Big] \\
&= \sum_{\substack{\{\mathrm{Y}_{i}\}, \\ \text{disjoint}}} \prod_{i} K(\mathrm{Y}_{i}, \Psi)
\end{aligned}
\end{equation}
where, $\{\mathrm{Y}_{i}\}$ are polymers and for a particular $\mathrm{Y}_{i}$,
\begin{equation}
K(\mathrm{Y}, \Psi) = \sum_{\cup \mathrm{X}_{i} = \mathrm{Y}} \prod_{i} \Big[e^{H^{\#}(\mathrm{X}_{i}, \Psi)} - 1\Big].
\end{equation}

\textbf{Lemma 3.1} Given an analytic function $H^{\#}$
as defined earlier with $|H^{\#}(Z)| \leqslant c\hspace{0.1 cm} (c_{0} \hspace{0.05 cm} e_{0}^{1-2 \epsilon})^{\frac{1}{2}}
\hspace{0.1 cm} e^{- m t(Z)}$. Then for a polymer $\mathrm{X} = \mathrm{X}_{Z}$ containing $|\mathrm{X}|$ blocks, 
$e^{-m t(Z)} \leqslant  e^{-\kappa |\mathrm{X}|}$
and thus,
\begin{equation} 
|H^{\#}(\mathrm{X})| \leqslant c\hspace{0.1 cm}(c_{0} \hspace{0.05 cm} e_{0}^{1-2 \epsilon})^{\frac{1}{2}}
\hspace{0.1 cm} e^{-\kappa |\mathrm{X}|}.
\end{equation}
 \textit{Proof} By definition the polymer $\mathrm{X} = \mathrm{X}_{Z}$, therefore, 
$t(Z) \geqslant t(\mathrm{X}) = (t(\mathrm{X}) +1) - 1$ implies
$e^{-m t(Z)} \leqslant e^{m} e^{-m (t(\mathrm{X})+1)}$. 
Using the inequality (see for example \cite{D}) $t(\mathrm{X}) \leqslant |\mathrm{X}| \leqslant c (t(\mathrm{X}) +1)$, it follows that
$e^{-m t(Z)} \leqslant e^{m} e^{-\kappa |\mathrm{X}|}$. 
Thus, the result.  

\textbf{Lemma 3.2} Given a disjoint collection of polymers $\mathrm{Y} = \{\mathrm{Y}_{i}\}$ and 
$K(\mathrm{Y}_{i}, \Psi)$ as defined above, then for a new constant $\kappa^{\prime}$, $|K(\mathrm{Y})| \leqslant 
c \hspace{0.1 cm} (c_{0} \hspace{0.05 cm} e_{0}^{1-2 \epsilon})^{\frac{1}{2}} \hspace{0.1 cm}  e^{-\kappa^{\prime} |\mathrm{Y}|}$.

\textit{Proof} First write the unordered collection $\{\mathrm{X}_{i}\}$ as ordered sets 
$(\mathrm{X}_{1}, \mathrm{X}_{2}, \cdots, \mathrm{X}_{n})$.
Then for a particular $\mathrm{Y}_{i}$,
\begin{equation}
K(\mathrm{Y}, \Psi) = \sum_{n=1}^{\infty}\frac{1}{n!} \sum_{\cup \mathrm{X}_{i} = \mathrm{Y}}
\prod_{i} \Big[e^{H^{\#}(\mathrm{X}_{i}, \Psi)} - 1\Big].
\end{equation}
Now using lemma 3.1, 
\begin{equation}
\begin{aligned}
\abs{e^{H^{\#}(\mathrm{X}_{i})} - 1} = \abs{\int_{0}^{1}\frac{d}{ds} e^{s H^{\#}(\mathrm{X}_{i})} ds}
&= \abs{\int_{0}^{1} H^{\#}(\mathrm{X}_{i}) e^{s H^{\#}(\mathrm{X}_{i})} ds} \\
&\leqslant c\hspace{0.1 cm} (c_{0} \hspace{0.05 cm} e_{0}^{1-2 \epsilon})^{\frac{1}{2}} \hspace{0.1 cm} e^{-\kappa |\mathrm{X}_{i}|}
\end{aligned}
\end{equation}
and
\begin{equation}
\begin{aligned}
\abs{K(\mathrm{Y})} &\leqslant  \sum_{n=1}^{\infty}\frac{1}{n!} \sum_{\cup \mathrm{X}_{i} = \mathrm{Y}}
\prod_{i=1}^{n} c\hspace{0.1 cm} (c_{0} \hspace{0.05 cm} e_{0}^{1-2 \epsilon})^{\frac{1}{2}} \hspace{0.1 cm} e^{-\kappa |\mathrm{X}_{i}|} \\ 
&\leqslant (c_{0} \hspace{0.05 cm} e_{0}^{1-2 \epsilon})^{\frac{1}{2}} \hspace{0.1 cm} e^{-\frac{\kappa}{2} |\mathrm{Y}|}
  \sum_{n=1}^{\infty}\frac{1}{n!} \sum_{\{\mathrm{X}_{1},\cdots, \mathrm{X}_{n}\} \subset \mathrm{Y}}
 \prod_{i=1}^{n} c  \hspace{0.1 cm}(c_{0} \hspace{0.05 cm} e_{0}^{1-2 \epsilon})^{\frac{1}{4}} \hspace{0.1 cm} 
 e^{-\frac{\kappa}{2} |\mathrm{X}_{i}|}  \\  
&\leqslant (c_{0} \hspace{0.05 cm} e_{0}^{1-2 \epsilon})^{\frac{1}{2}} \hspace{0.1 cm} e^{-\frac{\kappa}{2} |\mathrm{Y}|} \sum_{n=1}^{\infty}\frac{1}{n!} 
\Big[\sum_{\mathrm{X} \subset \mathrm{Y}} c  \hspace{0.1 cm}(c_{0} \hspace{0.05 cm} e_{0}^{1-2 \epsilon})^{\frac{1}{4}} \hspace{0.1 cm} 
e^{-\frac{\kappa}{2} |\mathrm{X}|}\Big]^{n} \\
&\leqslant (c_{0} \hspace{0.05 cm}e_{0}^{1-2 \epsilon})^{\frac{1}{2}}\hspace{0.1 cm} e^{-\frac{\kappa}{2} |\mathrm{Y}|} \sum_{n=1}^{\infty}\frac{1}{n!} 
(c  \hspace{0.1 cm}(c_{0} \hspace{0.05 cm} e_{0}^{1-2 \epsilon})^{\frac{1}{4}} \hspace{0.1 cm} |\mathrm{Y}|)^{n} \\
&\leqslant (c_{0} \hspace{0.05 cm} e_{0}^{1-2 \epsilon})^{\frac{1}{2}} \hspace{0.1 cm} 
e^{-\frac{\kappa}{2}|\mathrm{Y}| + c  \hspace{0.1 cm} (c_{0} \hspace{0.05 cm} e_{0}^{1-2 \epsilon})^{\frac{1}{4}}|\mathrm{Y}|} \\
&\leqslant  (c_{0} \hspace{0.05 cm} e_{0}^{1-2 \epsilon})^{\frac{1}{2}} \hspace{0.1 cm}  e^{-\kappa^{\prime} |\mathrm{Y}|},
\end{aligned}
\end{equation}
where, $\sum_{\mathrm{X} \subset \mathrm{Y}} c  \hspace{0.1 cm}(c_{0} \hspace{0.05 cm} e_{0}^{1-2 \epsilon})^{\frac{1}{4}} \hspace{0.1 cm} 
e^{-\frac{\kappa}{2} |\mathrm{X}|} = c  \hspace{0.1 cm}(c_{0} \hspace{0.05 cm} e_{0}^{1-2 \epsilon})^{\frac{1}{4}}
\hspace{0.1 cm} |\mathrm{Y}|$ follows from lemma 25 in \cite{D}.

\section{Large field region}
Recall that we have divided the whole lattice $\Lambda$ into $[r_{\lambda}]$ blocks $\Box$. 
Now there are two large field regions,
\begin{enumerate}
 \item  intermediate large field region $\Lambda_{1} - \Omega_{1}$ such that $\forall \hspace{0.05 cm} \Box \in \Lambda_{1} - \Omega_{1},
 \exists$ at least one $\xi \in \Box,$ with $p_{0,\lambda} < |\Phi (\xi)|$ and $\forall \xi \in \Box,  |\Phi (\xi)| < p_{\lambda}$,
  \item  large field region $\Lambda_{0}^{c}$ such that $\forall \hspace{0.05 cm} \Box \in \Lambda_{0}^{c}, \exists$
 at least one $\xi \in \Box,$ with  $|\Phi (\xi)| \geqslant p_{\lambda}$ and $v \neq 0$.
 \end{enumerate}
Let $\mathrm{P} \equiv \Lambda_{1} - \Omega_{1}$ and $\tilde{\mathrm{Q}} \equiv \Lambda_{1}^{c}$. 
Let $\{\mathrm{P}_{l}\}$ and $\{\tilde{\mathrm{Q}}_{k}\}$ denote  connected  components of $\mathrm{P}$ and $\tilde{\mathrm{Q}}$ respectively,
that is, $\mathrm{P}_{l}$ and $\tilde{\mathrm{Q}}_{k}$ are polymers. 
For any $\tilde{\mathrm{Q}}_{k} \subset \Lambda_{1}^{c}$, let $\bigcup_{i}\{\mathrm{Q}_{k, i}\} \subset \tilde{\mathrm{Q}}_{k}$ such that
each $\mathrm{Q}_{k, i}$ is a connected component of $\Lambda_{0}^{c}$. Denote $\tilde{\mathrm{Q}}_{k} \supset
\bigcup_{i}\{\mathrm{Q}_{k, i}\} \equiv \mathrm{Q}_{k}$. Rewrite the large field quadratic part in the generating functional Eq 2.29,
\begin{equation}
\begin{aligned}
-\frac{1}{2}\langle\Phi, \text{T}_{\Lambda^{c}_{1}} \Phi\rangle + \frac{1}{2}
\langle \Phi, \text{T}_{\Lambda^{c}_{1}\Lambda_{1}} C_{\Lambda_{1}} \text{T}_{\Lambda_{1}\Lambda^{c}_{1}}
\Phi \rangle 
&= -\frac{1}{2} \sum_{i} \langle\Phi, (\text{T}_{\tilde{\mathrm{Q}}_{i}} - \text{T}_{\tilde{\mathrm{Q}}_{i}\Lambda_{1}}
C_{\Lambda_{1}} \text{T}_{\Lambda_{1}\tilde{\mathrm{Q}}_{i}})\Phi\rangle \\ & + \frac{1}{2} 
\sum_{\substack{i,j \\ i\neq j}} \langle \Phi, \text{T}_{\tilde{\mathrm{Q}}_{i}\Lambda_{1}}
C_{\Lambda_{1}} \text{T}_{\Lambda_{1}\tilde{\mathrm{Q}}_{j}} \Phi\rangle.
\end{aligned}
\end{equation}
The first term on the right hand side can be represented by $\{\tilde{\mathrm{Q}}_{k}\}$. For the second term we construct 
polymers in the following manner. Consider two disjoint blocks $\Box$ and $\Box^{\prime}$ such that
$d(\Box, \Box^{\prime}) \geqslant [r_{\lambda}]$. Let X be a polymer connecting $\Box$ and $\Box^{\prime}$. For example,
  
$\text{X} (\Box, \Box^{\prime}) : \text{all d-dimensional rectangular paths connecting}\hspace{0.1 cm} \Box, \Box^{\prime}$.
 
(By rectangular paths we mean starting at $\Box$ and selecting coordinate axis one at a time and traveling along it until the
coordinate of $\Box^{\prime}$ is reached.)
 Let $1_{\Box}$ and $1_{\Box^{\prime}}$ denote the characteristic functions restricting operators to 
  $\Box$ and $\Box^{\prime}$ respectively. Then,
 \begin{equation}
 \begin{aligned}
\frac{1}{2}\sum_{\substack{i,j \\ i\neq j}}  \langle \Phi, \text{T}_{\tilde{\mathrm{Q}}_{i}\Lambda_{1}}
C_{\Lambda_{1}} \text{T}_{\Lambda_{1}\tilde{\mathrm{Q}}_{j}} \Phi\rangle &= \frac{1}{2}\sum_{\Box, \Box^{\prime}} \sum_{i,j}
\langle \Phi, 1_{\Box} \text{T}_{\tilde{\mathrm{Q}}_{i}\Lambda_{1}} 
C_{\Lambda_{1}} \text{T}_{\Lambda_{1}\tilde{\mathrm{Q}}_{j}} 1_{\Box^{\prime}} \Phi\rangle \\  
&= \frac{1}{2}\sum_{\text{X}} \sum_{\Box, \Box^{\prime} \rightarrow \text{X}} \sum_{i,j} \langle \Phi, 1_{\Box} \text{T}_{\tilde{\mathrm{Q}}_{i}\Lambda_{1}} C_{\Lambda_{1}} \text{T}_{\Lambda_{1}\tilde{\mathrm{Q}}_{j}} 1_{\Box^{\prime}} \Phi\rangle \\
&= \sum_{\text{X}} \sigma (\text{X}),
\end{aligned}
\end{equation}
where, only disjoint $\Box, \Box^{\prime}$ contribute and
\begin{equation}
\sigma (\text{X}) = \frac{1}{2}\sum_{\Box, \Box^{\prime} \rightarrow \text{X}} \sum_{i,j}\langle \Phi, 1_{\Box} \text{T}_{\tilde{\mathrm{Q}}_{i}\Lambda_{1}} C_{\Lambda_{1}} \text{T}_{\Lambda_{1}\tilde{\mathrm{Q}}_{j}} 1_{\Box^{\prime}} \Phi\rangle.
\end{equation}
Thus, $e^{\sum_{i \neq j}  \langle \Phi, \text{T}_{\tilde{\mathrm{Q}}_{i}\Lambda_{1}}
C_{\Lambda_{1}} \text{T}_{\Lambda_{1}\tilde{\mathrm{Q}}_{j}} \Phi\rangle} =  e^{\sum_{\text{X}} \sigma (\text{X})}$.
Using Mayer expansion, rewrite 
\begin{equation}
 e^{\sum_{\text{X}} \sigma (\text{X})} = \prod_{\text{X}} \Big[ (e^{\sigma(\text{X})} - 1) + 1\Big]
= \sum_{\substack{\{\text{X}_{i}\} \\ \text{distinct}}} \prod_{i} \Big[ e^{\sigma(\text{X}_{i})} - 1\Big] = 
\sum_{\substack{\{\text{X}_{i}\} \\ \text{disjoint}}} \prod_{i} f(\text{X}_{i})
\end{equation}
where, in the last step we group the $\{\text{X}_{i}\}$ into connected sets and define for X connected,
\begin{equation}
f(\text{X}) = \sum_{\cup \text{X}_{i} = \text{X}} \prod_{i} \Big[ e^{\sigma(\text{X}_{i})} - 1\Big].
\end{equation}
Recall from Eq 2.27, that $\mathrm{Z}_{0}(\Omega_{1}) = \int \mathcal{D}\Phi^{\prime}_{\Omega_{1}} 
e^{-||\Phi^{\prime}||_{\Omega_{1}}^{2}}$, rewrite 
\begin{equation}
\begin{aligned}
\mathrm{Z}_{0}(\Omega_{1}) =  
\mathrm{Z}_{0}(\Lambda) \hspace{0.05 cm} \mathrm{Z}_{0}(\Omega_{1}^{c})^{-1}  
= \mathrm{Z}_{0}(\Lambda) \hspace{0.05 cm} \mathrm{Z}_{0}(\tilde{\mathrm{Q}})^{-1} \hspace{0.05 cm} \mathrm{Z}_{0}(\mathrm{P})^{-1}.
\end{aligned}
\end{equation}
Then the generating functional is
\begin{equation}
\begin{aligned}
\mathrm{Z}[\text{J}] = & \Big(\frac{2\pi}{e_{0}}\Big)^{-|\Lambda| + 1} (\text{det}\hspace{0.1 cm}C^{\frac{1}{2}}) 
\hspace{0.1 cm} \mathrm{Z}_{0}(\Lambda) \hspace{0.1 cm}
 \sum_{v: dv=0} \sum_{\{\tilde{\mathrm{Q}}_{k}\}} \sum_{\{\text{X}_{i}\}}
\sum_{\{\mathrm{P}_{l}\}} \sum_{\{\text{Y}_{j}\}}  \hspace{0.1 cm} \prod_{k} e^{W_{2}(\tilde{\mathrm{Q}}_{k})} \\ &
\int \prod_{k} \mathcal{D}\Phi_{\tilde{\mathrm{Q}}_{k}}
 e^{-\frac{1}{2} \langle\Phi, (\text{T}_{\tilde{\mathrm{Q}}_{k}} - \text{T}_{\tilde{\mathrm{Q}}_{k}\Lambda_{1}}
C_{\Lambda_{1}} \text{T}_{\Lambda_{1}\tilde{\mathrm{Q}}_{k}})\Phi\rangle - V(\tilde{\mathrm{Q}}_{k}, \Phi)}
 \zeta_{\mathrm{Q}_{k}} \hspace{0.05 cm} \chi_{\tilde{\mathrm{Q}}_{k} \backslash \mathrm{Q}_{k}}(\Phi) 
 \prod_{k}  \mathrm{Z}_{0}(\tilde{\mathrm{Q}}_{k})^{-1} \\ & \prod_{i}  f(\text{X}_{i}, \text{J}) \hspace{0.1 cm}    
\prod_{l} e^{W_{2}(\mathrm{P}_{l})} \int \prod_{l} \mathcal{D}\Phi^{\prime}_{\mathrm{P}_{l}} e^{-\frac{1}{2}||\Phi^{\prime}||_{\mathrm{P}_{l}}^{2} - 
 V(\mathrm{P}_{l}, C^{\frac{1}{2},\text{loc}}_{\Lambda_{1}} \Phi^{\prime} - 
 C_{\Lambda_{1}}\text{T}_{\Lambda_{1}\Lambda^{c}_{1}} \Phi_{\Lambda^{c}_{1}})}
 \hat{\zeta}_{\mathrm{P}_{l}}(\Phi^{\prime})  \\ &   \hspace{0.1 cm}  
\chi_{\mathrm{P}_{l}}(C^{\frac{1}{2},\text{loc}}_{\Lambda_{1}} \Phi^{\prime} - C_{\Lambda_{1}}\text{T}_{\Lambda_{1}\Lambda^{c}_{1}}\Phi_{\Lambda^{c}_{1}}) \prod_{l} \mathrm{Z}_{0}(\mathrm{P}_{l})^{-1}  \hspace{0.1 cm}  \prod_{j} K(\text{Y}_{j}, \text{J}).
\end{aligned}
\end{equation}
where, $\zeta_{\mathrm{Q}_{k}} = \prod_{i}\zeta_{\mathrm{Q}_{k, i} : \cup_{i} \mathrm{Q}_{k, i} \subset \tilde{\mathrm{Q}}_{k}}(\Phi)$. 
Denote 
\begin{equation}
\begin{aligned}
\rho(\tilde{\mathrm{Q}}_{k}, \text{J}) &= e^{W_{2}(\tilde{\mathrm{Q}}_{k})} \hspace{0.1 cm}
 e^{-\frac{1}{2} \langle\Phi, (\text{T}_{\tilde{\mathrm{Q}}_{k}} - \text{T}_{\tilde{\mathrm{Q}}_{k}\Lambda_{1}}
C_{\Lambda_{1}} \text{T}_{\Lambda_{1}\tilde{\mathrm{Q}}_{k}})\Phi\rangle - V(\tilde{\mathrm{Q}}_{k}, \Phi)}\zeta_{\mathrm{Q}_{k}}(\Phi)
\hspace{0.05 cm} \chi_{\tilde{\mathrm{Q}}_{k} \backslash \mathrm{Q}_{k}}(\Phi)  \hspace{0.1 cm} \mathrm{Z}_{0}(\tilde{\mathrm{Q}}_{k})^{-1} \\
\rho^{\prime}(\mathrm{P}_{l}, \text{J}) &= e^{W_{2}(\mathrm{P}_{l})}  
 e^{-\frac{1}{2}||\Phi^{\prime}||_{\mathrm{P}_{l}}^{2} - V(\mathrm{P}_{l}, C^{\frac{1}{2},\text{loc}}_{\Lambda_{1}} \Phi^{\prime} - 
 C_{\Lambda_{1}}\text{T}_{\Lambda_{1}\Lambda^{c}_{1}} \Phi_{\Lambda^{c}_{1}})}
 \hat{\zeta}_{\mathrm{P}_{l}}(\Phi^{\prime})  \\ &
\hspace{0.4 cm}  \chi_{\mathrm{P}_{l}} (C^{\frac{1}{2},\text{loc}}_{\Lambda_{1}} \Phi^{\prime} - 
C_{\Lambda_{1}}\text{T}_{\Lambda_{1}\Lambda^{c}_{1}}\Phi_{\Lambda^{c}_{1}}) \hspace{0.1 cm} \mathrm{Z}_{0}(\mathrm{P}_{l})^{-1}
\end{aligned}
\end{equation}
and rewrite the generating functional as
\begin{equation}
\begin{aligned}
\mathrm{Z}[\text{J}] &= \Big(\frac{2\pi}{e_{0}}\Big)^{-|\Lambda| + 1} (\text{det}\hspace{0.1 cm}C^{\frac{1}{2}})
\hspace{0.1 cm} \mathrm{Z}_{0}(\Lambda) \hspace{0.1 cm}
 \sum_{v: dv=0} \sum_{\{\tilde{\mathrm{Q}}_{k}\}} \sum_{\{\text{X}_{i}\}}
\sum_{\{\mathrm{P}_{l}\}} \sum_{\{\text{Y}_{j}\}} \\ &  \prod_{k} \int  \mathcal{D}\Phi_{\tilde{\mathrm{Q}}_{k}} \rho(\tilde{\mathrm{Q}}_{k}, \text{J})  
 \hspace{0.1 cm} \prod_{i}  f(\text{X}_{i}, \text{J}) \hspace{0.1 cm}     
  \prod_{l} \int \mathcal{D}\Phi^{\prime}_{\mathrm{P}_{l}} \rho^{\prime}(\mathrm{P}_{l}, \text{J}) \hspace{0.1 cm} \prod_{j} K(\text{Y}_{j}, \text{J}).
\end{aligned}
\end{equation}
\textbf{Lemma 4.1} $\abs{f(\text{X}, \text{J})} \leqslant c \hspace{0.05 cm} e_{0} \hspace{0.05 cm} e^{-\gamma^{\prime} |\text{X}|}$,
for some $c, \gamma^{\prime} > 0$.

\textit{Proof} Let $\xi \in \Box$ and $\xi^{\prime} \in \Box^{\prime}$.
Define a metric $d(\xi, \xi^{\prime}) = \sup_{\mu} |\xi_{\mu} - \xi^{\prime}_{\mu}| $ and
$d(\Box, \Box^{\prime}) = \inf_{\xi, \xi^{\prime}} d(\xi, \xi^{\prime})$. Then
$|C(\xi, \xi^{\prime})| \leqslant c \hspace{0.05 cm} e^{-\gamma d(\xi, \xi^{\prime})} \leqslant 
c \hspace{0.05 cm} e^{-\gamma d(\Box, \Box^{\prime})}$.  
For any $(\tilde{\mathrm{Q}}_{i}, \tilde{\mathrm{Q}}_{j})$ with $\Box \in \tilde{\mathrm{Q}}_{i}$ and $\Box^{\prime} \in \tilde{\mathrm{Q}}_{j}$,
X is a polymer joining $\Box, \Box^{\prime}$ such that
\begin{equation}
d(\Box, \Box^{\prime}) = \inf_{\substack{\Box \in \tilde{\mathrm{Q}}_{i} \\ \Box^{\prime} \in \tilde{\mathrm{Q}}_{j}}} d(\Box, \Box^{\prime}). 
\end{equation}
We assume that the number of $(\Box, \Box^{\prime})$ satisfying the above condition is bounded by $\mathcal{O}(|\text{X}|)$
and the number of $(\tilde{\mathrm{Q}}_{i}, \tilde{\mathrm{Q}}_{j})$ is also bounded by $\mathcal{O}(|\text{X}|)$. Now there are $d!$ ways 
to construct X. Thus, $d(\Box, \Box^{\prime}) \geqslant \frac{|\text{X}|}{d!}$. Also, as $\Box, \Box^{\prime}$ are disjoint
$d(\Box, \Box^{\prime}) \geqslant [r_{\lambda}]$. Note that $\Phi$ is at most $p_{\lambda}$ near the boundary $\partial \Lambda_{1}$ 
since $|\Phi(\xi)| \leqslant p_{\lambda}, \forall \xi \in \Lambda_{0}$. 
From the definition, Eq 4.3 and using $e^{-\frac{\gamma}{2} [r_{\lambda}]} = \mathcal{O}(e_{0}^{2})$, it follows
 \begin{equation}
 \begin{aligned}
\abs{\sigma (\text{X})} &\leqslant  \frac{1}{2} \mathcal{O}(|\text{X}|^{2}) p_{\lambda}^{2} \hspace{0.05 cm} 
e^{-\frac{\gamma}{2} d(\Box, \Box^{\prime})} \hspace{0.05 cm} e^{-\frac{\gamma}{2} [r_{\lambda}]} \\ 
&\leqslant \frac{1}{2}  e_{0}^{2} \hspace{0.05 cm} |\text{X}|^{2} \hspace{0.05 cm}
p_{\lambda}^{2}  e^{-\frac{\gamma}{2 d!} |\text{X}|} \\
&\leqslant c \hspace{0.05 cm} e_{0} \hspace{0.05 cm} e^{-\frac{\gamma}{2 d!} |\text{X}|}
\end{aligned}
\end{equation}
and following Eq 3.60 for some $\gamma^{\prime} < \frac{\gamma}{2 d!}$,  
\begin{equation}
\abs{f(\text{X}, \text{J})} \leqslant c \hspace{0.05 cm} e_{0} \hspace{0.05 cm} e^{-\gamma^{\prime} |\text{X}|}.
\end{equation}

\textbf{Lemma 4.2} Let $m_{\text{min}} = \text{min} (\mu^{2}, m_{A}^{2})$. Then for sufficiently large $m_{\text{min}}$ and 
some  constant $\gamma_{1} > 0$, the large field action is bounded below as
\begin{equation}
\langle\Phi, (\text{T}_{\tilde{\mathrm{Q}}_{k}} - \text{T}_{\tilde{\mathrm{Q}}_{k}\Lambda_{1}}
C_{\Lambda_{1}} \text{T}_{\Lambda_{1}\tilde{\mathrm{Q}}_{k}})\Phi\rangle + V(\tilde{\mathrm{Q}}_{k}, \Phi)
\geqslant \gamma_{1} \langle\Phi, \text{T} \Phi\rangle_{\tilde{\mathrm{Q}}_{k}} + c \hspace{0.05 cm} ||\Phi||^{2}_{\tilde{\mathrm{Q}}_{k}}.
\end{equation}
\textit{Proof} Rewrite
 \begin{equation}
 \begin{aligned}
\langle \Phi, (\text{T}_{\tilde{\mathrm{Q}}_{k}} - 
\text{T}_{\tilde{\mathrm{Q}}_{k}\Lambda_{1}} C_{\Lambda_{1}} \text{T}_{\Lambda_{1}\tilde{\mathrm{Q}}_{k}}) \Phi\rangle
+ V(\tilde{\mathrm{Q}}_{k}, \Phi) &= \frac{1}{2} \langle \Phi, (\mathrm{T}_{\tilde{\mathrm{Q}}_{k}} - 
 2 \hspace{0.05 cm} \text{T}_{\tilde{\mathrm{Q}}_{k}\Lambda_{1}} C_{\Lambda_{1}} \text{T}_{\Lambda_{1}\tilde{\mathrm{Q}}_{k}}) \Phi\rangle  \\
& +  \frac{1}{2} \langle\Phi, \text{T} \Phi\rangle_{\tilde{\mathrm{Q}}_{k}} + V(\tilde{\mathrm{Q}}_{k}, \Phi).
\end{aligned}
\end{equation}
From the definition of $\text{T} = C^{-1}$ (Eq 2.8), note that for some constant 
$c > 0, ||C_{\Lambda_{1}}||_{\infty} < \frac{c}{m_{\text{min}}}$ and due to locality in $\text{T}_{\tilde{\mathrm{Q}}_{k}\Lambda_{1}}$,
there is no mass term in $\text{T}_{\tilde{\mathrm{Q}}_{k}\Lambda_{1}}$. Using the positivity of $-\Delta$ and $\delta d$,
\begin{equation}
\begin{aligned}
\langle \Phi, (\mathrm{T}_{\tilde{\mathrm{Q}}_{k}} - 
2 \hspace{0.05 cm} \text{T}_{\tilde{\mathrm{Q}}_{k}\Lambda_{1}} C_{\Lambda_{1}} \text{T}_{\Lambda_{1}\tilde{\mathrm{Q}}_{k}}) \Phi\rangle 
&\geqslant (m_{\text{min}} -  \frac{c}{m_{\text{min}}}) ||\Phi||^{2}_{\tilde{\mathrm{Q}}_{k}}.
\end{aligned}
\end{equation}
The constant on the right is positive for $m_{\text{min}}$ sufficiently large. And the result then follows from the 
stability lemma 11.1 in \cite{BBIJ}.

\textbf{Proposition 4.1} Let $\tilde{\mathrm{Q}} = \{\tilde{\mathrm{Q}}_{k}\}$ with $\tilde{\mathrm{Q}}_{k} \supset \mathrm{Q}_{k}$ 
be polymers as defined and the function $\rho(\tilde{\mathrm{Q}}_{k}, \text{J})$ be as defined in Eq 4.8. Then 
\begin{equation}
\abs{\sum_{v \hspace{0.05 cm} \text{on} \hspace{0.05 cm} \tilde{\mathrm{Q}}_{k} : dv=0} \rho (\tilde{\mathrm{Q}}_{k}, \text{J})} 
\leqslant c \hspace{0.05 cm} e_{0}^{2} \hspace{0.05 cm} 
e^{-\frac{3}{8} p_{\lambda} |\mathrm{Q}_{k}|} e^{- (m_{\text{min}} \gamma_{1} - p_{\lambda}^{-1} + c) ||\Phi||_{\tilde{\mathrm{Q}}_{k}}^{2}}.
\end{equation}
\textit{Proof} Note that for $\Box \in \mathrm{Q}_{k}, \Phi(\xi) > p_{\lambda}$, for at least one $\xi \in \Box$ and so
\begin{equation}
\zeta(\Box) \leqslant  e^{-p_{\lambda} + p_{\lambda}^{-1}||\Phi||_{\Box}^{2}}.
\end{equation}
Therefore, for $\bigcup_{i} \mathrm{Q}_{k, i} = \mathrm{Q}_{k} \subset \tilde{\mathrm{Q}}_{k}$,
\begin{equation}
\zeta(\mathrm{Q}_{k}) = \prod_{i} \prod_{\Box \in \mathrm{Q}_{k, i}} \hat{\zeta}(\Box) \leqslant  
e^{-p_{\lambda}\sum_{i}|\mathrm{Q}_{k, i}| + p_{\lambda}^{-1}\sum_{i} ||\Phi||_{\mathrm{Q}_{k, i}}^{2}}
\leqslant e^{-p_{\lambda}|\mathrm{Q}_{k}| + p_{\lambda}^{-1}||\Phi||_{\mathrm{Q}_{k}}^{2}}.
\end{equation}
From lemma 2.7, $W_{2}(\Box) \leqslant \mathcal{O}([r_{\lambda}]^{d})$.
Note that $\abs{\mathrm{Z}_{0}(\tilde{\mathrm{Q}}_{k})^{-1}} = \pi^{\frac{-|\tilde{\mathrm{Q}}_{k}|}{2}} \leqslant 
e^{-\frac{1}{2} |\tilde{\mathrm{Q}}_{k}|}$. Thus,
\begin{equation}
\begin{aligned}
e^{W_{2}(\tilde{\mathrm{Q}}_{k})} \hspace{0.1 cm} \mathrm{Z}_{0}(\tilde{\mathrm{Q}}_{k})^{-1} \zeta_{\mathrm{Q}_{k}}(\Phi)\hspace{0.1 cm}
\chi_{\tilde{\mathrm{Q}}_{k} \backslash \mathrm{Q}_{k}}(\Phi) 
&\leqslant e^{(\mathcal{O}([r_{\lambda}]^{d}) -\frac{1}{2}) \hspace{0.05 cm} |\tilde{\mathrm{Q}}_{k}|}
 \hspace{0.1 cm} e^{-p_{\lambda} \hspace{0.05 cm} |\mathrm{Q}_{k}| + p_{\lambda}^{-1}||\Phi||_{\mathrm{Q}_{k}}^{2}} \\
 &\leqslant e^{3^{d} (\mathcal{O}([r_{\lambda}]^{d}) -\frac{1}{2}) \hspace{0.05 cm} |\mathrm{Q}_{k}|}
 \hspace{0.1 cm} e^{-p_{\lambda} \hspace{0.05 cm} |\mathrm{Q}_{k}| + p_{\lambda}^{-1}||\Phi||_{\tilde{\mathrm{Q}}_{k}}^{2}}
\end{aligned}
\end{equation} 
where, we have used the fact that $|\Lambda_{1}^{c}| \leqslant 3^{d} |\Lambda_{0}^{c}|$ and therefore, 
$|\tilde{\mathrm{Q}}_{k}| \leqslant 3^{d}|\mathrm{Q}_{k}|$ and $||\Phi||_{\mathrm{Q}_{k}^{2}} \leqslant ||\Phi||_{\tilde{\mathrm{Q}}_{k}^{2}}$.
Now using lemma 4.2,
\begin{equation}
\abs{\sum_{v \hspace{0.05 cm} \text{on} \hspace{0.05 cm} \tilde{\mathrm{Q}}_{k}  : dv=0} \rho (\tilde{\mathrm{Q}}_{k}, \text{J})} 
\leqslant e^{3^{d} (\mathcal{O}([r_{\lambda}]^{d}) -\frac{1}{2}) \hspace{0.05 cm} |\mathrm{Q}_{k}|}
 \hspace{0.1 cm} e^{-p_{\lambda} \hspace{0.05 cm} |\mathrm{Q}_{k}|} \hspace{0.1 cm} \sum_{v \in \tilde{\mathrm{Q}}_{k}} 
 e^{- \frac{1}{2} \gamma_{1} \langle\Phi, \text{T} \Phi\rangle_{\tilde{\mathrm{Q}}_{k}}  - \langle dA + v, \text{J} \rangle +
(p_{\lambda}^{-1}- c) ||\Phi||_{\tilde{\mathrm{Q}}_{k}}^{2}}. 
\end{equation}
The bound of $|\text{J}| < \alpha$, where, $\alpha < 1$, implies 
$\abs{e^{- \langle dA + v, \text{J} \rangle}} \leqslant e^{ \sum_{p} \alpha |(dA + v)(p)|}$. 
From the definition $\langle \Phi, \text{T} \Phi\rangle = (dA + v)^{2}(p) + m_{A}^{2} |A(b)|^{2} + \mathcal{O}(\rho^{2})$.
Rewrite,
\begin{equation}
\begin{aligned}
\abs{\sum_{v \hspace{0.05 cm} \text{on} \hspace{0.05 cm} \mathrm{Q}_{k} : dv=0} \rho (\tilde{\mathrm{Q}}_{k}, \text{J})} 
&\leqslant e^{3^{d}(\mathcal{O}([r_{\lambda}]^{d}) - \frac{1}{2}) \hspace{0.05 cm} |\mathrm{Q}_{k}|}
 \hspace{0.1 cm} e^{-p_{\lambda} \hspace{0.05 cm} |\mathrm{Q}_{k}|}
 \hspace{0.1 cm} \sum_{v \in \tilde{\mathrm{Q}}_{k}} e^{-\frac{1}{2} \gamma_{1} \langle\Phi, \text{T}\Phi\rangle_{\tilde{\mathrm{Q}}_{k}} 
  + \sum_{p} \alpha |(dA + v)(p)| + (p_{\lambda}^{-1}- c) ||\Phi||_{\tilde{\mathrm{Q}}_{k}}^{2}} \\
   &\leqslant e^{3^{d}(\mathcal{O}([r_{\lambda}]^{d}) -\frac{1}{2}) \hspace{0.05 cm} |\mathrm{Q}_{k}|}
 \hspace{0.1 cm} e^{-p_{\lambda} \hspace{0.05 cm} |\mathrm{Q}_{k}|}
 \hspace{0.1 cm}  e^{ (p_{\lambda}^{-1}- c) ||\Phi||^{2}_{\tilde{\mathrm{Q}}_{k}}} e^{- \sum_{b \in \tilde{\mathrm{Q}}_{k}} 
 \gamma_{1} m_{A}^{2} |A(b)|^{2} - \gamma_{1} \mathcal{O}(\rho^{2})}  \\
 & \hspace{0.4 cm} \sum_{v(p), p \in \tilde{\mathrm{Q}}_{k}} e^{\sum_{p \in \tilde{\mathrm{Q}}_{k}} 
 - \frac{1}{2}  \gamma_{1} |dA(p) + v(p)|^{2} + \alpha |dA(p) + v(p)|}
\end{aligned}
\end{equation}
using $v(p) \in \frac{2\pi}{e_{0}}\mathbb{Z}$ and $|dA| < \frac{4\pi}{e_{0}}$, only writing the terms containing \textit{v},
\begin{equation}
\begin{aligned}
\sum_{v(p), p \in \mathrm{Q}_{k}}& e^{\sum_{p\in \tilde{\mathrm{Q}}_{k}} - \frac{1}{2} \gamma_{1} |dA(p) + v(p)|^{2}  + \alpha |dA(p) + v(p))|}
 = \prod_{p \in \tilde{\mathrm{Q}}_{k}} \sum_{v(p)} e^{-\frac{1}{2}\gamma_{1} |dA(p) + v(p)|^{2} + \alpha |dA(p) + v(p))|} \\  
& \leqslant
\prod_{p \in \tilde{\mathrm{Q}}_{k}} \Big(5 + \sum_{|v| \geqslant \frac{6\pi}{e_{0}}} 
e^{-\frac{1}{2} \gamma_{1} |dA(p) + v(p)|^{2} + \alpha |dA(p) + v(p)|}\Big) \\
&\leqslant \prod_{p \in\tilde{ \mathrm{Q}}_{k}}  \Big(5 + \sum_{|v| \geqslant \frac{6\pi}{e_{0}}} 
 e^{-\frac{1}{2}\gamma_{1} |dA(p) + v(p)|^{2} + \frac{\alpha^{2}}{\gamma^{2}}}\Big).
\end{aligned}
\end{equation}
Since, $|dA + v| > \frac{2\pi}{e_{0}} = \frac{v}{3}$, therefore,
\begin{equation}
\begin{aligned}
\sum_{v(p), p \in \tilde{\mathrm{Q}}_{k}} e^{\sum_{p\in \tilde{\mathrm{Q}}_{k}} - \frac{1}{2} \gamma_{1} |dA(p) + v(p)|^{2}  + \alpha |dA(p) + v(p))|}
&\leqslant \prod_{p \in \tilde{\mathrm{Q}}_{k}}  \Big(5 +  e^{\frac{\alpha^{2}}{\gamma_{1}^{2}}} \sum_{|v| \geqslant \frac{6\pi}{e_{0}}} 
 e^{-\frac{1}{18}\gamma_{1} v^{2}}\Big) \\
&\leqslant \prod_{p \in \tilde{\mathrm{Q}}_{k}} \text{c} = e^{\text{ln c}\hspace{0.05 cm} |\tilde{\mathrm{Q}}_{k}|}
\leqslant  e^{\text{ln c}\hspace{0.05 cm}3^{d} |\mathrm{Q}_{k}|}
\end{aligned}
\end{equation}
Note that $\sum_{b \in \tilde{\mathrm{Q}}_{k}} \gamma_{1} m_{A}^{2} |A(b)|^{2} + \gamma_{1} \mathcal{O}(\rho^{2}) \geqslant
m_{\text{min}} \gamma_{1} ||\Phi||_{\tilde{\mathrm{Q}}_{k}}^{2}$.
Also note that $ |\tilde{\mathrm{Q}}_{k}|$ is getting cancelled by $ |\mathrm{Q}_{k}|$ since
we can assume that $3^{d} (\mathcal{O}([r_{\lambda}]^{d}) -\frac{1}{2}) + 3^{d} \text{ln c}  < \frac{p_{\lambda}}{8}$. Thus,
\begin{equation}
\begin{aligned}
\abs{\sum_{v \hspace{0.05 cm} \text{on} \hspace{0.05 cm} \tilde{\mathrm{Q}}_{k} : dv=0} \rho (\tilde{\mathrm{Q}}_{k}, \text{J})}
 &\leqslant e^{3^{d}(\mathcal{O}([r_{\lambda}]^{d}) -\frac{1}{2} + \text{ln c}) \hspace{0.05 cm} |\mathrm{Q}_{k}|}
 \hspace{0.1 cm} e^{-p_{\lambda} \hspace{0.05 cm} |\mathrm{Q}_{k}|}
 \hspace{0.1 cm} e^{- (m_{\text{min}}  \gamma_{1}  - p_{\lambda}^{-1} + c) ||\Phi||^{2}_{\tilde{\mathrm{Q}}_{k}}} \\
&\leqslant c \hspace{0.05 cm} 
e^{-\frac{7}{8} p_{\lambda} |\mathrm{Q}_{k}|} e^{- (m_{\text{min}} \gamma_{1}  - p_{\lambda}^{-1} + c) 
||\Phi||_{\tilde{\mathrm{Q}}_{k}}^{2}} \\ 
&\leqslant c \hspace{0.05 cm} e_{0}^{2} \hspace{0.05 cm} 
e^{-\frac{3}{8} p_{\lambda} |\mathrm{Q}_{k}|} e^{- (m_{\text{min}}\gamma_{1}  - p_{\lambda}^{-1} + c) ||\Phi||_{\tilde{\mathrm{Q}}_{k}}^{2}}.
\end{aligned}
\end{equation}
\textbf{Lemma 4.3}  In region $\{\mathrm{P}_{l}\}$, the following relation holds
\begin{equation}
e^{- V(\mathrm{P}_{l}, C^{\frac{1}{2},\text{loc}}_{\Lambda_{1}} \Phi^{\prime} - C_{\Lambda_{1}}\text{T}_{\Lambda_{1}\Lambda^{c}_{1}} 
 \Phi_{\Lambda^{c}_{1}})} \leqslant e^{ c \hspace{0.05 cm} (c_{0} \hspace{0.05 cm} e_{0}^{1-2 \epsilon})^{\frac{1}{2}} |\mathrm{P}_{l}|}. 
\end{equation}
\textit{Proof} Note that $\forall \hspace{0.05 cm} \Box \in \mathrm{P} \hspace{0.05 cm} \exists$ at least one $\xi \in \Box,$ with 
$p_{0,\lambda} < |\Phi (\xi)|$ and $\forall \xi \in \Box,  |\Phi (\xi)| < p_{\lambda}$, 
thus we do not need a stability condition of $\tilde{\mathrm{Q}}$. Now from 3.48,
$|V(\mathrm{P}_{l}, C^{\frac{1}{2},\text{loc}}_{\Lambda_{1}} \Phi^{\prime} - C_{\Lambda_{1}}\text{T}_{\Lambda_{1}\Lambda^{c}_{1}} 
\Phi_{\Lambda^{c}_{1}})| \leqslant c \hspace{0.05 cm} (c_{0} \hspace{0.05 cm} e_{0}^{1-2 \epsilon})^{\frac{1}{2}} |\mathrm{P}_{l}|$ 
and the result follows using $|e^{-V}| \leqslant e^{|V|}$.

\textbf{Proposition 4.2} Let $\mathrm{P} = \{\mathrm{P}_{l}\}$ be a collection of  polymers
and the function $\rho^{\prime}(\mathrm{P}_{l}, \text{J})$ be as defined in Eq 4.8. Then 
\begin{equation}
\abs{\rho^{\prime}(\mathrm{P}_{l}, \text{J})} \leqslant c \hspace{0.05 cm} e_{0}^{2} \hspace{0.05 cm} 
e^{-\frac{3}{8} p_{0,\lambda} |\mathrm{P}_{l}|} e^{- (\frac{1}{2}- p_{0, \lambda}^{-1}) ||\Phi^{\prime}||_{\mathrm{P}_{l}}^{2}}.
\end{equation}
\textit{Proof} From lemma 4.3,
\begin{equation}
e^{-\frac{1}{2}||\Phi^{\prime}||_{\mathrm{P}_{l}}^{2} - 
 V(\mathrm{P}_{l}, C^{\frac{1}{2},\text{loc}}_{\Lambda_{1}} \Phi^{\prime} - 
 C_{\Lambda_{1}}\text{T}_{\Lambda_{1}\Lambda^{c}_{1}} \Phi_{\Lambda^{c}_{1}})}
\leqslant e^{ c \hspace{0.05 cm} (c_{0} \hspace{0.05 cm} e_{0}^{1-2 \epsilon})^{\frac{1}{2}} |\mathrm{P}_{l}|} 
e^{- \frac{1}{2}||\Phi^{\prime}||_{\mathrm{P}_{l}}^{2}}. 
\end{equation}
Note that $\Phi(\xi) > p_{0,\lambda}, \forall \xi \in \Box$ and so
\begin{equation}
\hat{\zeta}(\Box) \leqslant  
e^{-p_{0, \lambda} + p_{0, \lambda}^{-1}||\Phi^{\prime}||_{\Box}^{2}}.
\end{equation}
Therefore,
\begin{equation}
\hat{\zeta}(\mathrm{P}_{l}) = \prod_{\Box \in \mathrm{P}_{l}} \hat{\zeta}(\Box) \leqslant  
e^{-p_{0, \lambda}|\mathrm{P}_{l}| + p_{0, \lambda}^{-1}||\Phi^{\prime}||_{\mathrm{P}_{l}}^{2}}.
\end{equation}
From lemma 2.7, $W_{2}(\Box) \leqslant \mathcal{O}([r_{\lambda}]^{d})$. We can assume by making $p_{0, \lambda}$  
big enough that,  $[r_{\lambda}]^{d} + c \hspace{0.05 cm} (c_{0} \hspace{0.05 cm} e_{0}^{1-2 \epsilon})^{\frac{1}{2}}
 < \frac{p_{0,\lambda}}{8}$. Note that
$\abs{ \mathrm{Z}_{0}(\mathrm{P}_{l})^{-1}} = \pi^{\frac{-|\mathrm{P_{l}}|}{2}} \leqslant e^{-\frac{1}{2}|\mathrm{P}_{l}|}$ and
$\chi_{\mathrm{P}_{l}} (C^{\frac{1}{2},\text{loc}}_{\Lambda_{1}} \Phi^{\prime} - 
C_{\Lambda_{1}}\text{T}_{\Lambda_{1}\Lambda^{c}_{1}}\Phi_{\Lambda^{c}_{1}}) \leqslant 1$. 
Thus,
\begin{equation}
\begin{aligned}
\abs{\rho^{\prime}(\mathrm{P}_{l}, \text{J})} &\leqslant c \hspace{0.05 cm} e^{(\mathcal{O}([r_{\lambda}]^{d}) - \frac{1}{2} + 
c \hspace{0.05 cm} (c_{0} \hspace{0.05 cm} e_{0}^{1-2 \epsilon})^{\frac{1}{2}})
 \hspace{0.05 cm} |\mathrm{P}_{l}|} \hspace{0.05 cm} e^{-p_{0, \lambda} \hspace{0.05 cm} |\mathrm{P}_{l}|} \hspace{0.1 cm}  
e^{- \frac{1}{2} ||\Phi^{\prime}||_{\mathrm{P}_{l}}^{2} + p_{0, \lambda}^{-1}||\Phi^{\prime}||_{\mathrm{P}_{l}}^{2}} \\
&\leqslant c \hspace{0.05 cm} e^{-\frac{7}{8} p_{0, \lambda} \hspace{0.05 cm} |\mathrm{P}_{l}|} \hspace{0.05 cm} 
e^{- (\frac{1}{2} - p_{0, \lambda}^{-1}) ||\Phi^{\prime}||_{\mathrm{P}_{l}}^{2}} \\
&\leqslant c \hspace{0.05 cm} e_{0}^{2} \hspace{0.1 cm} e^{-\frac{3}{8} p_{0, \lambda} \hspace{0.05 cm} |\mathrm{P}_{l}|}
 \hspace{0.1 cm}  
e^{- (\frac{1}{2} - p_{0, \lambda}^{-1}) ||\Phi^{\prime}||_{\mathrm{P}_{l}}^{2}}.
\end{aligned}
\end{equation}
where, we have used $e^{-\frac{p_{0,\lambda}}{2}} = \mathcal{O}(e_{0}^{2})$.

\section{Convergence}
Lattice $\Lambda$ is composed of disjoint polymers $\{\mathrm{Y}_{j}\}$ of $\Omega_{1}$, $\{\mathrm{P}_{l}\}$ of
$\Lambda_{1} - \Omega_{1}$, $\{\mathrm{X}_{i}\}$ and $\{\tilde{\mathrm{Q}}_{k}\}$ of $\Lambda_{1}^{c}$.
These connected components from various regions overlap near the boundaries of those regions. 
We combine these overlapping parts into connected components in two steps. 
\begin{enumerate}
  \item In small field region $\Lambda_{1}$,
define connected components $\{Z_{m}\}$, such that any $Z \in \{Z_{m}\}$, 
\begin{equation}
Z = \cup_{j}\{Y_{j}\} \cup_{l} \{\mathrm{P}_{l}\}
\end{equation}
where, $\text{Y}_{j}$ and $\mathrm{P}_{l}$ overlap and rewrite
\begin{equation}
\begin{aligned}
\sum_{\{Y_{j}\}} \sum_{\{\mathrm{P}_{l}\}} &
\hspace{0.05 cm} \prod_{l} \int \mathcal{D}\Phi^{\prime}_{P_{l}} \hspace{0.05 cm} \rho^{\prime}(\mathrm{P}_{l}, \text{J}) \hspace{0.1 cm} 
 \prod_{j} K(\text{Y}_{j}, \text{J})\\ &= \sum_{\{Z_{m}\}} \sum_{\cup_{j}\{Y_{j}\} \cup_{l} \{\mathrm{P}_{l}\} \rightarrow Z_{m}}  
 \hspace{0.05 cm} \prod_{l} \int \mathcal{D}\Phi^{\prime}_{P_{l}} \hspace{0.05 cm} \rho^{\prime}(\mathrm{P}_{l}, \text{J}) \hspace{0.1 cm} 
 \prod_{j} K(\text{Y}_{j}, \text{J}) \\ &= \sum_{\{Z_{m}\}} \prod_{m} K^{\prime} (Z_{m}, \text{J})
\end{aligned}
\end{equation}
where, for connected $Z$,
\begin{equation}
K^{\prime} (Z, \text{J}) = \sum_{\cup_{j}\{Y_{j}\} \cup_{l} \{\mathrm{P}_{l}\} \rightarrow Z}  
 \hspace{0.05 cm} \prod_{l} \int \mathcal{D}\Phi^{\prime}_{P_{l}} \hspace{0.05 cm} \rho^{\prime}(\mathrm{P}_{l}, \text{J}) \hspace{0.1 cm} 
 \prod_{j} K(\text{Y}_{j}, \text{J}).
\end{equation}
Rewrite the generating functional
\begin{equation}
\begin{aligned}
\mathrm{Z}[\text{J}] &= \Big(\frac{2\pi}{e_{0}}\Big)^{-|\Lambda| + 1} (\text{det}\hspace{0.1 cm}C^{\frac{1}{2}})
\hspace{0.1 cm} \mathrm{Z}_{0}(\Lambda) \hspace{0.1 cm}
 \sum_{v: dv=0} \sum_{\{\tilde{\mathrm{Q}}_{k}\}} \sum_{\{\text{X}_{i}\}}
 \sum_{\{Z_{m}\}} \\ &  \prod_{k} \int  \mathcal{D}\Phi_{\tilde{\mathrm{Q}}_{k}} \rho(\tilde{\mathrm{Q}}_{k}, \text{J})  
 \hspace{0.1 cm} \prod_{i}  f(\text{X}_{i}, \text{J}) \hspace{0.1 cm} \prod_{m} K^{\prime}(Z_{m}, \text{J}).
\end{aligned}
\end{equation}

  \item In entire lattice $\Lambda$, define connected components $\{\mathrm{C}_{l}\}$, such that any $\mathrm{C} \in \{\mathrm{C}_{l}\}$,
\begin{equation}
\mathrm{C} =  \cup_{i}\{\text{X}_{i}\} \cup_{k}\{\tilde{\mathrm{Q}}_{k}\}\cup_{m}\{Z_{m}\}
\end{equation}
where, $X_{i}$, $\tilde{\mathrm{Q}}_{k}$ and $Z_{m}$ overlap and rewrite
\begin{equation}
\begin{aligned}
& \sum_{\{\text{X}_{i}\}} \sum_{\{\tilde{\mathrm{Q}}_{k}\}} \sum_{\{Z_{m}\}} 
 \prod_{k}  \int \mathcal{D}\Phi_{\tilde{\mathrm{Q}}_{k}} \sum_{v \hspace{0.05 cm} \text{on} \hspace{0.05 cm} \tilde{\mathrm{Q}}_{k} : dv=0}
 \rho(\tilde{\mathrm{Q}}_{k}, \text{J})  \hspace{0.05 cm} \prod_{i} f(\text{X}_{i}, \text{J}) 
 \hspace{0.05 cm} \prod_{m} K^{\prime}(Z_{m}, \text{J}) \\  
&=  \sum_{\{\mathrm{C}_{l}\}} \sum_{\cup_{i}\{\text{X}_{i}\} \cup_{k}\{\tilde{\mathrm{Q}}_{k}\} \cup_{m}\{Z_{m}\} \rightarrow \{\mathrm{C}_{l}\}} 
 \prod_{k} \int \mathcal{D}\Phi_{\tilde{\mathrm{Q}}_{k}} \sum_{v \hspace{0.05 cm} \text{on} \hspace{0.05 cm} \tilde{\mathrm{Q}}_{k} : dv=0}
 \rho(\tilde{\mathrm{Q}}_{k}, \text{J})  \hspace{0.05 cm} \prod_{i} f(\text{X}_{i}, \text{J}) 
 \hspace{0.05 cm} \prod_{m} K^{\prime}(Z_{m}, \text{J}) \\
&= \sum_{\{\mathrm{C}_{l}\}} \prod_{l} K^{\#} (\mathrm{C}_{l}, \text{J}),  
\end{aligned}
\end{equation}
where, for any connected $\mathrm{C}$, 
\begin{equation}
 K^{\#} (\mathrm{C}, \text{J}) = \sum_{\cup_{i}\{\text{X}_{i}\} \cup_{k}\{\tilde{\mathrm{Q}}_{k}\} \cup_{m}\{Z_{m}\}  \rightarrow \mathrm{C}} 
 \prod_{k} \int \mathcal{D}\Phi_{\mathrm{Q}_{k}} \sum_{v \hspace{0.05 cm} \text{on} \hspace{0.05 cm} \tilde{\mathrm{Q}}_{k} : dv=0}
 \rho(\tilde{\mathrm{Q}}_{k}, \text{J})  \hspace{0.05 cm} \prod_{i} f(\text{X}_{i}, \text{J}) 
 \hspace{0.05 cm} \prod_{m} K^{\prime}(Z_{m}, \text{J}). 
\end{equation}
Rewrite the generating functional
\begin{equation}
\mathrm{Z}[\text{J}] = \Big(\frac{2\pi}{e_{0}}\Big)^{-|\Lambda| + 1} (\text{det}\hspace{0.1 cm}C^{\frac{1}{2}})
\hspace{0.1 cm} \mathrm{Z}_{0}(\Lambda) \hspace{0.1 cm} \sum_{\{\mathrm{C}_{l}\}} \prod_{l} K^{\#} (\mathrm{C}_{l}, \text{J}).
\end{equation}
\end{enumerate}
To take the logarithm of the generating functional, first using standard procedure, see for example \cite{D}, write
\begin{equation}
\sum_{\{\mathrm{C}_{l}\}} \prod_{l} K^{\#}(\mathrm{C}_{l}, \text{J}) = e^{\sum_{\mathrm{C}} E(\mathrm{C}, \text{J})}
\end{equation}
where,
\begin{equation}
 E(\mathrm{C}, \text{J}) = \sum_{n=1}^{\infty} \sum_{\{\mathrm{C}_{1}, \cdots, \mathrm{C}_{n}\} : \cup_{l}\mathrm{C}_{l} = \mathrm{C}} 
 \rho^{T}(\mathrm{C}_{1}, \cdots, \mathrm{C}_{n}) \prod_{l} K^{\#}(\mathrm{C}_{l}, \text{J})
\end{equation}
and $\rho^{T}(\mathrm{C}_{1}, \cdots, \mathrm{C}_{n}) = 0$ if $\mathrm{C}_{l}$ can be divided into disjoint sets and from \cite{D}
\begin{equation}
\abs{E(\mathrm{C}, \text{J})} \leqslant \mathcal{O}(1) \abs{K^{\#}(\mathrm{C}, \text{J})}.
\end{equation}
Rewrite the generating functional  
\begin{equation}
\mathrm{Z}[\text{J}] = \Big(\frac{2\pi}{e_{0}}\Big)^{-|\Lambda| + 1} (\text{det}\hspace{0.1 cm}C^{\frac{1}{2}})
\hspace{0.1 cm} \mathrm{Z}_{0}(\Lambda) \hspace{0.1 cm}e^{\sum_{\mathrm{C}} E(\mathrm{C}, \text{J})}
\end{equation}
and take the logarithm
\begin{equation}
\text{log} \hspace{0.05 cm} \mathrm{Z}[\text{J}] = \sum_{\mathrm{C}} E(\mathrm{C}, \text{J}) +
\text{log} \hspace{0.05 cm} \Big[\Big(\frac{2\pi}{e_{0}}\Big)^{-|\Lambda|+1}  \hspace{0.05 cm}(\text{det} \hspace{0.1 cm} C^{\frac{1}{2}})
\hspace{0.05 cm} \mathrm{Z}_{0}(\Lambda)\Big]. 
\end{equation}

\textbf{Lemma 5.1} There exists constant $\kappa_{1} > 0$ such that    
$|K^{\prime}(Z)| \leqslant (c_{0}e_{0}^{5-2\epsilon})^{\frac{1}{2}}  \hspace{0.05 cm} e^{-\kappa_{1} |Z|}$.

\textit{Proof} From the definition, 
\begin{equation}
\begin{aligned}
\abs{K^{\prime}(Z, \text{J})} &\leqslant  \sum_{\cup_{j}\{Y_{j}\} \cup_{l} \{\mathrm{P}_{l}\} = Z}  \prod_{l} 
\abs{\int \mathcal{D}\Phi^{\prime}_{P_{l}} \hspace{0.05 cm} \rho^{\prime} (\mathrm{P}_{l}, \text{J})} \hspace{0.1 cm}
 \prod_{j}\abs{K(\text{Y}_{j}, \text{J})} \\
 &\leqslant  \sum_{\cup_{j}\{Y_{j}\} \cup_{l} \{\mathrm{P}_{l}\} = Z}  
\prod_{j} (c_{0}e_{0}^{1-2\epsilon})^{\frac{1}{2}} \hspace{0.05 cm}e^{- \kappa^{\prime} |Y_{j}|}
\hspace{0.1 cm} \prod_{l} c \hspace{0.05 cm} e_{0}^{2}\hspace{0.05 cm} e^{-\frac{3}{8} p_{0, \lambda} \hspace{0.05 cm} |\mathrm{P}_{l}|}
 \hspace{0.1 cm}  \int \mathcal{D}\Phi^{\prime}_{P_{l}} \hspace{0.05 cm} e^{- (\frac{1}{2}- p_{0, \lambda}^{-1}) ||\Phi^{\prime}||_{\mathrm{P}_{l}}^{2}} \\  
 &\leqslant e^{- \frac{\kappa^{\prime}}{2} |Z|} \sum_{\cup_{j}\{Y_{j}\} \cup_{l} \{\mathrm{P}_{l}\} \subset Z}  
\prod_{j} (c_{0}e_{0}^{1-2\epsilon})^{\frac{1}{2}} \hspace{0.05 cm}e^{- \frac{\kappa^{\prime}}{2} |Y_{j}|}
\hspace{0.1 cm} \prod_{l} c \hspace{0.05 cm} e_{0}^{2}\hspace{0.05 cm} e^{-\frac{3}{16} p_{0, \lambda} \hspace{0.05 cm} |\mathrm{P}_{l}|}
 \hspace{0.1 cm}  \int \mathcal{D}\Phi^{\prime}_{P_{l}} \hspace{0.05 cm} e^{- (\frac{1}{2}- p_{0, \lambda}^{-1}) ||\Phi^{\prime}||_{\mathrm{P}_{l}}^{2}} \\  
 &\leqslant e^{- \frac{\kappa^{\prime}}{2} |Z|}  \Big(\frac{\pi}{1 - p_{0,\lambda}^{-1}}\Big)^{\frac{|Z|}{2}}  \Big[\sum_{\cup_{j}\{Y_{j}\} \subset Z} 
\prod_{j} (c_{0}e_{0}^{1-2\epsilon})^{\frac{1}{2}} \hspace{0.05 cm}e^{- \frac{\kappa^{\prime}}{2} |Y_{j}|}\Big]
\hspace{0.05 cm}\Big[\sum_{\cup_{l} \{\mathrm{P}_{l}\} \subset Z} \prod_{l} c \hspace{0.05 cm} e_{0}^{2}\hspace{0.05 cm} 
e^{-\frac{3}{16} p_{0, \lambda} \hspace{0.05 cm} |\mathrm{P}_{l}|}\Big]. 
\end{aligned}
\end{equation}
Rewrite $\{Y_{j}\}$ and $\{\mathrm{P}_{l}\}$ as ordered collection $\{Y_{1}, \cdots, Y_{n}\}$ and 
$\{\mathrm{P}_{1}, \cdots, \mathrm{P}_{m}\}$,
\begin{equation}
\begin{aligned}
\abs{K^{\prime}(Z, \text{J})} &\leqslant (c_{0}e_{0}^{5-2\epsilon})^{\frac{1}{2}} \hspace{0.05 cm} e^{- (\frac{\kappa^{\prime}}{2}-1) |Z|}
 \Big[\sum_{n=1}^{\infty} \frac{1}{n!} \sum_{\{Y_{1}, \cdots, Y_{n}\} \subset Z}  
\prod_{j=1}^{n} (c_{0}e_{0}^{1-2\epsilon})^{\frac{1}{4}} \hspace{0.05 cm}e^{- \frac{\kappa^{\prime}}{2} |Y_{j}|}\Big] \\
& \hspace{0.05 cm}\Big[\sum_{m=1}^{\infty} \frac{1}{m!} \sum_{\{\mathrm{P}_{1}, \cdots, \mathrm{P}_{m}\} \subset Z} 
\prod_{l=1}^{m} c \hspace{0.05 cm} e_{0}\hspace{0.05 cm} e^{-\frac{3}{16} p_{0, \lambda} \hspace{0.05 cm} |\mathrm{P}_{l}|}\Big]
\end{aligned}
\end{equation}
following Eq 3.60, the first bracketed term is bounded by $e^{c (c_{0}e_{0}^{1-2\epsilon})^{\frac{1}{4}} |Z|}$ and the 
second bracketed term is bounded by $e^{c \hspace{0.05 cm} e_{0} |Z|}$. Then for some $\kappa_{1} < \frac{\kappa^{\prime}}{2} - 1$,
\begin{equation}
\abs{K^{\prime}(Z, \text{J})} \leqslant (c_{0}e_{0}^{5-2\epsilon})^{\frac{1}{2}} \hspace{0.05 cm} e^{-\kappa_{1}|Z|}. 
\end{equation}
\textbf{Lemma 5.2} There exist constants $\beta > 0$ and $\kappa_{2}^{\prime} > 0$ such that
$|K^{\#} (\mathrm{C}, \text{J})| \leqslant c \hspace{0.05 cm} e_{0}^{\beta} \hspace{0.05 cm} e^{-\kappa_{2}^{\prime}|\mathrm{C}|}$.

\textit{Proof} From the definition,  
\begin{equation}
\begin{aligned}
|K^{\#}(\mathrm{C}, \text{J})| &\leqslant \sum_{\cup_{i}\{\text{X}_{i}\} \cup_{k}\{\tilde{\mathrm{Q}}_{k}\} \cup_{m}\{Z_{m}\} = \mathrm{C}} 
\prod_{i} \abs{ f(\text{X}_{i}, \text{J})} \hspace{0.1 cm}   \prod_{m} \abs{ K^{\prime}(Z_{m}, \text{J})}
\prod_{k}\abs{\int \mathcal{D}\Phi_{\tilde{\mathrm{Q}}_{k}}
\sum_{v \hspace{0.05 cm} \text{on} \hspace{0.05 cm} \tilde{\mathrm{Q}}_{k} : dv=0} \rho(\tilde{\mathrm{Q}}_{k}, \text{J})} \\
&\leqslant  \sum_{\cup_{i}\{\text{X}_{i}\} \cup_{m}\{Z_{m}\} \cup_{k}\{\tilde{\mathrm{Q}}_{k}\} = \mathrm{C}} 
\hspace{0.1 cm} \prod_{i} c \hspace{0.05 cm} e_{0} \hspace{0.05 cm} e^{- \gamma^{\prime} |\text{X}_{i}|} \hspace{0.1 cm}
 \prod_{m} (c_{0}e_{0}^{5-2\epsilon})^{\frac{1}{2}} \hspace{0.05 cm} e^{- \kappa_{1} |Z_{m}|}
 \prod_{k} c \hspace{0.05 cm} e_{0}^{2} \hspace{0.05 cm} e^{-\frac{3}{8} p_{\lambda} |\mathrm{Q}_{k}|} \hspace{0.05 cm}  \\
&\hspace{2.1 cm}  
\int  \mathcal{D}\Phi_{\tilde{\mathrm{Q}}_{k}} \hspace{0.05 cm} e^{- (m_{\text{min}} \gamma_{1} - p_{\lambda}^{-1} + c)
 ||\Phi^{\prime}||_{\tilde{\mathrm{Q}}_{k}}^{2}} \\
&\leqslant (c_{0}e_{0}^{11-2\epsilon})^{\frac{1}{2}} \hspace{0.05 cm} e^{- \frac{\kappa_{1}}{2} |\mathrm{C}|}
\sum_{\cup_{i}\{\text{X}_{i}\} \cup_{m}\{Z_{m}\} \cup_{k}\{\tilde{\mathrm{Q}}_{k}\} \subset \mathrm{C}} 
\hspace{0.1 cm} \prod_{i} c \hspace{0.05 cm} e_{0} \hspace{0.05 cm} e^{- \frac{\gamma^{\prime}}{2} |\text{X}_{i}|} \hspace{0.1 cm}
 \prod_{m} (c_{0}e_{0}^{5-2\epsilon})^{\frac{1}{2}} \hspace{0.05 cm} e^{- \frac{\kappa_{1}}{2} |Z_{m}|} \\
&\hspace{2 cm}  \prod_{k} c \hspace{0.05 cm} e_{0}^{2} \hspace{0.05 cm} e^{-\frac{3}{16} p_{\lambda} |\mathrm{Q}_{k}|} \hspace{0.05 cm}  
\int  \mathcal{D}\Phi_{\tilde{\mathrm{Q}}_{k}} \hspace{0.05 cm} e^{- (m_{\text{min}} \gamma_{1} - p_{\lambda}^{-1} + c)
 ||\Phi^{\prime}||_{\tilde{\mathrm{Q}}_{k}}^{2}} \\
&\leqslant (c_{0}e_{0}^{11-2\epsilon})^{\frac{1}{2}} \hspace{0.05 cm} e^{- \frac{\kappa_{1}}{2} |\mathrm{C}|}
\Big(\frac{\pi}{m_{\text{min}} \gamma_{1} - p_{\lambda}^{-1}}\Big)^{\frac{|\mathrm{C}|}{2}} \Big[\sum_{\cup_{i}\{\text{X}_{i}\} \subset \mathrm{C}} 
 \prod_{i} c \hspace{0.05 cm} e_{0} \hspace{0.05 cm} e^{- \frac{\gamma^{\prime}}{2} |\text{X}_{i}|}\Big] \\ 
& \hspace{1 cm} \Big[\sum_{\cup_{m}\{Z_{m}\} \subset \mathrm{C}} \prod_{m} (c_{0}e_{0}^{5-2\epsilon})^{\frac{1}{2}} \hspace{0.05 cm} 
 e^{- \frac{\kappa_{1}}{2} |Z_{m}|}\Big] 
\hspace{0.1 cm} \Big[\sum_{ \cup_{k}\{\mathrm{Q}_{k}\} \subset \mathrm{C}} \prod_{k} c \hspace{0.05 cm} e_{0}^{2} \hspace{0.05 cm} 
e^{-\frac{3}{16} p_{\lambda} |\mathrm{Q}_{k}|}\Big].
\end{aligned}
\end{equation} 
Following the steps of lemma 5.1, the first bracketed term is bounded by $e^{c \hspace{0.05 cm} e_{0}^{\frac{1}{2}} |\mathrm{C}|}$, the
second bracketed term is bounded by $e^{c  (c_{0}e_{0}^{5-2\epsilon})^{\frac{1}{4}} |\mathrm{C}|}$ and the third bracketed term is bounded
by $e^{c \hspace{0.05 cm} e_{0} |\mathrm{C}|}$ . Then for some $\kappa_{2} < \frac{\kappa_{1}}{2} - 1$ and $\beta = \frac{11}{2} - \epsilon$,
\begin{equation}
|K^{\#}(\mathrm{C}, \text{J})| \leqslant c \hspace{0.05 cm} e_{0}^{\beta} \hspace{0.05 cm} e^{-\kappa_{2}|\mathrm{C}|}.
\end{equation}
 
\section{Mass gap}
Let $p_{1}, p_{2} \in \mathrm{C}$ and let $d(x,y) = \sup_{\mu} |x_{\mu} - y_{\mu}|$ be a metric. 
Define $d(p_{1}, p_{2}) = \inf_{\substack{x\in p_{1} \\ y\in p_{2}}} d(x,y)$. Let $|\mathrm{C}| : \# [r_{\lambda}] \hspace{0.1 cm} \Box$
in $\mathrm{C}$. Then
\begin{equation}
|\mathrm{C}| \geqslant \frac{d(p_{1}, p_{2})}{[r_{\lambda}]}.
\end{equation}
\subsection{proof of theorem 2}
Consider two plaquettes $p_{1}$ and $p_{2}$. From Eq 1.19
\begin{equation}
\begin{aligned}
\langle (dA)(p_{1}) (dA)(p_{2})\rangle - & \langle (dA)(p_{1})\rangle \langle (dA)(p_{2})\rangle \\
&= \left. \frac{\partial}{\partial \text{J}_{p_{1}}}\frac{\partial}{\partial \text{J}_{p_{2}}}  
\text{log} \hspace{0.05 cm}  \mathrm{Z}[\text{J}]\right\rvert_{\text{J} = 0} \\
&= \sum_{\mathrm{C}} \left. \frac{\partial}{\partial \text{J}_{p_{1}}}\frac{\partial}{\partial \text{J}_{p_{2}}}  
E(\mathrm{C}, \text{J}) \right\rvert_{\text{J} = 0}.
\end{aligned}
\end{equation}
Now $E(\text{J}) = \sum_{\mathrm{C}} E(\mathrm{C}, \text{J})$ is analytic in J. 
First note that the potential term is linear in J due to which the functions $\rho^{\prime}(\mathrm{P}, \text{J}), \rho(\mathrm{Q}, \text{J})$ 
and $f(X, \text{J})$ are analytic. Then as stated below theorem 3.2, the function $H(Z,\Psi)$ converges in a unit disc in complex plane of J 
and therefore, the function $K(Y, \text{J})$ is also analytic. Then from Eq 5.3 and 5.7 each $E(\mathrm{C}, \text{J})$ is analytic in J,
so when we sum it up, the absolutely convergent series is also analytic in J. The right hand side only depends on
$\text{J}(p), p \in \mathrm{C}$, we can take the derivative term by term but only those terms with non zero support of J
in $\mathrm{C}$ contribute. Thus,
\begin{equation}
\left. \frac{\partial}{\partial \text{J}_{p_{1}}}\frac{\partial}{\partial \text{J}_{p_{2}}}  
E(\text{J}) \right\rvert_{\text{J} = 0} = 
\sum_{\mathrm{C} \supset \{p_{1},p_{2}\}} \left. \frac{\partial}{\partial \text{J}_{p_{1}}}\frac{\partial}{\partial \text{J}_{p_{2}}}  
E(\mathrm{C}, \text{J}) \right\rvert_{\text{J} = 0}. 
\end{equation}
Let $\alpha_{p} = e_{0}^{3 \epsilon} < 1$ denote the bound of $\text{J}_{p}$.
Define a contour $\Gamma_{p}$ to be a circle of radius $\mathcal{O}(e_{0}^{3 \epsilon})$ with center as origin in
complex plane containing $\text{J}_{p}$. By Cauchy integral formula
\begin{equation}
E(\text{J}) = \frac{1}{2\pi i} \oint_{\Gamma_{p}}
\frac{E(\text{J}_{\sim p}, \text{J}^{\prime}_{p})}{(\text{J}_{p} - \text{J}^{\prime}_{p})} \hspace{0.1 cm} d\text{J}^{\prime}_{p}
\end{equation}
where, p denotes a plaquette $p_{1}$ or $p_{2}$ and $\text{J}_{\sim p} = \{\text{J}_{p^{\prime}}\}_{p^{\prime} \neq p}$.
By differentiating the above
\begin{equation}
\frac{\partial}{\partial \text{J}_{p}} E(\text{J}) = 
 \frac{-1}{2\pi i} \oint_{\Gamma_{p}} 
 \frac{E(\text{J}_{\sim p}, \text{J}^{\prime}_{p})}{(\text{J}_{p} - \text{J}^{\prime}_{p})^{2}} \hspace{0.1 cm} d\text{J}^{\prime}_{p}
\end{equation}
and for $\text{J}_{p} = 0$, we get
\begin{equation}
\frac{\partial}{\partial \text{J}_{p}} E(\text{J}) = 
 \frac{-1}{2\pi i} \oint_{\Gamma_{p}} 
 \frac{E(\text{J}_{\sim p}, \text{J}^{\prime}_{p})}{(\text{J}^{\prime}_{p})^{2}} \hspace{0.1 cm} d\text{J}^{\prime}_{p}
\end{equation}
Therefore for plaquette $p_{1}$ and $p_{2}$, we have
\begin{equation}
\frac{\partial}{\partial \text{J}_{p_{1}}}\frac{\partial}{\partial \text{J}_{p_{2}}} E(\text{J}) =  \frac{1}{(2\pi i)^{2}}
\oint_{\Gamma_{p_{1}}} \frac{ d\text{J}^{\prime}_{p_{1}}}{(\text{J}^{\prime}_{p_{1}})^{2}}
\oint_{\Gamma_{p_{2}}} \frac{ d\text{J}^{\prime}_{p_{2}}}{(\text{J}^{\prime}_{p_{2}})^{2}}
\hspace{0.1 cm} E(\text{J}_{\sim (p_{1}p_{2})}, \text{J}^{\prime}_{p_{1}}, \text{J}^{\prime}_{p_{2}}).
\end{equation}
From Eq 6.2, 6.3 and 6.7,
\begin{equation}
\begin{aligned}
\langle (dA)(p_{1}) (dA)(p_{2})\rangle & -  \langle (dA)(p_{1})\rangle \langle (dA)(p_{2})\rangle \\
&= \sum_{\mathrm{C} \supset \{p_{1},p_{2}\}} \frac{1}{(2\pi i)^{2}}
\oint_{\Gamma_{p_{1}}} \frac{ d\text{J}^{\prime}_{p_{1}}}{(\text{J}^{\prime}_{p_{1}})^{2}}
\oint_{\Gamma_{p_{2}}} \frac{ d\text{J}^{\prime}_{p_{2}}}{(\text{J}^{\prime}_{p_{2}})^{2}}
\hspace{0.1 cm} E(\mathrm{C}, \text{J}) \\  
&\leqslant \sum_{\mathrm{C} \supset \{p_{1},p_{2}\}} \frac{1}{|\alpha_{p_{1}}|}\frac{1}{|\alpha_{p_{2}}|} 
\abs{E(\mathrm{C}, \text{J})}  \\
&\leqslant  \frac{1}{|\alpha_{p_{1}}|}\frac{1}{|\alpha_{p_{2}}|} e^{- \frac{\kappa_{2}}{2 [r_{\lambda}]} \hspace{0.05 cm} d(p_{1},p_{2})}
\sum_{\mathrm{C} \supset \{p_{1},p_{2}\}} c \hspace{0.05 cm} e_{0}^{\beta} \hspace{0.05 cm} e^{-\frac{\kappa_{2}}{2}|\mathrm{C}|} \\
&\leqslant c \hspace{0.1 cm} e_{0}^{\beta- 6\epsilon} \hspace{0.1 cm}  e^{- \frac{\kappa_{2}}{2 [r_{\lambda}]} \hspace{0.05 cm} d(p_{1},p_{2})}
\end{aligned}
\end{equation}
where, we have used Eq 5.11 and lemma 5.2. The second last step follows from Eq 6.1 and the last step follows from lemma 25 in \cite{D}.
\subsection{n point truncated correlation}
Let $\mathcal{P} = \{p_{1}, p_{2}, \cdots, p_{n}\}$ be a set of plaquettes. Each plaquette $p_{i} \in \mathcal{P}$ 
carries a current $\text{J}_{p_{i}}$. Denote $(dA) (p_{i}) \equiv dA_{p_{i}}$.

\textbf{Theorem 6.1} Define the n - point truncated correlation function as
\begin{equation}
\begin{aligned}
\left. \frac{\partial^{n}}{\partial\text{J}_{p_{1}} \partial\text{J}_{p_{2}} \cdots \partial\text{J}_{p_{n}}} 
\text{log} \hspace{0.05 cm}  \mathrm{Z}[\text{J}]\right\rvert_{\text{J}_{p_{i}} = 0} &=
\langle dA_{p_{1}}, dA_{p_{2}}, \cdots, dA_{p_{n}}\rangle^{T} \\
&= \langle dA_{p_{1}}, dA_{p_{2}}, \cdots, dA_{p_{n}}\rangle \hspace{0.1 cm} - \\ &
 \sum_{\substack{ \text{partitions} \\ P_{i} \text{of}  1,\cdots , n \\ (\pi_{1}, \pi_{2}, \cdots , \pi_{k})}} 
\bigg\langle \prod_{P_{i} \in \pi_{i}} dA_{p_{i}} \bigg\rangle \cdots  \bigg\langle \prod_{P_{i} \in \pi_{k}} dA_{p_{i}} \bigg\rangle
 \end{aligned}
\end{equation}
Then 
\begin{equation}
\langle dA_{p_{1}}, dA_{p_{2}}, \cdots, dA_{p_{n}}\rangle^{T} \leqslant c \hspace{0.1 cm} 
 e_{0}^{\beta - 3n\epsilon} e^{-\frac{\kappa_{2}}{2 [r_{\lambda}]} \hspace{0.5 mm} t(p_{1}, p_{2},\cdots, p_{n})}
\end{equation}
where $t(p_{1}, p_{2},\cdots, p_{n})$ is the length of the shortest tree connecting all the plaquettes.
\textit{Proof}  
\begin{equation}
\left. \frac{\partial^{n}}{\partial\text{J}_{p_{1}} \partial\text{J}_{p_{2}} \cdots \partial\text{J}_{p_{n}}} 
\text{log} \hspace{0.05 cm}  \mathrm{Z}[\text{J}]\right\rvert_{\text{J}_{p_{i}} = 0}
= \left. \sum_{\mathrm{C}}\frac{\partial^{n}}{\partial\text{J}_{p_{1}} \partial\text{J}_{p_{2}} \cdots \partial\text{J}_{p_{n}}} 
 E(\mathrm{C}, \text{J})\right\rvert_{\text{J}_{p_{i}} = 0}.
\end{equation}
Set the bound of  $\text{J}_{p_{i}}$ to be $\alpha_{p_{i}} = e_{0}^{3\epsilon} < 1$.
Define a contour $\Gamma_{p_{i}}$ to be a circle of radius $e_{0}^{3\epsilon}$ with center as origin in
complex plane containing $\text{J}_{p_{i}}$. Following Eq 6.3 and 6.7
\begin{equation}
\begin{aligned}
\langle dA_{p_{1}}, dA_{p_{2}}, \cdots, dA_{p_{n}}\rangle^{T} 
&= \sum_{\mathrm{C} \supset \{p_{1}, \cdots, p_{n}\}} \frac{1}{(2\pi i)^{n}}
\oint_{\Gamma_{p_{1}}} \frac{ d\text{J}^{\prime}_{p_{1}}}{(\text{J}^{\prime}_{p_{1}})^{2}} \cdots 
\oint_{\Gamma_{p_{n}}} \frac{ d\text{J}^{\prime}_{p_{n}}}{(\text{J}^{\prime}_{p_{n}})^{2}}
\hspace{0.1 cm} E(\mathrm{C}, \text{J}) \\  
&\leqslant \sum_{\mathrm{C} \supset \{p_{1},\cdots, p_{n}\}} \frac{1}{|\alpha_{p_{1}}|} \cdots \frac{1}{|\alpha_{p_{n}}|} 
\abs{E(\mathrm{C}, \text{J})}  \\
&\leqslant  \frac{1}{|\alpha_{p_{1}}|} \cdots \frac{1}{|\alpha_{p_{n}}|} 
e^{- \frac{\kappa_{2}}{2[r_{\lambda}]} \hspace{0.05 cm} t(p_{1},\cdots, p_{n})}
\sum_{\mathrm{C} \supset \{p_{1}, \cdots, p_{n}\}}   e_{0}^{\beta} \hspace{0.05 cm} e^{- \frac{\kappa_{2}}{2} |\mathrm{C}|} \\
&\leqslant  c \hspace{0.1 cm} e_{0}^{\beta - 3n\epsilon} \hspace{0.1 cm}  
e^{- \frac{\kappa_{2}}{2[r_{\lambda}]} \hspace{0.05 cm} t(p_{1}, \cdots, p_{n})}.
\end{aligned}
\end{equation}
where, we have used $|\mathrm{C}| \geqslant \frac{t(p_{1}, \cdots, p_{n})}{[r_{\lambda}]}$. 

\textit{Remark 6}. Let $\gamma_{1}$ and $\gamma_{2}$ be two closed curves composed of lattice bonds and 
$W_{\gamma_{i}}(A)$ be a Wilson loop. Let $t_{i}$ be a source function defined on bonds forming $\gamma_{i}$. Then theorem 2
is also true for the observable $e^{\sum_{i=1}^{2} t_{i}W_{\gamma_{i}}(A)}$.

\textit{Acknowledgement}. I would like to thank my advisor Jonathan Dimock for his guidance.


\begin{thebibliography}{99}

\bibitem{BBIJ}
  
  {Balaban, T., Imbrie, J., Jaffe, A. and Brydges, D.},
  {The Mass Gap for Higgs Models on a Unit Lattice }
  {Annals of Physics},
    {158},
    {2},
    {281-319},
    {13},
   {(1984)},
    {0003-4916}.
   
   
 \bibitem{BFKT}

{Balaban, T., Feldman, J., Kn$\ddot{\text{o}}$rrer, H. and Trubowitz, E.},
{Power Series Representations for Bosonic Effective Actions}, {J Stat Phys (2009)}, {134: 839}.    


\bibitem{BIJ2}
 {Balaban, T., Imbrie, J. and Jaffe, A.}, {Renormalization of the Higgs Model: Minimizers, Propagators, and the Stability of Mean Field Theory},
 {Commun. Math. Phys. 97, 299-329 (1985)}.

\bibitem{BIJ3}
{Balaban, T., Imbrie, J. and Jaffe, A.} {Effective Action and Cluster Properties of the Abelian Higgs Model}, 
 {Commun. Math. Phys. 114, 257-315 (1988)}.

\bibitem{KK} 

 {Kennedy, Tom and King, Chris}, {Spontaneous Symmetry Breakdown in the Abelian Higgs Model},
 {Commun. Math Phys.},
 {104},
 {2},
 {327--347}, {(1986)},
 {1432-0916}.

\bibitem{D1} 

 {Dimock, J.}, {The Renormalization Group according to Balaban - II. Large Fields}
 {Journal of Mathematical Physics},
 {(2013)},
 {54},
 {092301}.
  
 \bibitem{B}  
 {Balaban, T.}
 {Localization expansions I. function of the background configurations.}
  {Commun. Math. Phys.} {182: 33-82}, {(1996)}.
 
 \bibitem{D} 

  {Dimock, J.}, {The Renormalization Group according to Balaban - I. Small Fields}
 {Rev. Math. Phys},
 {(2013)},
 {25},
 {1330010}.
  
 
 \end{thebibliography}
\end{document}